\documentclass[a4paper,fleqn,usenatbib,useAMS]{mnras}

\usepackage{graphicx}	
\usepackage{amsmath}	
\usepackage{amssymb}	
\usepackage{multicol}        
\usepackage{bm}		

\newcommand{\msun}{\,M$_{\sun}$}

\usepackage[T1]{fontenc}
\usepackage{ae,aecompl}

\title[L1188]{The young star population of L1188}

\author[E. Szegedi-Elek et al.]{E. Szegedi-Elek$^{1}$\thanks{Contact e-mail: \href{mailto:elek.elza@mcsfk.mta.hu}{elek.elza@csfk.mta.hu}}, M. Kun$^{1}$, A. Mo\'or$^{1}$, G. Marton$^{1}$ and B. Reipurth$^{2}$
\\
$^{1}$Konkoly Observatory, Research Centre for Astronomy and Earth Sciences, Hungarian Academy of Sciences,\\ H-1121 Budapest, Konkoly Thege \'ut 15--17, Hungary\\
$^{2}$Institute for Astronomy, University of Hawaii at Manoa, 640 N. Aohoku Place, Hilo, HI 96720, USA}
\date{Last updated 2015 May 22; in original form 2013 September 5}

\pubyear{2017}

\begin{document}
\label{firstpage}
\pagerange{\pageref{firstpage}--\pageref{lastpage}}
\maketitle

\begin{abstract}
We present new results on the young star population of the Lynds~1188 molecular cloud,  associated with the Cepheus Bubble, a giant interstellar shell around the association Cep~OB\,2. In order to reveal the star-forming scenario of the molecular cloud located on the supershell, and understand the history of star formation in the region, we identified and characterized young star candidates based on an H$\alpha$ emission survey and various published photometric datasets. Using Gaia DR2 astrometry we studied the spatial distribution of the young star
candidates and isolated three groups based on their distances. We constructed spectral energy distributions of our target stars, based on Pan-STARRS, 2MASS, {\it Spitzer}  and {\it WISE}  photometric data, estimating their spectral types, extinctions, and luminosities. We estimated masses by means of pre-main-sequence evolutionary models, and derived accretion rates from the equivalent width of the H$\alpha$ line. We studied the structure of the cloud by constructing a new extinction map, based on Pan-STARRS data. Our results show that the distribution of low-mass young stars in L1188 is well correlated with that of the dust and mole\-cular gas. We identified two small, compact clusters and a loose aggregate of young stars. We found that star formation in L1188 started about 5 million years ago. The apparent age gradient of young stars across the cloud and  the ammonia cores located to the east of the optically visible young stellar groups  support
the scenario of star formation propagating away from the centre of the Cepheus Bubble.
\end{abstract}





\section{Introduction}
Massive stars affect their environment during their lifetime by ionizing radiation, stellar winds, and at the end of their lives by supernova explosions. During these energetic  processes fast-moving shock waves move into the surrounding interstellar medium,  creating a giant bubble, filled with hot and tenuous gas, and bordered by the ambient interstellar clouds, interacting with the expanding shock wave. The comp\-ression from the shock wave may trigger formation of new stars in the clouds of the bubble shells on timescales of a few tens of million years \citep{ab1}. Giant ring-like structures are common in the cold interstellar matter of our Galaxy \citep[e.g.][]{KMT2004,EP2013,simpson2012}, indicating long-term effects of high-mass stars on their large-scale environments. Propagation of star formation due to expanding bubbles have been observed in several nearby OB associations \citep[e.g.][]{PZ1999,Bally2008}.

The Cepheus Bubble \citep{bubi} is a ring-shaped infrared emission region with a diameter of 10\degr\ around the Cep~OB\,2 association, at a distance of 800--900~pc from the Sun \citep{cep-tav}. CO observations by \citet{patel} revealed some $10^{5}$\,\msun\ mole\-cular gas on the periphery of the Cepheus Bubble. They proposed that a supernova exploded some 1.7 million years ago in the Str\"omgren sphere--stellar-wind cavity of the oldest subsystem of Cep~OB\,2. Whereas the currently observed OB stars of the association \citep[e.g. the exciting stars of IC\,1396,][]{aurora} might have been born in clouds swept up by the expanding ionization front, the explosion of a high-mass star initiated the birth of the third generation of stars of Cep~OB\,2 in the molecular clouds of the bubble shell. 

Lynds~1188 (L1188) is  one of the star-forming mole\-cular clouds on the periphery of the Cepheus Bubble, at ($\it{l}$,$\it{b}$,)=(105.6$\degr$, 4.2$\degr$). Two red and nebulous objects, RNO~140 and RNO~141 \citep{cohen1980}, a reflection nebula, DG~180 \citep{dg}, as well as six \textit{IRAS\/} sources  with protostellar colour indices \citep{abraham} are associated with L1188. These objects indicate recent star formation in the direction of  the cloud. 

The $\sp{13}$CO observations of L1188 by \citet{abraham} revealed a total mass of some 1800\,M$\sb{\odot}$, distributed in six $\sp{13}$CO clumps. They discovered a dust filament stretching from S140/L1204 toward the L1188 on the 100-\micron~optical depth map constructed using {\it  IRAS} data, indicating that both regions are physically connected. Furthermore they identified fifteen H$\alpha$ emission objects.
Three dense NH$\sb{3}$ cores were identified in L1188 through an NH$\sb{3}$ survey presented by \citet{verebelyi}. These ammonia cores are encompassed by three known $\sp{13}$CO clumps. 

 \citet{alap} selected 73 potential young stellar objects (YSOs) in the mid-infrared images of the L1188 region, obtained by the {\it Spitzer Space Telescope} \citep{werner2004} during the {\it Spitzer} Galactic First Look Survey (\url{http://ssc.spitzer.caltech.edu/fls/galac/}). They identified three groups of potential YSOs, and proposed that two of them belong to L1188, whereas the third one is probably a distant star-forming region related to the Perseus spiral arm. 

Based on new $\sp{12}$CO, $\sp{13}$CO and C$\sp{18}$O observations, \citet{gong} revealed  that L1188 consists of two nearly orthogonal filamentary molecular clouds at two clearly separated velocities.  \citet{gong} found enhanced star formation activity in the intersection region, and proposed  that star formation is triggered by the collision of molecular clouds in L1188.

To identify and characterize the young star population of L1188, we performed a new H$\alpha$ survey to detect stars showing H$\alpha$ in emission. The earlier objective prism photographic plates were less sensitive \citep{abraham}. To examine spectral energy dist\-ributions (SEDs), we utilized available photometric data in the Pan-STARRS \citep{panstarrs} and IPHAS  databases \citep{iphas}, as well as 2MASS \citep{cutri2003},  {\it Spitzer}  \citep{werner2004} and  {\it WISE}  \citep{cutri2013}.  We selected further candidate YSOs based on 2MASS, {\it WISE}, and {\it Spitzer} colour indices. The \textit{Gaia}~DR2 data allowed us to distinguish members of the L1188 population from foreground and background YSOs. Our data reduction and target selection are described in Section \ref{Sect_2}. The applied methods are given in Section~\ref{sect_3}. The results and their discussions are presented in Section \ref{sect_4}. A brief summary is given in Section \ref{summary}.

\section{Observations, data reduction and YSO selection}
\label{Sect_2}
\subsection{Observations}
We observed L1188 with the Wide Field Grism Spectrograph 2 (WFGS2), installed on the University of Hawaii 2.2~m telescope, on 2012 July 27 and 29, and August 10 and 11. We used a 300 line mm$\sp{-1}$ grism, blazed at 6500~\AA\ and provi\-ding a dispersion of 3.8~\AA\,pixel$\sp{-1}$ and a resolving power of 820. The narrow band H$\alpha$ filter had a 500~\AA\ passband, centered near 6515~\AA. The detector for WFGS2 was a Tektronix 2048$\times$2048 CCD, whose pixel size of 24\,$\mu$m corresponded to 0.34~arcsec on the sky. The field of view was 11.5$\sp{\prime}\times11.5\sp{\prime}$. We covered an area of 60$\times$ 50 arcmin, centered on R.A.(2000) = 22$\mathrm{\sp{h}}$ 18$\mathrm{\sp{m}}$ and Dec(2000)= 61$\degr$ 42$\arcmin$, with a mosaic of 30 overlapping fields. During the grism observation we took a short, 60\,s exposure for each field in order to identify the H$\alpha$ line in the spectra of bright stars, and avoid saturation. Then we obtained three frames with 300\,s exposure time. Direct images with the same instrument were obtained through {\it $r\sp{\prime}$} and {\it $i\sp{\prime}$} filters before the spectroscopic exposures. One exposure was taken in each filter with integration time of 60\,s.

Bias subtraction and flat-field correction of the images were done in IRAF. Then we used the FITSH, a software package for astronomical image processing \citep{pal2012}, to remove cosmic rays, coadd long-exposure images, identify the stars on the images, and transform the pixel coordinates into equatorial coordinate system and determine equivalent width of H$\alpha$.


We examined the co-added image obtained with the slitless spectrograph visually to discover new young stars showing the H$\alpha$ line in emission. We identified 76 stars with H$\alpha$ emission in the observed region. We determined their equatorial coordinates by matching our images with a 2MASS \citep{cutri2003} image of the fields. All H$\alpha$ emission sources have 2MASS counterparts unambiguously within 1.4$\arcsec$, thus we use 2MASS designations of the stars for equatorial coordinates. The equivalent width of the H$\alpha$ emission line and its uncertainty were computed in the same way described by \citet{szegedi}. Due to the faint continuum or overlapping spectra we could measure EW(H$\alpha$) only in 26 cases. We could recover 13 of the 18 sources, listed in the IPHAS catalogue of H$\alpha$ emission-line sources \citep{witham2008} for the same area.


  \subsubsection{Photometric data}

To find candidate YSOs and then study the evolutionary stage of the selected candidate YSOs we supplemented our data with photometric data available in public databases.

The 2MASS survey \citep{2mass_ref} uniformly scanned the entire sky in three near-infrared bands {\it J, H, K$\mathrm{\sb{S}}$ } to detect and characterize point sources brighter than about 1 mJy in each band, with signal-to-noise ratio greater than 10, using a pixel size of 2.0$\arcsec$.

NASA's Wide-field Infrared Survey Explorer (WISE, \citep{wise_ref}) mapped the whole sky at 3.4, 4.6, 12, and 22 $\micron$ (W1, W2, W3, W4) in 2010 with an angular resolution of 6.1$\arcsec$, 6.4$\arcsec$, 6.5$\arcsec$, and 12.0$\arcsec$ in the four bands. WISE achieved 5$\sigma$ point source sensitivities better than 0.08, 0.11, 1 and 6 mJy in unconfused regions in the four bands. 

The Isaac Newton Telescope (INT) Photometric H$\alpha$ Survey of the Northern Galactic Plane (IPHAS) \citep{iphas} observed the Northern Milky Way in visible light (H$\alpha$, r$\sp{\prime}$, i$\sp{\prime}$) down to >20th magnitude, using the INT in La Palma. 

A small stripe of our target field (see Fig.~\ref{distribution}) was also observed by the {\it Spitzer Space Telescope}  \citep{werner2004} as part of the {\it Spitzer} First Look Survey (\url{http://ssc.spitzer.caltech.edu/fls/galac/}). The original goals of this survey were to characterize the cirrus and background source counts at low Galactic latitudes and the internal cirrus and background source counts toward a molecular cloud. 
Observations were performed on 2003 December 6 with the Infrared Array Camera \citep{irac} at 3.6, 4.5, 5.8 and 8\,$\micron$ and on 2003 December 8 at 24 and 70 $\micron$ with the Multiband Imaging Photometer \citep{mips} (aors 4959744 and 4961536).

{\it AKARI} was a satellite \citep{akari} dedicated for infrared astronomical observations. It was capable of observing across the 2-180 $\micron$ near-infrared to far-infrared spectral regions, with two focal-plane instruments: the InfraRed Camera (IRC) \citep{onaka2007} and the Far Infrared Surveyor (FIS) \citep{kawada2007}. 

The \textit{Midcourse Space Experiment (MSX)} \citep{msx2} surveyed the entire Galactic plane within |b| $>$ 5$\degr$ in four mid-infrared spectral bands between 6 and 25 $\micron$ at a spatial resolution of $\sim$ 18.$\arcsec$3.

\subsection{YSO selection}
\label{sect_3_2}

 Young stellar objects, bearing a circumstellar disc or embedded in a protostellar envelope, are located at specific regions in near- and mid-infrared colour-colour diagrams. Colour--colour diagrams can be used for identifying embedded protostars (Class~I infrared sources) and classical T~Tauri stars (Class~II sources), defined primarily on the basis of spect\-ral slopes over the 2--24\,\micron\ wavelength interval \citep{lada,greene}
The strong H$\alpha$ emission of T~Tauri stars originates from the interaction of the star and its accretion disc. To characterize the circumstellar environment of our H$\alpha$ emission stars, and find further candidate YSOs in the region we examined the infrared sources in the {\it Spitzer}, {\it WISE}, and 2MASS data bases.

\subsubsection{{\it Spitzer}}
We examined all sources within the field of view of our WFGS2 measurements to search for additional young stars. We downloaded {\it Spitzer} magnitudes from the  {\it Spitzer} Enhanced Imaging Products (SEIP) Source List. 

 \citet{gutermuth2009} described a classification scheme to distinguish YSOs based on {\it Spitzer} colour indices. In the selection method one performs straight cuts in colour-colour and colour-magnitude diagrams. During this method contaminants (PAH, AGN, shock, more evolved field stars ...) can be eliminated in multiple phases and Class I and Class II sources can be separated.


We applied the methods described by \citet{gutermuth2009} to remove probable  extragalactic, stellar, and interstellar contaminants and select candidate YSOs based on colour indices. We identified 32 Class~II and 5 Class~I sources, detected in all of the IRAC bands. 

\begin{figure}
 \includegraphics[width=\columnwidth]{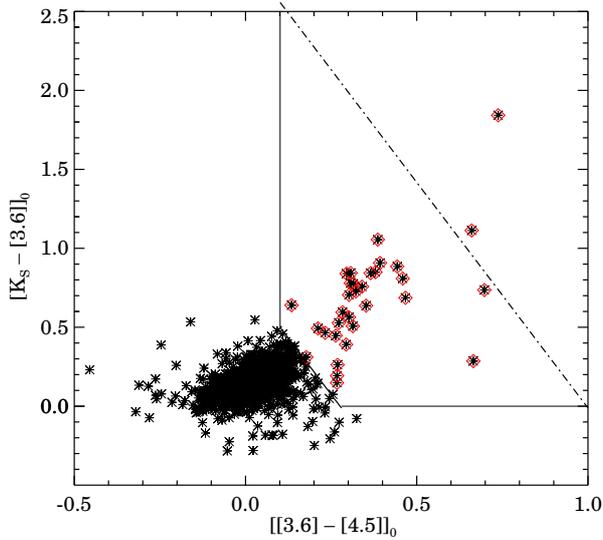}
 \caption{[$\mathrm{\it {K\sb{S}}}$-$[3.6]]\sb{0}$ vs. [$[3.6]-[4.5]]\sb{0}$ colour-colour diagram of all {\it Spitzer}  sources coinciding with 2MASS sources. All of the {\it Spitzer} sources are brighter than [3.6]=14.0 mag and [4.5]=14.0 mag in the studied area. Solid lines confine the area inhabited by candidate YSOs \citep{gutermuth2009}. The dash-dotted line distinguish   Class I protostars from Class II pre-main-sequence stars in accordance with \citet{gutermuth2009} \em Phase 2 \em method. Red diamonds represent candidate YSOs identified by this method.}
 \label{spitzer}
\end{figure}

We combined the 2MASS magnitudes with the  {\it Spitzer}  [3.6] and [4.5] magnitudes (Fig.~\ref{spitzer}) and applied also the \em Phase 2 \em criteria established by \citet{gutermuth2009}. To exclude the dim extragalactic contaminations we used a [3.6] and [4.5] $<$ 14 magnitude limit. With this method we recovered 32  additional candidate T~Tauri stars.

In the last step we added [24]~mag to the examination as mentioned by \citet{gutermuth2009} in \em Phase 3 \em method. With this step we found 13 additional young stars.
In total we added 35 new young star candidates to our list.
 We note that the high-sensitivity \textit{Spitzer\/} observations cover less than one third of the area observed by the WFGS2. Most of our H$\alpha$ emission stars are located outside the \textit{Spitzer\/} field of view, thus probably a large number of faint protostellar and pre-main-sequence members of L1188 is still undetected. 


At 70~$\micron$ photometry for the selected sources was carried out on the pipeline processed (post-BCD) filtered images that were downloaded from the {\it  Spitze}r Heritage Archive. In total 64 sources of all the young star candidates were situated in the field of view of  {\it Spitzer}. We detected point-like sources with a level of $\geq$4$\sigma$ at five out of our 64 targeted positions. In the case of SSTSL2\,J221653.13+614315.5 and SSTSL2\,J221722.25+614305.6 where the detected sources are blended by nearby bright objects we used PSF photometry to derive the 70 {\micron} flux densities. For the other three targets as well as for the undetected sources we performed 
aperture photometry using an aperture radius of 8{\arcsec} and a background annulus between 39{\arcsec} and 65{\arcsec}. For the detected sources the derived 
centroid was used as the centre of the aperture, while for non-detections the positions of the 2MASS or {\it WISE} counterparts were utilized as target coordinates. 
To account for the flux outside the aperture we applied aperture correction using a correction factor approp\-riate for a 60\,K blackbody. 
The final uncertainty of the photometry was derived as the quadratic sum of the measurement error and the absolute calibration uncertainty, (7\%, MIPS Instrument Handbook). 
For the undetected targets 3\,$\sigma$ upper limits were 
computed. For two targets (SSTSL2 J221643.39+612417.5 and SSTSL2 J221709.59+614111.5), which are associated with extended emission, no photo\-metry was derived. The results of the 70 $\micron$ photometry are listed in Table~\ref{phot}.

\begin{table}
\caption{{\it Spitzer} 70 $\micron$ photometry}
\label{phot}
\begin{tabular}{cc}
\hline
Name &70 $\micron$\\
&(mJy)\\
\hline
 SSTSL2 J221521.04+614827.3 & 423.0  $\pm$ 31.2 \\
 SSTSL2 J221619.80+612620.6 &  321.5 $\pm$  46.8 \\
 SSTSL2 J221653.13+614315.5 &  208.6 $\pm$  37.8  \\
 SSTSL2 J221702.70+611503.6 & 205.9  $\pm$ 33.7 \\
 SSTSL2 J221722.25+614305.6 & 1037.2 $\pm$ 119.6 \\

\hline
\end{tabular}
\end{table}

\subsubsection{{\it WISE}}
\citet{marton2016} conducted a comprehensive all-sky search for candidate YSOs in the All{\it WISE} data release applying the Support Vector Machine
method. Forty-eight sources of their candidate YSOs \citep{marton2016} can be found in our studied region. Twenty-seven of them coincide with sources classified as YSOs during the previous steps, thus twenty-one sources are new young star candidates.

\subsubsection{2MASS}

\begin{figure}
 \includegraphics[width=\columnwidth]{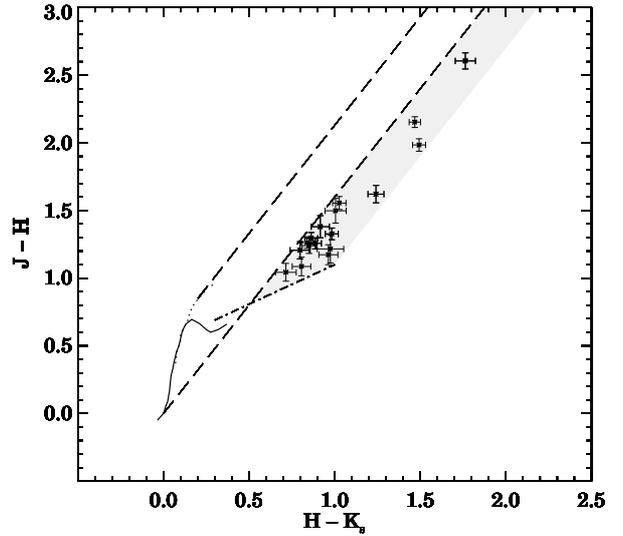}
 \caption{{\it J$-$H} vs. {\it H$-$K$\mathrm{\sb{S}}$} colour-colour diagram of all 2MASS young star candidates which have at least B quality magnitudes in the 60$\arcmin$ $\times$ 50$\arcmin$ vicinity of L1188.}
 \label{2mass}
\end{figure}

We used the 2MASS All-Sky Catalog \citep{2mass_ref} to identify classical T~Tauri stars in the observed region. Figure~\ref{2mass} shows the ({\it J$-$H}) vs. ({\it H$-$K$\sb{\mathrm{S}}$}) colour-colour diagram for all 2MASS young star candidates in the studied region. 
The solid curve shows the colours of the zero-age main sequence, and the dotted line represents the giant branch \citep{loci}. The long-dashed lines delimit the area occupied by the reddened normal stars \citep*{cardelli}. The dash-dotted line is the locus of unreddened T~Tauri stars \citep{meyer1997}, and the gray shaded band  borders the area of the reddened {\it K$\sb{\mathrm{S}}$}-excess stars. YSO candidates in this diagram are sources having photometric quality flags A or B and whose error bars are entirely in the grey region. Four sources show unusually high infrared excess, so we examined these sources individually (namely 2MASS J22161983+6126204, 2MASS J22182711+6141567, 2MASS J22190554+6136156 and 2MASS 22142611+6127246). 2MASS J22182711+6141567 \citep{szen1,szen2}, and 2MASS J22161983+6126204 (H$\alpha$42) are known carbon stars  so we rejected them from the YSO candidate list. Taking into account also the overlapping among the methods we got three new young star candidates in total. 

\subsubsection{{\it Gaia} DR2}
\citet{dist} derived distances for 1.3 billion stars of {\it Gaia} DR2 \citep{gaia1,gaia2} using a weak distance prior that varies smoothly as a function of Galactic longitude and latitude. We adopted distances for the young star candidates from \citet{dist}. Based on these data the sources can be divided into three main groups: there is a small foreground  population of 9 young stars,  45 sources are much farther than the published distance of L1188, and  66 stars are in the published distance of L1188.  No distance is available for 15 stars, so we did not include them directly in the young star candidate list. The average distance of the stars associated with L1188 is 855$\sp{+208}\sb{-107}$ pc.

\subsubsection{Final list of candidate YSOs}

 In the last step we investigated H$\alpha$ sources in the distance of L1188, whose EW are unknown or their SED classification is impossible because of the lack of inf\-rared data. Three of the H$\alpha$ stars are classified as galaxies based on IPHAS images (J22143936+6126174, J22150668+6132270 and J22193369+6137391), so we rejected them as YSO candidates. The other stars without infrared excess seem to be possible YSO candidates based on IPHAS colour-magnitude diagram. We also checked Class~I sources without unknown distances and based on the surface distribution three of them maybe associated with L1188 (see Fig~\ref{diagram}).

Table \ref{halfa_tab} lists the H$\alpha$ sources related to L1188. In the first column the identifier, in the second column the 2MASS designations can be found and measured EW(H$\alpha$) e\-qui\-va\-lent widths are in the third column. For comparison, we list EWs estimated from the IPHAS {\it $r\sp{\prime}$} $-$ [H$\alpha$] vs. {\it $r\sp{\prime}$ $-$ $i\sp{\prime}$} \citep{barentsen} in the fourth column. {\it$r\sp{\prime}$} magnitudes are listed in column 5. The last column of Table \ref{halfa_tab} shows the shape of the SED (see Sect. \ref{sect_3}).

Table \ref{infra_table} lists all of the sources identified based on inf\-ra\-red excess. 2MASS designations,  {\it Spitzer}  or  {\it WISE} names can be found in the first column. The second column of Table \ref{infra_table} shows the shape of the SED (see Sect. \ref{sect_3}).

Table \ref{possible} lists three young star candidates, whose distance is unknown, but their location suggests a possible connection to L1188.

Table \ref{kimaradt} lists all of the young star candidates, which do not belong to L1188 based on their distances, or their distance is unknown.

Hereinafter we only work with 63 sources associated with L1188 (41 H$\alpha$ sources and 22 infrared excess stars).

\begin{figure}
\begin{center}

 \includegraphics[width=\columnwidth]{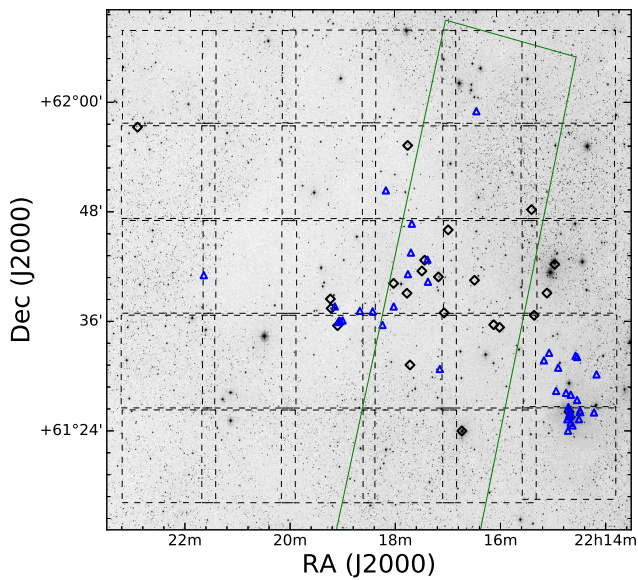}
 \caption{Candidate young stars identified during the present
survey, overplotted on the DSS2 red image. Blue triangles represent H$\alpha$ emission stars, while black diamonds indicate candidate PMS stars exhibiting excess emission in the infrared regime. The area  bordered  by  the dashed squares was the target of our H$\alpha$ survey, while the solid green rectangular indicates the boundary of the region studied by the  {\it Spitzer} .}\label{distribution}

\end{center}

\end{figure}
The surface distribution of the candidate YSOs overplotted on DSS2 red image, can be seen in Fig. \ref{distribution}. Blue triangles show H$\alpha$ emission sources, while black diamonds represent sources identified based on infrared excess. The clustering of H$\alpha$ emission stars near the edge of the observed area suggests a possible more widely spread young star population. 

 
\section{Analysis}
\label{sect_3}
\subsection{Spectral Energy Distribution} 

The available photometric data allowed us to construct the spectral energy distributions (SEDs) of all selected candidate young  stars over the 0.36$-$24 (70)~$\micron$ wavelength region. 

We estimated the spectral type and the extinction of each selected young star, which had measured magnitudes with error in each Pan-STARRS band ({\it u,g,r,i,z}) and the 2MASS {\it J} magnitude had at least B quality. We applied a reduced $\chi\sp{2}$ method,\begin{equation}
   \chi\sp{2} =\sum_{j}\frac{(F\sb{\rm{observed}}-F\sb{\rm{model}})\sp{2}}{\sigma\sp{2}},
	\label{eq:chi}
\end{equation}
  comparing the optical-near infrared SED (from the {\it B} to the {\it J}-band (F$\sb{\rm{observed}}$)) with those of a grid of reddened photospheres, using the reddening-free colour indices of \citet{mamajack} (F$\sb{\rm{model}}$), the extinction law of \citet{cardelli}, and the {\it A$\sb{V}$} $>$  1~mag restriction \citep{gong}.   We disregarded the effect of veiling because \citet{veiling} [see their  Fig. 4] have shown that stars with low accretion-rate (see \ref{accretion}) show very low veiling values.
To compare Pan-STARRS colour indices with reddening-free colour indices of \citet{mamajack}, we transformed Pan-STARRS magnitudes ({\it u,g,r,i,z}) into the Johnson/Cousins system ({\it B,V,R$\rm{\sb{C}}$, I$\rm{\sb{C}}$}) using the transformation equations presented by \citet{pan}. We found the best fit of the photometry and photospheric colours using the reddening law {\it R$\sb{V}$} = 3.1 for {\it A$\sb{V}$} $<$ 2.0, and  {\it R$\sb{V}$} = 5.5 for {\it A$\sb{V}$} $>$  2.0. 
The estimated spectral types are ranging from F1 to M2. The most common spectral type is K4.
We estimate the accuracy of the resulting spectral type and extinction as $\pm$2 subclass and $\pm$0.5 mag, respectively.

\subsection{Extinction mapping}

\begin{figure*}
\begin{center}

 \includegraphics[scale=0.8]{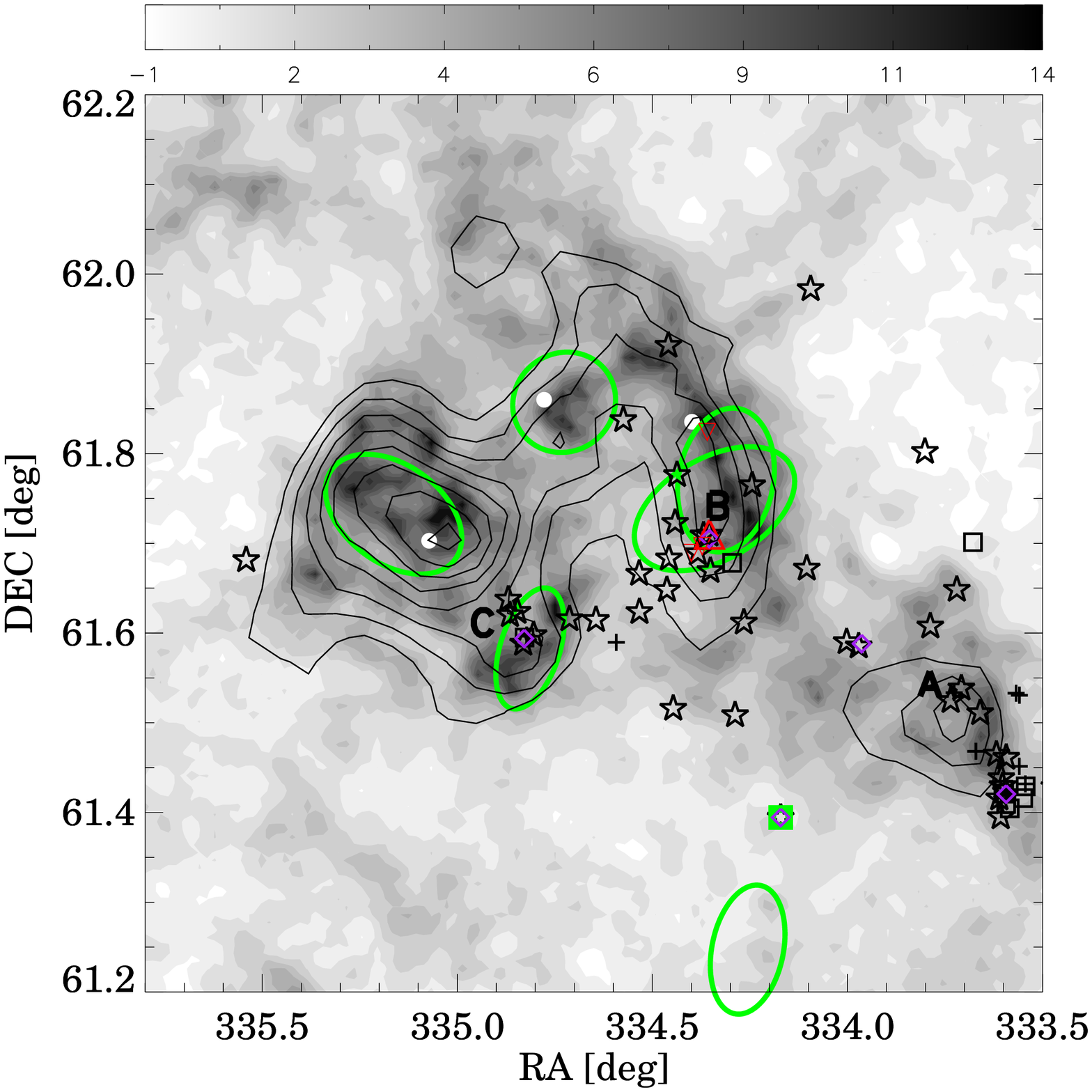}\vspace{3mm}
  \caption{ Distribution of candidate young stellar objects and prestellar clumps and cores, overplotted on the visual extinction map of L1188, based on star counts in the Pan-STARRS DR1. Thin solid black contours show the integrated instensity contours of the $\sp{13}$CO \citep{abraham}. The lowest contour corresponds to 2\,K$\times$km\,s$^{-1}$, and the contour interval is 1\,K$\times$km\,s$^{-1}$. White circles show the positions of NH$\sb{3}$ cores, and green ellipses show {\it Planck} Galactic cold  clumps.  Purple diamonds indicate {\it IRAS} sources, red triangle shows Class~I object, Class~II YSOs are marked by star symbols, Flat YSOs are indicated by black squares, while the remaining H$\alpha$ sources without SED classification are drawn by plus signs. Green square indicates the Be star. Red downward triangles indicate Class~I sources without known distances. Letters indicate the A, B, C $\sp{13}$CO clumps identified by \citet{abraham}.}

\label{ext_map}
\end{center}

\end{figure*}

The high sensitivity Pan-STARRS data allow us to refine the picture of the dust column density structure of L1188, compared
to available extinction maps of the region \citep{rowles, dobashi}. We constructed an extinction map, applying the classical method of star counts \citep{dickman} on the Pan-STARRS  data set. We counted the stars within of 60-arcsec radii, whose centres were distributed on a regular grid with step of 30$\arcsec$. The number of stars with V $\la$ 25 mag within a 1 square degree area was 55709. We removed each star with V$\la$ 14 mag as probable foreground object. The off-cloud reference area was an 1$\sp{\prime}$ $\times$ 1$\sp{\prime}$ field centered at RA(2000) =333.67$\sp{\circ}$  D(2000) =62.35$\sp{\circ}$, containing 86 stars. 
The uncertainty of the extinction of a pixel, derived by the formula given in \citet{dickman} and depending on the pixel value itself, extends from $\sim$ 0.2  mag in the low extinction areas (between 0 $-$ 1mag) to 1.1  mag near the extinction peaks. The resulting A$\sb{V}$ map can be seen in Fig.~\ref{ext_map}.

\subsection{Colour-magnitude diagram}

\begin{figure}
\begin{center}

 \includegraphics[width=\columnwidth]{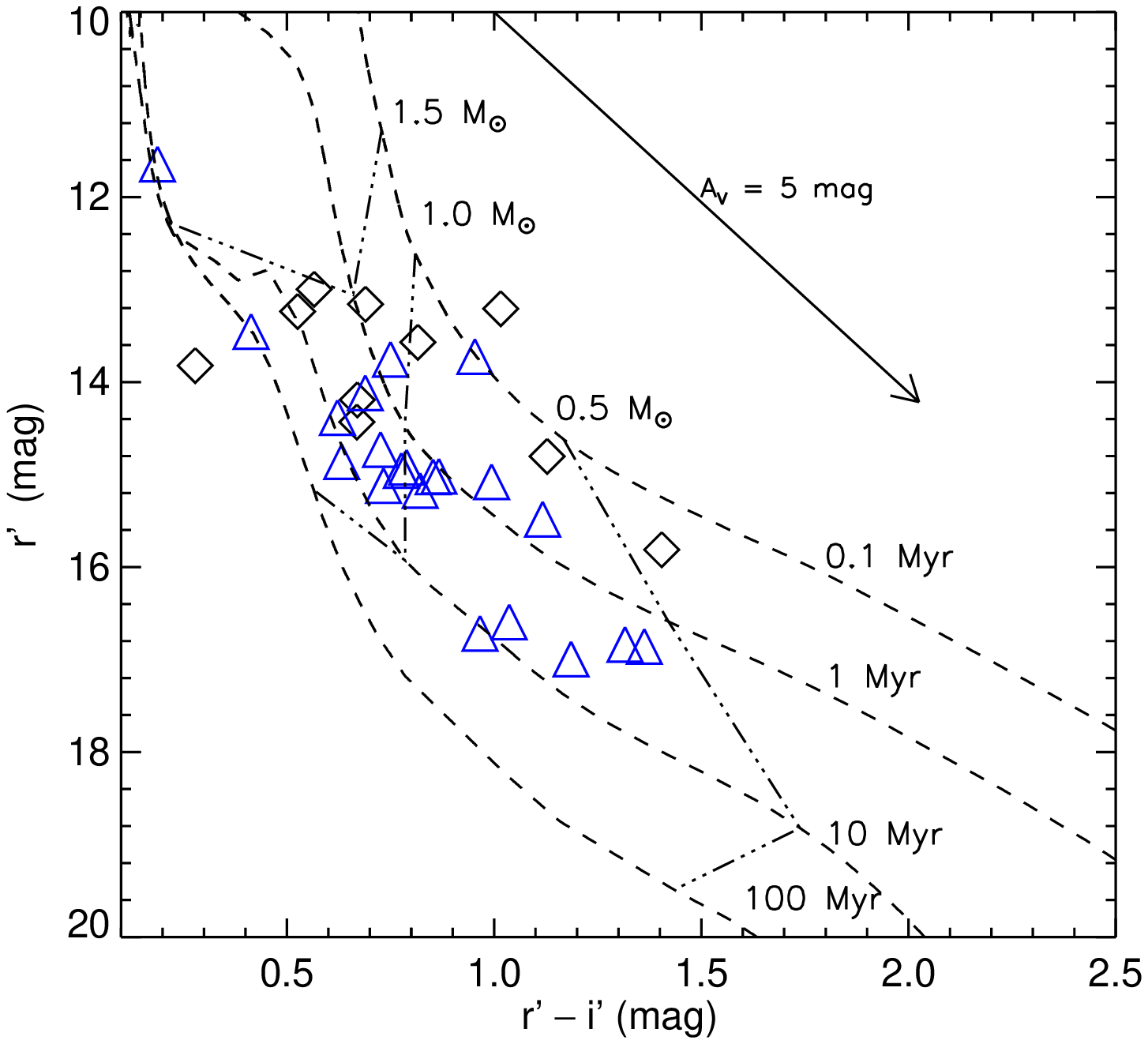}
 \caption{Positions of the YSO candidates in the colour-magnitude diagram. Blue triangles show H$\alpha$ emission stars, while black diamonds symbol infrared-excess stars. Dashed and dash-three dots lines show isochrones and mass tracks from semi-empirical PMS isochrones presented by \citet{bell} for the IPHAS bands, based on the Pisa PMS tracks and
isochrones \citep{tognelli} and BT-Settl \citep[and references therein]{baraffe} atmosphere models. The tracks have been placed at the distance modulus of the L1188, the stars have been corrected for the {\it r$\sp{\prime}$} band excess affected by H$\alpha$ emission line and extinction. The arrow shows the reddening vector for {\it A$\sb{V}$}=5 due to \citet{cardelli}.}
\label{diagram}

\end{center}
\end{figure}

We plotted the {\it $r\sp{\prime}$} vs. {\it $r\sp{\prime}-i\sp{\prime}$} colour-magnitude diagram using the IPHAS photometry for the candidate young stars, which have the signal to noise ratio $>$10 in all bands and the object looks like a single, unconfused point source. {\it $r\sp{\prime}$} magnitudes may be affected by the presence of the H$\alpha$ emission line within the {\it r$\sp{\prime}$} band \citep{drew2005,barentsen}. We applied the correction to the {\it r$\sp{\prime}$}
magnitudes following \citet{barentsen} and plotted the {\it $r\sp{\prime}$} vs. {\it $r\sp{\prime}-i\sp{\prime}$}  colour-magnitude diagram of the candidate YSOs in Fig.~\ref{diagram}. 

We compared the distribution of the sources with semi-empirical PMS isochrones presented by \citet{bell} for the IPHAS bands, based on the Pisa PMS tracks and
isochrones \citep{tognelli} and BT-Settl \citep[and references therein]{baraffe} atmosphere models.



We shifted the model tracks and isochrones  according to the average distance of 855 pc, but we did not correct them for the extinction. Instead  we corrected individually all of the stars for the extinction with the previously determined individual values using the relationships {\it A$\sb{r\sp{\prime}}$}=0.843 {\it A$\sb{V}$} and {\it $A\sb{i\sp{\prime}}$}=0.639 {\it A$\sb{V}$} by \citet{schlegel}.

Most of the YSO candidates in this diagram are  scattered between the 10$\sp{5}$\,yr  and 10$\sp{7}$\,yr isochrones plotted for the average distance. 

Based on Fig.~\ref{diagram} a few of the YSO candidates are seen to fall near or below the 100~Myr isochrone. Taking into account that extinction can be underestimated, the positions of these stars could move towards slightly younger isochrones. 
\section{Results and discussion}
\label{sect_4}
\subsection{Cloud Structure}

The new extinction map of L1188 can be seen in Fig.~\ref{ext_map}. To compare the distribution of gas and dust, $\sp{13}$CO contours \citep{abraham}, NH$\sb{3}$ cores \citep{verebelyi}, distribution of the  young stellar objects, and Planck Galactic cold cores (PGCCs, Planck Collaboration 2015) are overplotted on the extinction map. Class~I sources with unknown distances are also overplotted with smaller red downward triangles. Their positions suggest physical connection to L1188. A slight difference can be seen between the $\sp{13}$CO contours and extinction. This effect may be caused by the different angular resolution.
The new extinction map of L1188 shows filamentary structure with dark knots. 
The presence of NH$\sb{3}$ cores suggests to existence of so high column density regions, which can be future stellar nurseries.

Our extinction map saturates at {\it A$\sb{V,\rm{max}}$}=13.7~mag,  corresponding to a hydrogen column density of 3.03$\times$ 10$\sp{22}$ cm$\sp{-2}$ \citep{extav}. The exctinction exceeds this value on a few pixels in the densest part of the cloud. The extinction at these positions exceeds the critical lower limit of {\it A$\sb{V}$} $\approx$ 8~mag found by \citet{molinari} for star-forming clumps.

The structure and also the mean {\it A$\sb{V}$} of our extinction map are very similar to those determined by \citet{rowles} and \citet{dobashi} for the same region based on 2MASS colour excesses. We find the mean extinction {\it A$\sb{V}$}=2.74 for the cloud within the {\it A$\sb{V}$} = 1.0~mag contour. The same average is 2.94 for \citet{rowles} extinction map, and 2.44 for the extinction map of \citet{dobashi}. A remarkable difference among the extinction maps is that the extended dark areas on the 2MASS based images fragment into smaller concentrated spots on our extinction map based on classical star-count method.
 
Six {\it Planck} Galactic cold clumps are  projected over L1188. Five of these cold clumps are projected near extinction peaks, suggesting their physical connection to the cloud. 

\subsection{The Young Star Candidates}



 Our selection criteria included detection by {\it Gaia} at optical wavelengths, thus both the H$\alpha$ emission line sources and infrared excess stars, listed in Tables~\ref{halfa_tab} and \ref{infra_table} are most probably classical T~Tauri stars (CTTSs) born in L1188, inc\-lu\-ding the only Class~I source. Three stars of Table~\ref{halfa_tab} were included in spectroscopic follow-up observations of IPHAS by \citet{drew2005}. They classified two of them as T~Tauri stars, and another one as an M3Ve (J22142723+6129433). This star shows strong H$\alpha$ emission, which can not be explained as chromospheric activity, so we propose  it as a YSO candidate.  

We could measure EW(H$\alpha$) only in 17 cases. To compare these results with another method, we estimated the EW(H$\alpha$) based on the IPHAS {\it r$\sp{\prime}-$H$\alpha$} vs. {\it r$\sp{\prime}-$i$\sp{\prime}$} colour-colour diagram \citep{barentsen} for stars, having detection with S/N~$>$10 in each band. The comparison explored no systematic difference between the equivalent widths measured by different methods. Most of the measured equivalent widths are between 10 and 100~\AA, typical of classical T~Tauri stars  \citep[e.g.][]{ew2}. 
The highest values ($>$ 100~\AA) belong to J22174025+6147025 and J22190442+6136250.

\begin{figure}
\begin{center}

\includegraphics[width=\columnwidth]{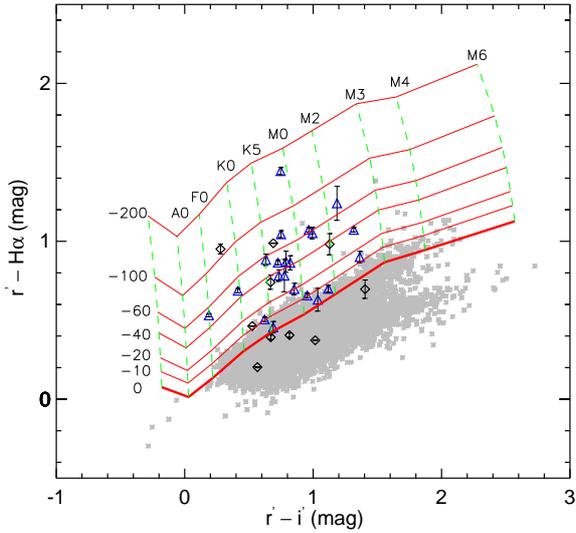}
\caption{IPHAS {\it r$\sp{\prime}$-H$\alpha$} vs. {\it r$\sp{\prime}$-i$\sp{\prime}$} colour-colour diagram of the candidate
YSOs within the observed region. Blue triangles indicate H$\alpha$ emission stars identified in our WFGS2 images, whereas black diamonds indicate infrared excess stars.  Small grey stars show the colour distribution of the IPHAS sources with S/N $>$ 10 in each band within the same area. Simulated colours of main sequence stars with H$\alpha$ emission, computed by \citet{barentsen} are indicated by the solid red lines. Dashed green lines show the colour variations of the spectral types indicated at the upper ends, due to increasing EW(H$\alpha$).}\label{cc}
\end{center}

\end{figure}

Figure \ref{cc} shows the IPHAS colour-colour diagram for the candidate young stars detected in each IPHAS band. Blue triangles show H$\alpha$ emission sources, while black diamonds are infrared excess stars. Synthetic colours of main sequence stars and H$\alpha$ emission objects are taken from table~A1 in \citet{barentsen}. The thick red solid line indicates the normal main sequence, and thin solid lines show the main sequence colours modified by H$\alpha$ emission. Each colour of H$\alpha$ emission objects and infrared excess stars were dereddened by {\it A$\sb{V}$} determined during the analysis. It can be seen that not all of the  candidate YSOs appear as H$\alpha$ emission stars in this diagram. 
Five of the infrared excess stars seem to be also H$\alpha$ emission sources. These sources either were too faint to detect as H$\alpha$ sources during our H$\alpha$ survey or were too bright and saturated in the images.

Presumably the most massive star associated with L1188 in our sample is J221643.39+612417.5 ([ADM95]~IRAS3) with B3-spectral type \citep{bestar}, a known H$\alpha$ emission star LS~III~+61~9 \citep{wackerling}. We identified this star based on its infrared excess, because it was too bright and saturated in our H$\alpha$ images. This star is located at the outskirts of L1188. Supposedly this star was born at the same time as e.g. the stars of IC 1396.
Based on the spectral type estimate most of our young star candidates are K-type stars. It is expected that an important part of the young population is below the detection threshold of the slitless grism spectrograph. The lower mass limit is $\sim$ 0.3 M$\sb{\odot}$, corresponding to a spectral type of $\sim$ M3.


\subsubsection{Spectral energy distribution}
SEDs of the H$\alpha$ emission stars are shown in Fig. \ref{halfa_mamajack_sed}, while the SEDs of the infrared excess stars are displayed in Fig. \ref{infra_mamajack_sed}. For the Be star we adopted optical magnitudes from \citet{bestar}. The dereddened SED and that of the best fitting photosphere are also plotted.  The photometry-based effective temperature and extinction are indicated in each plot. We also constructed the Taurus median SED using \citet{furlan} data, established for K5$-$M2 type stars over the 1.25 $\micron$ $\la$ $\alpha$ $\la$ 34.00 $\micron$ region, and those of \citet{dalessio} for optical and far-infrared wavelengths.  According to the
classification scheme \citep{lada,greene} our list of H$\alpha$ emission stars contains one Class I source, 7 Flat SED and 23 Class II. 
The stars selected based on infrared excess contain 3 Flat, 19 Class II sources.

\begin{figure*}
\begin{center}

\centerline{\includegraphics[scale=0.2]{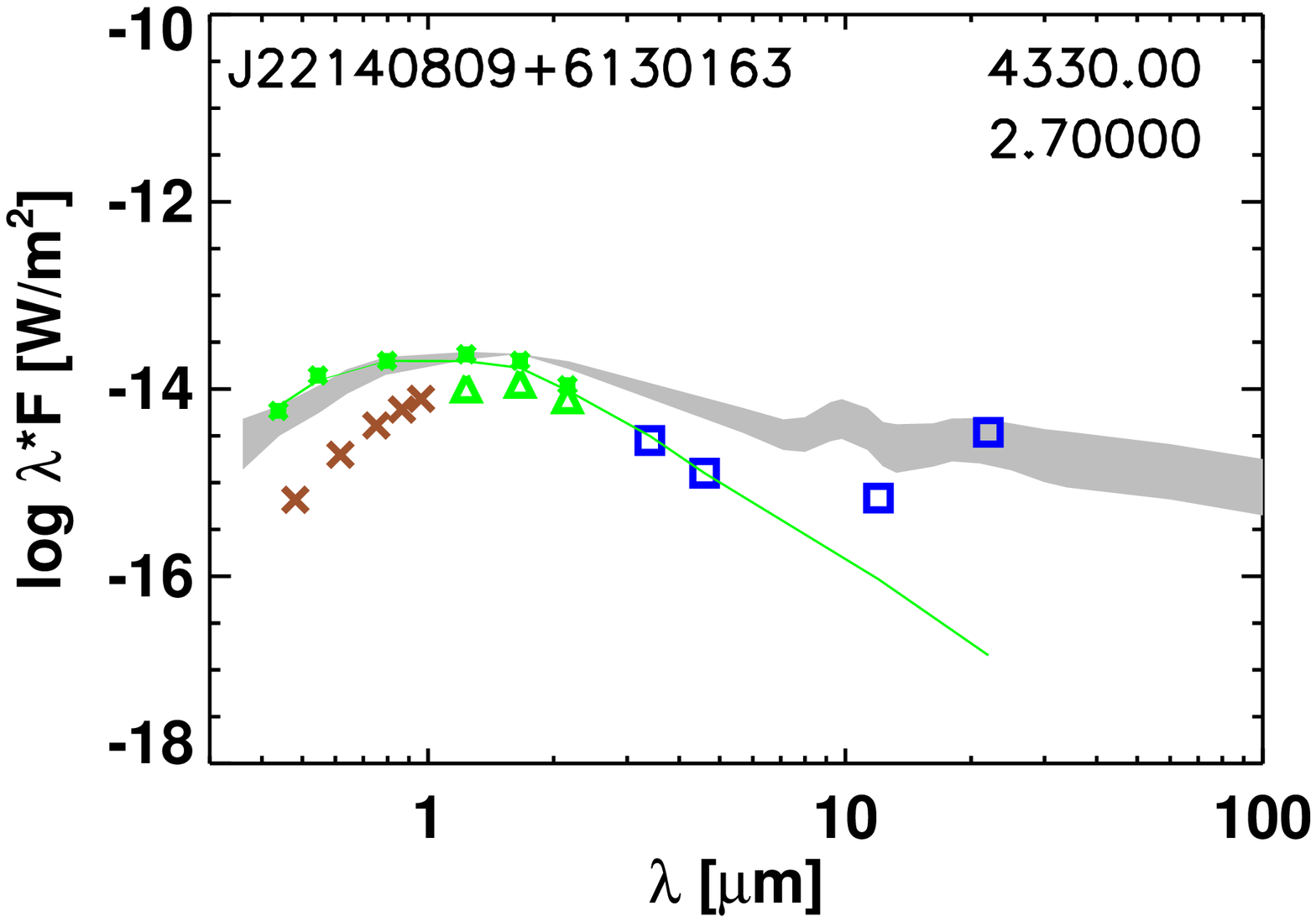}\includegraphics[scale=0.2]{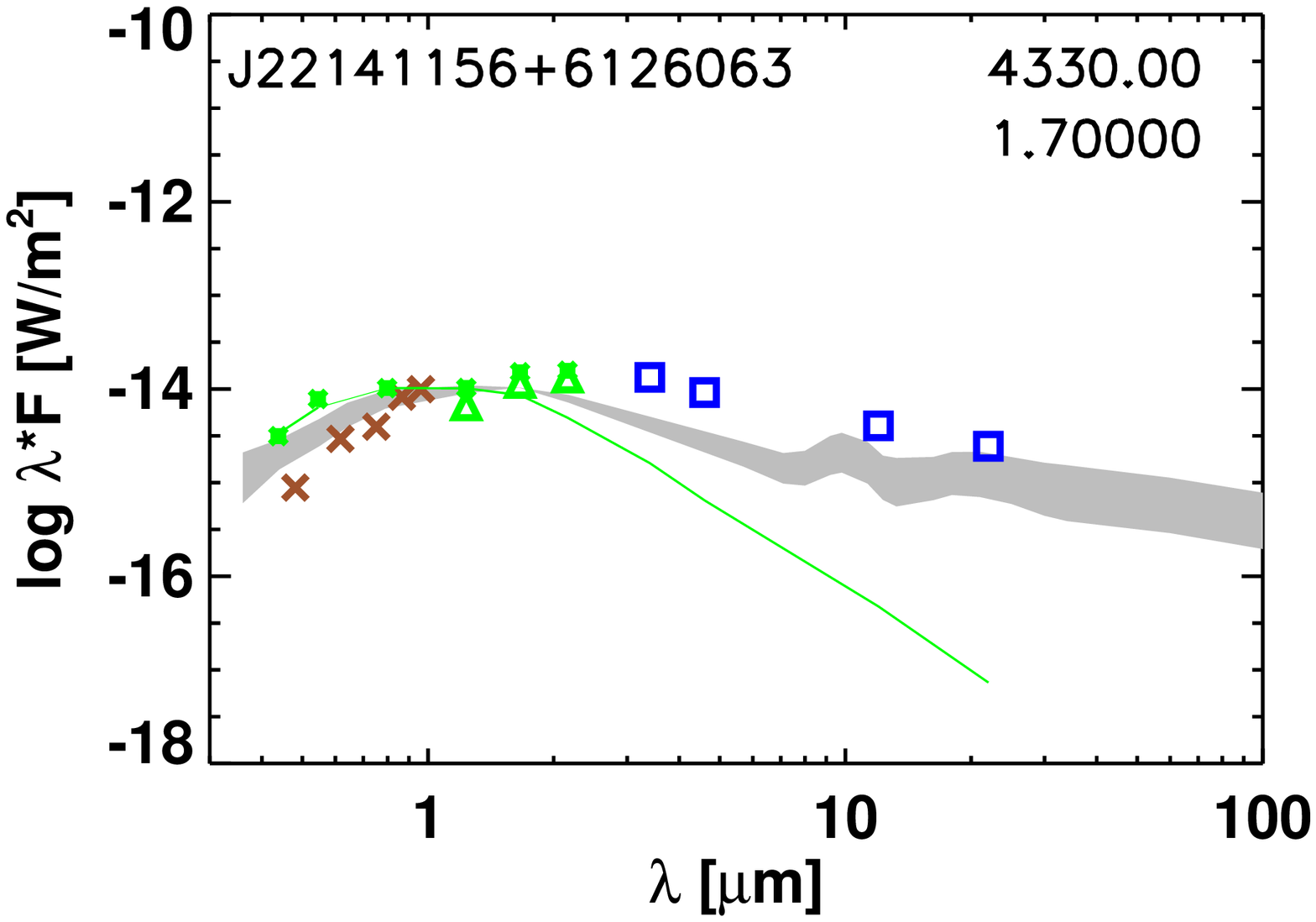}\includegraphics[scale=0.2]{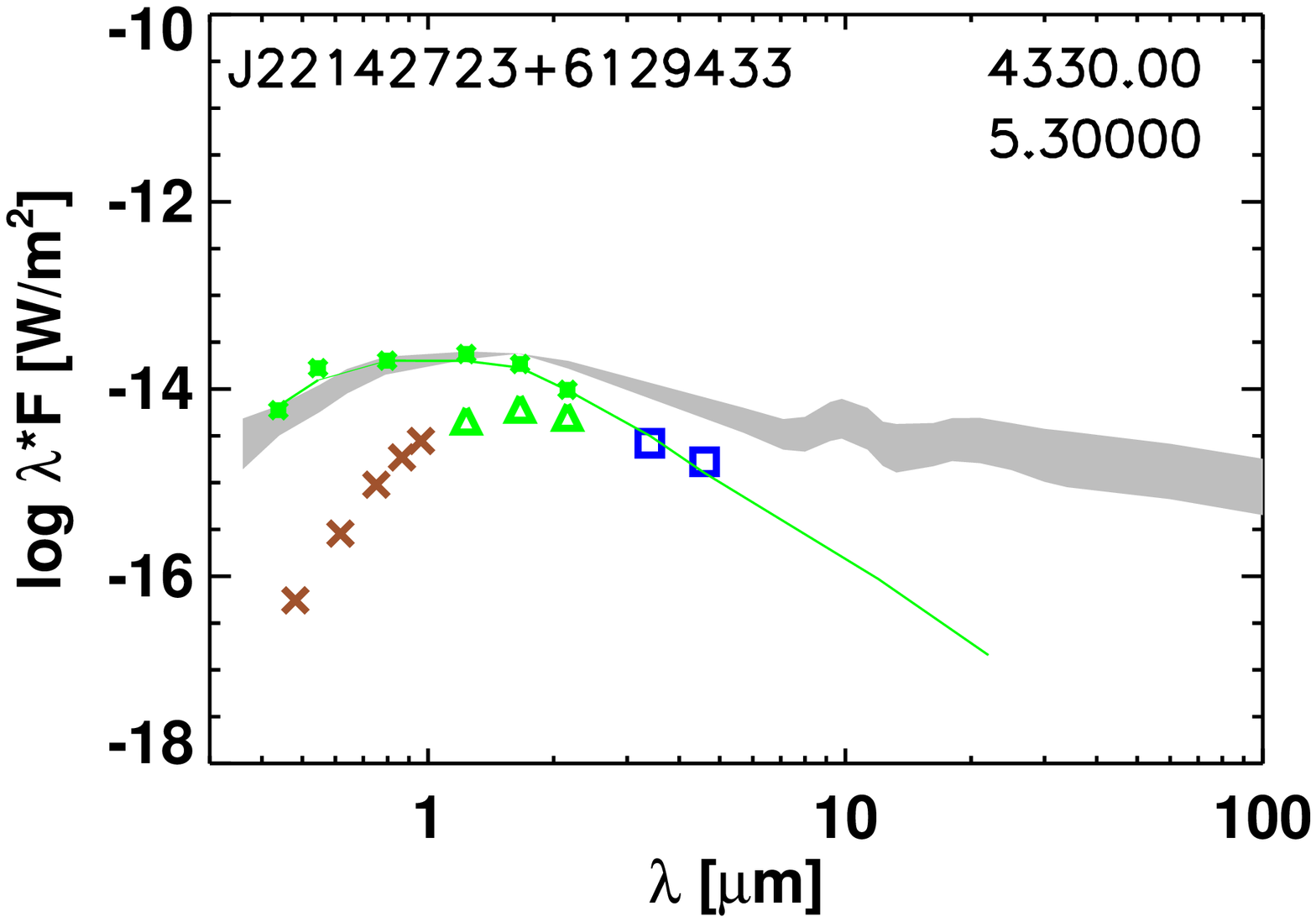}\includegraphics[scale=0.2]{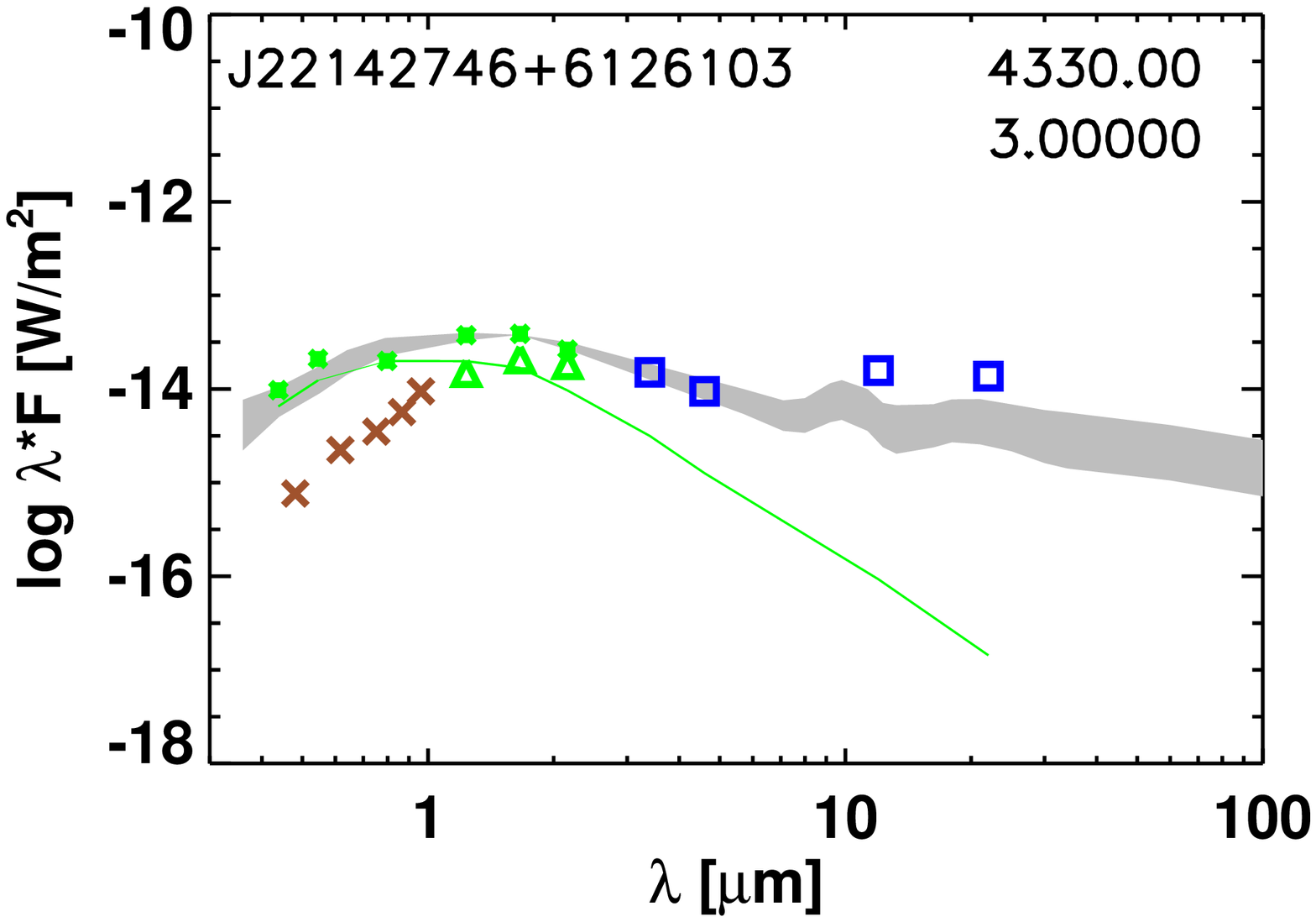}}

\centerline{\includegraphics[scale=0.2]{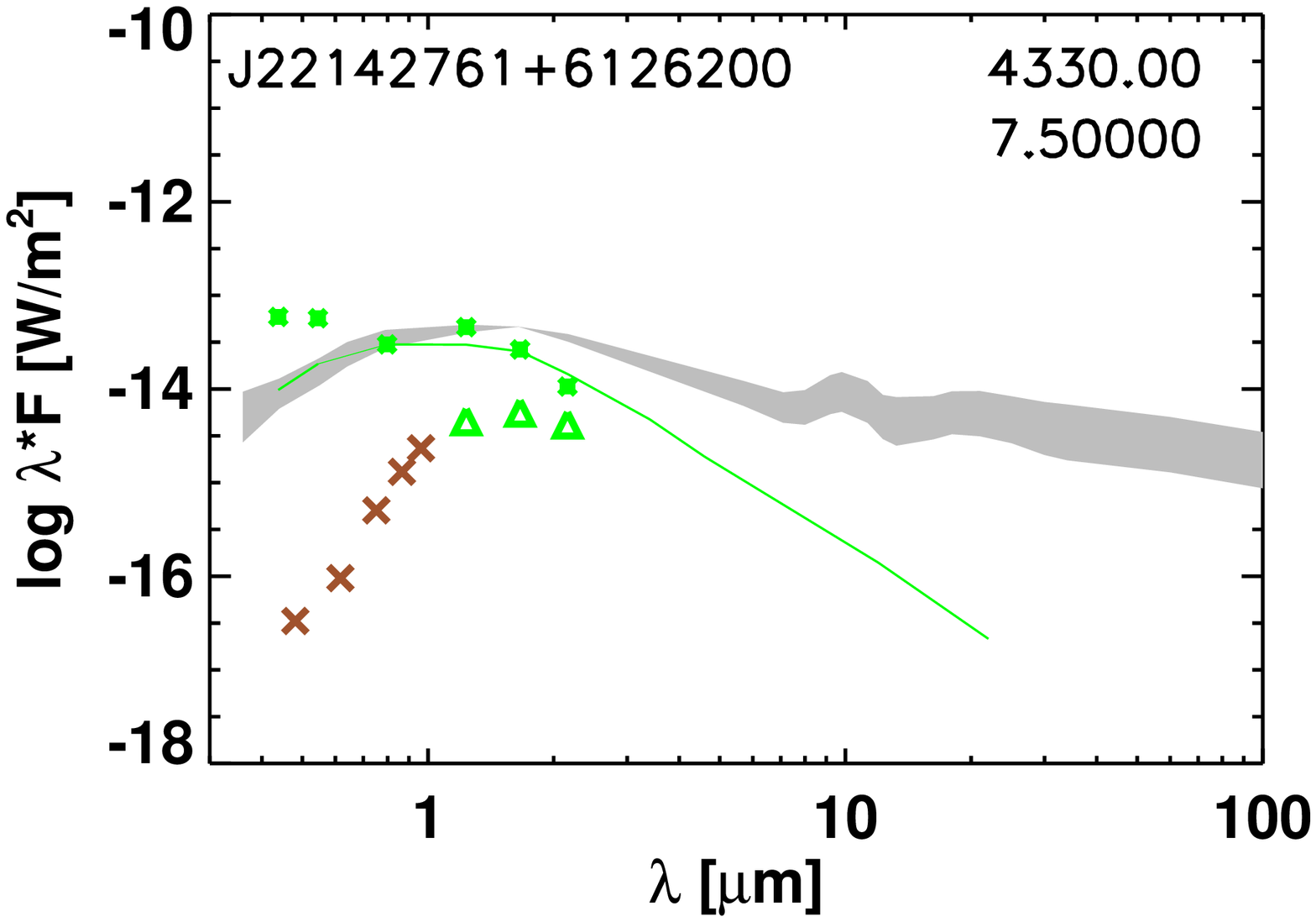}\includegraphics[scale=0.2]{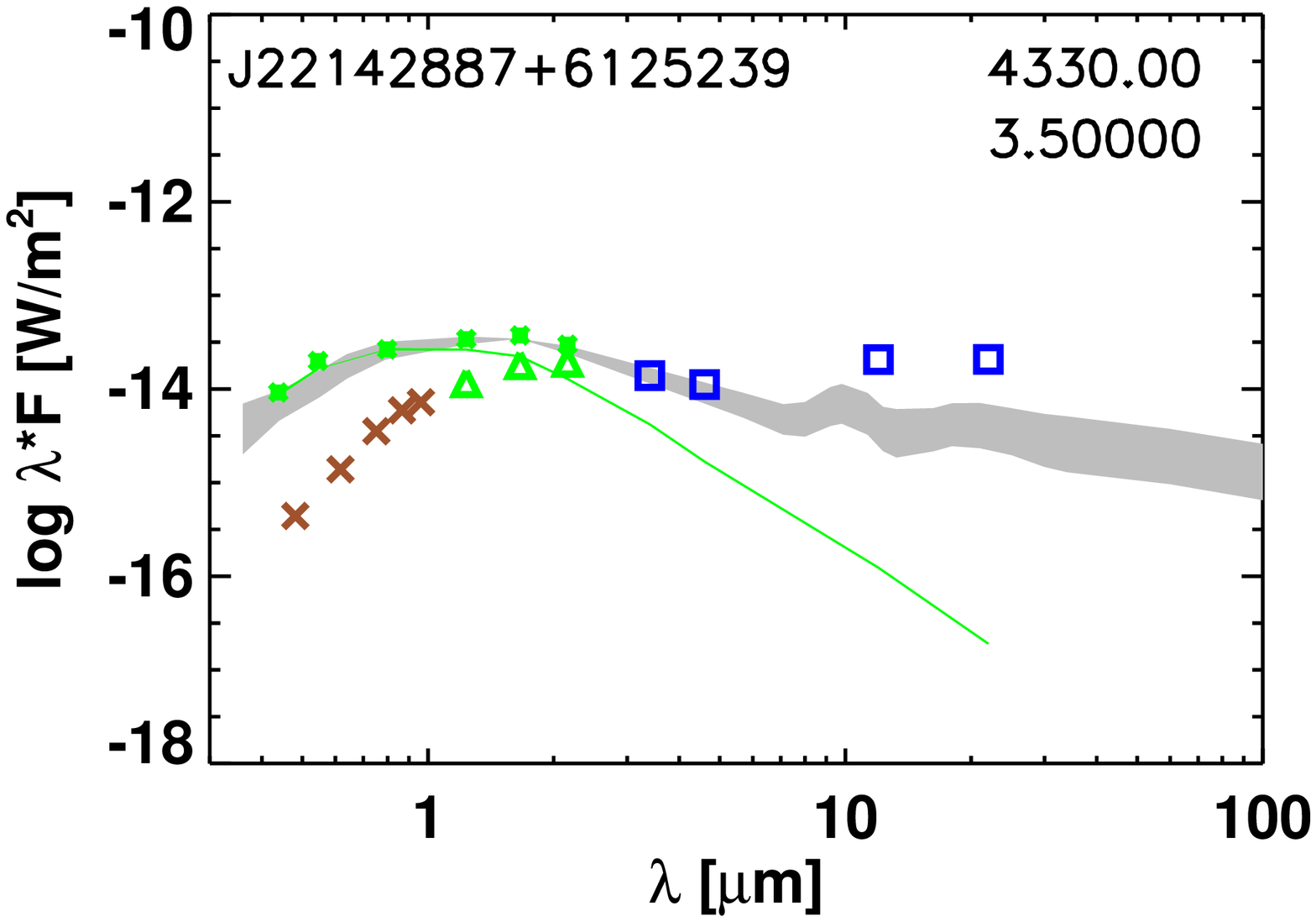}\includegraphics[scale=0.2]{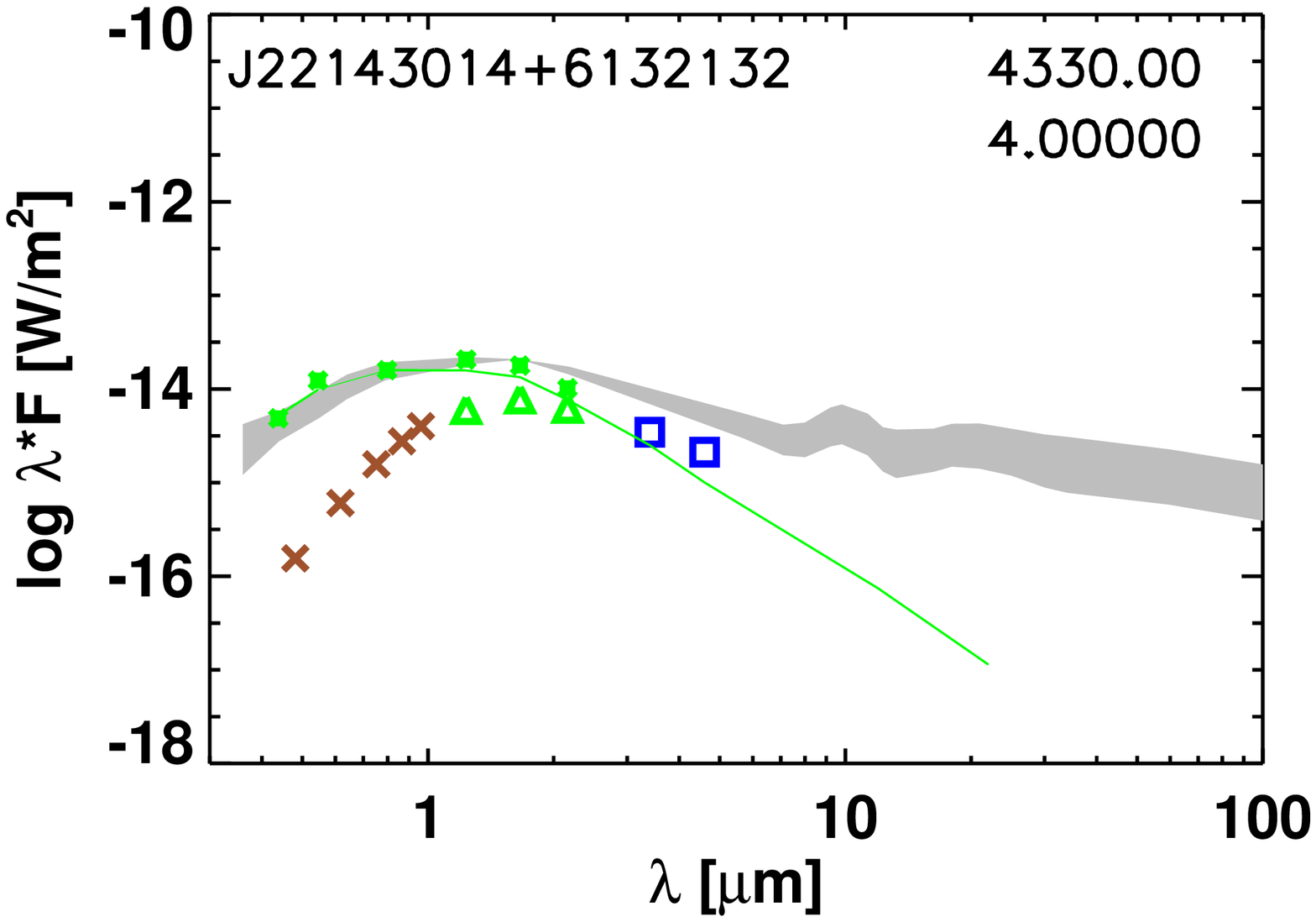}\includegraphics[scale=0.2]{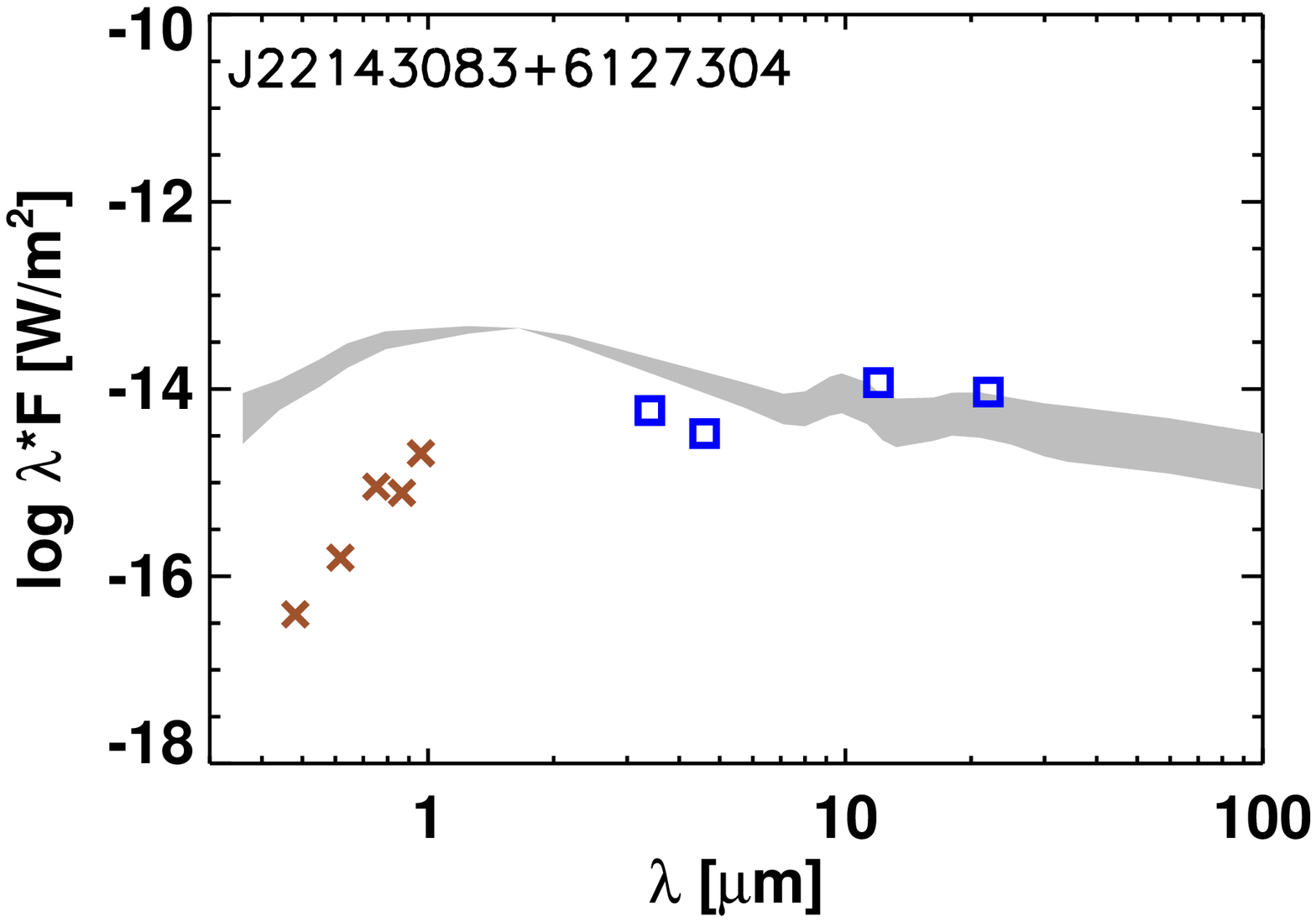}}

\centerline{\includegraphics[scale=0.2]{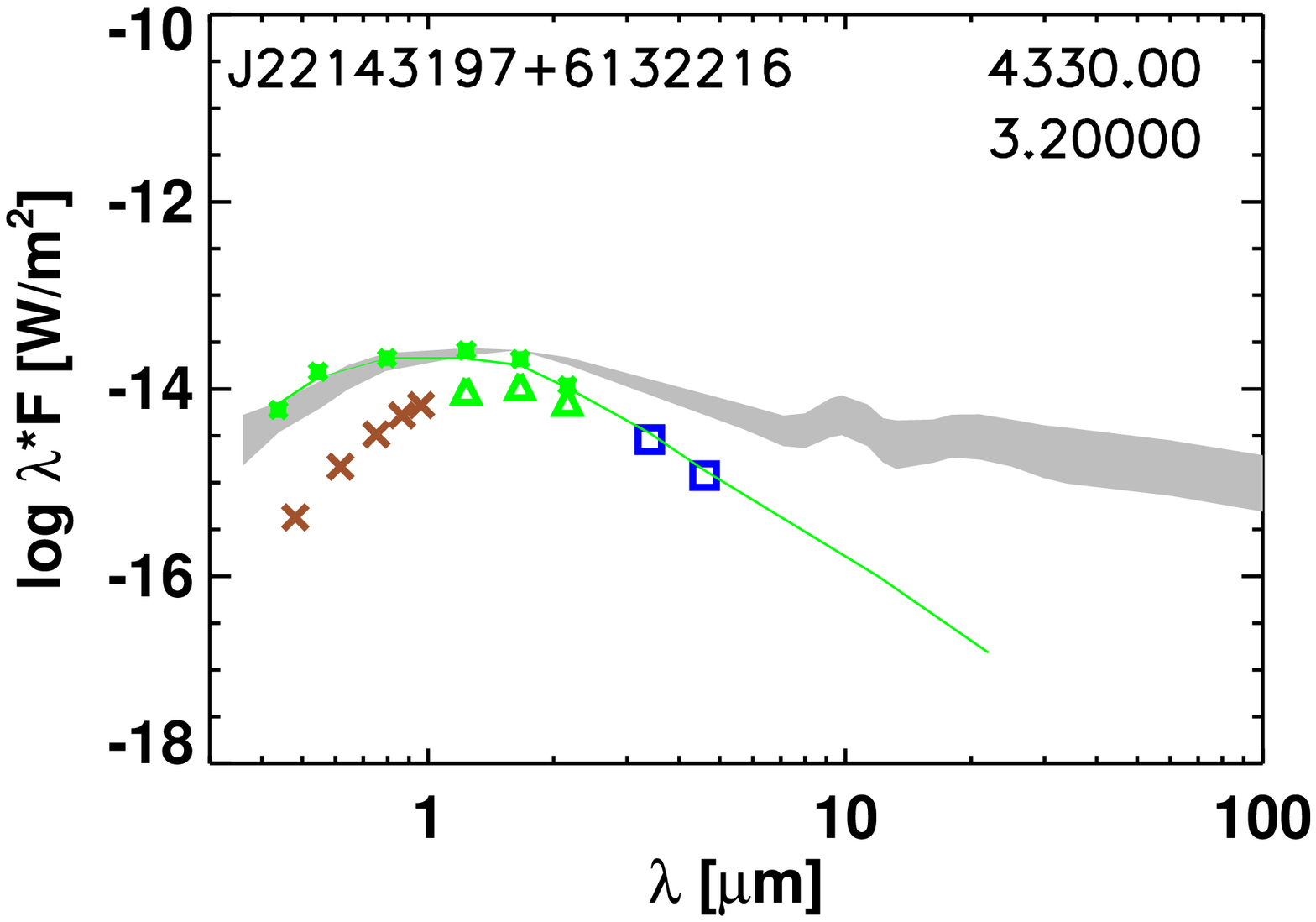}\includegraphics[scale=0.2]{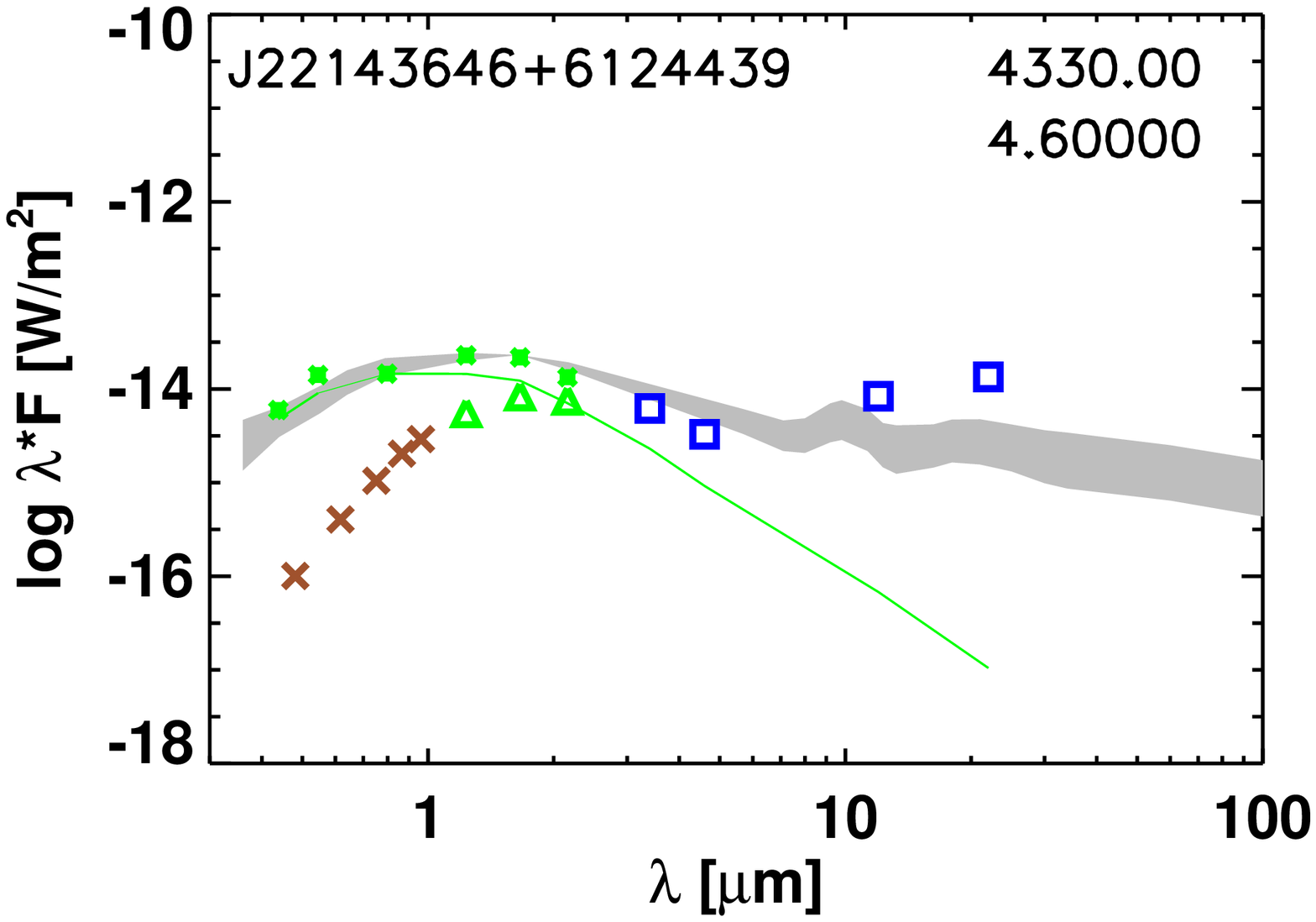}\includegraphics[scale=0.2]{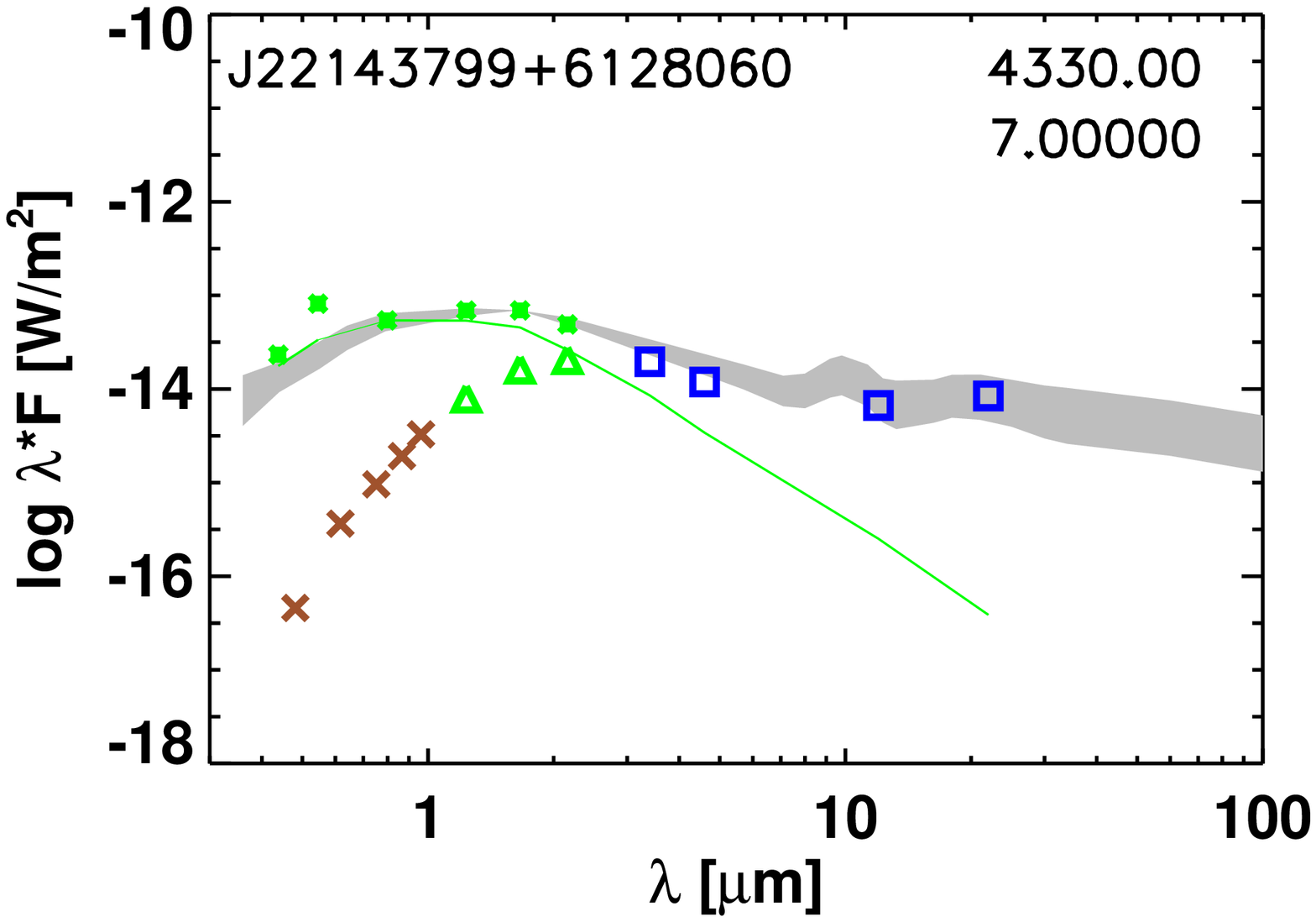}
\includegraphics[scale=0.2]{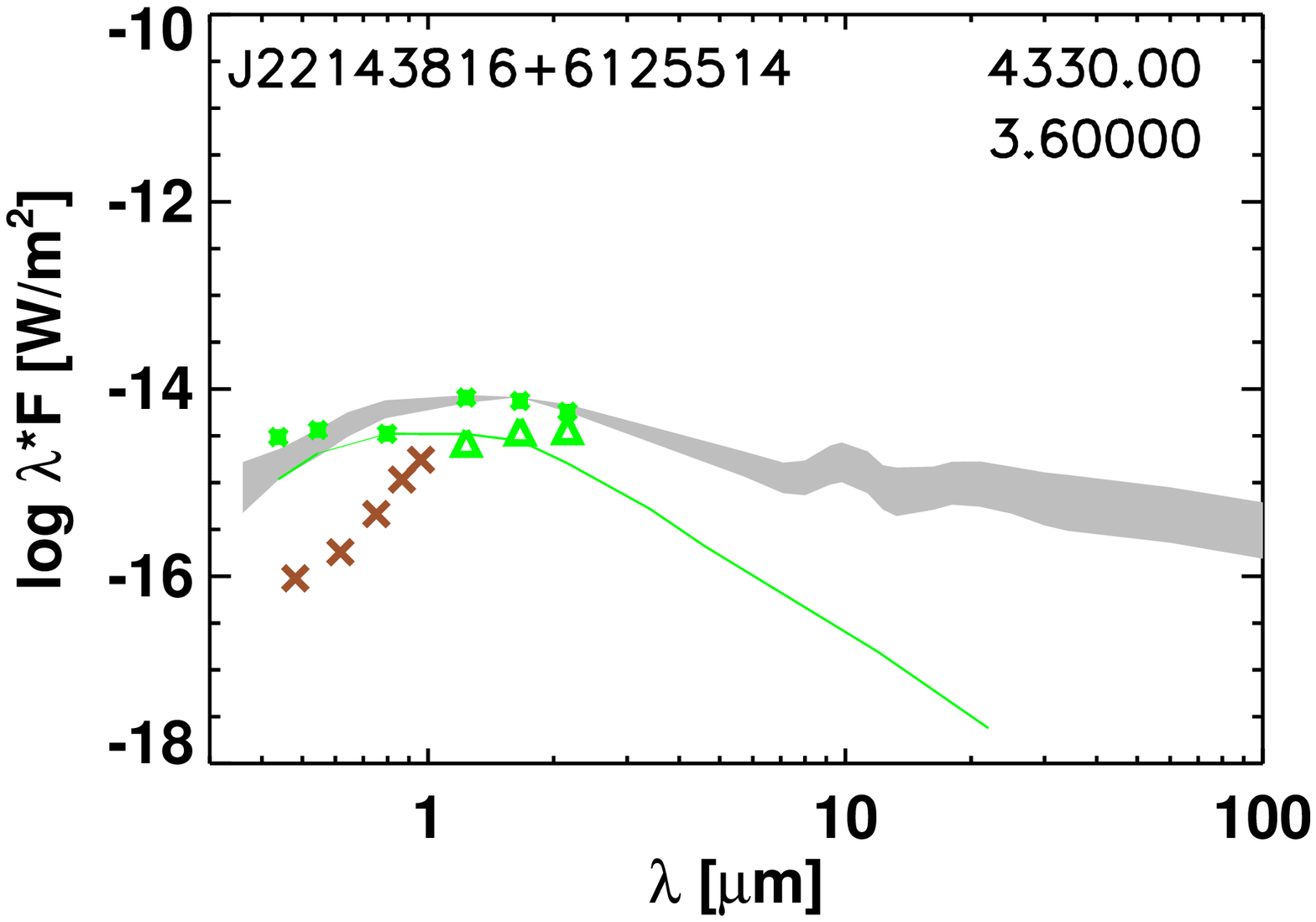}}

\centerline{\includegraphics[scale=0.2]{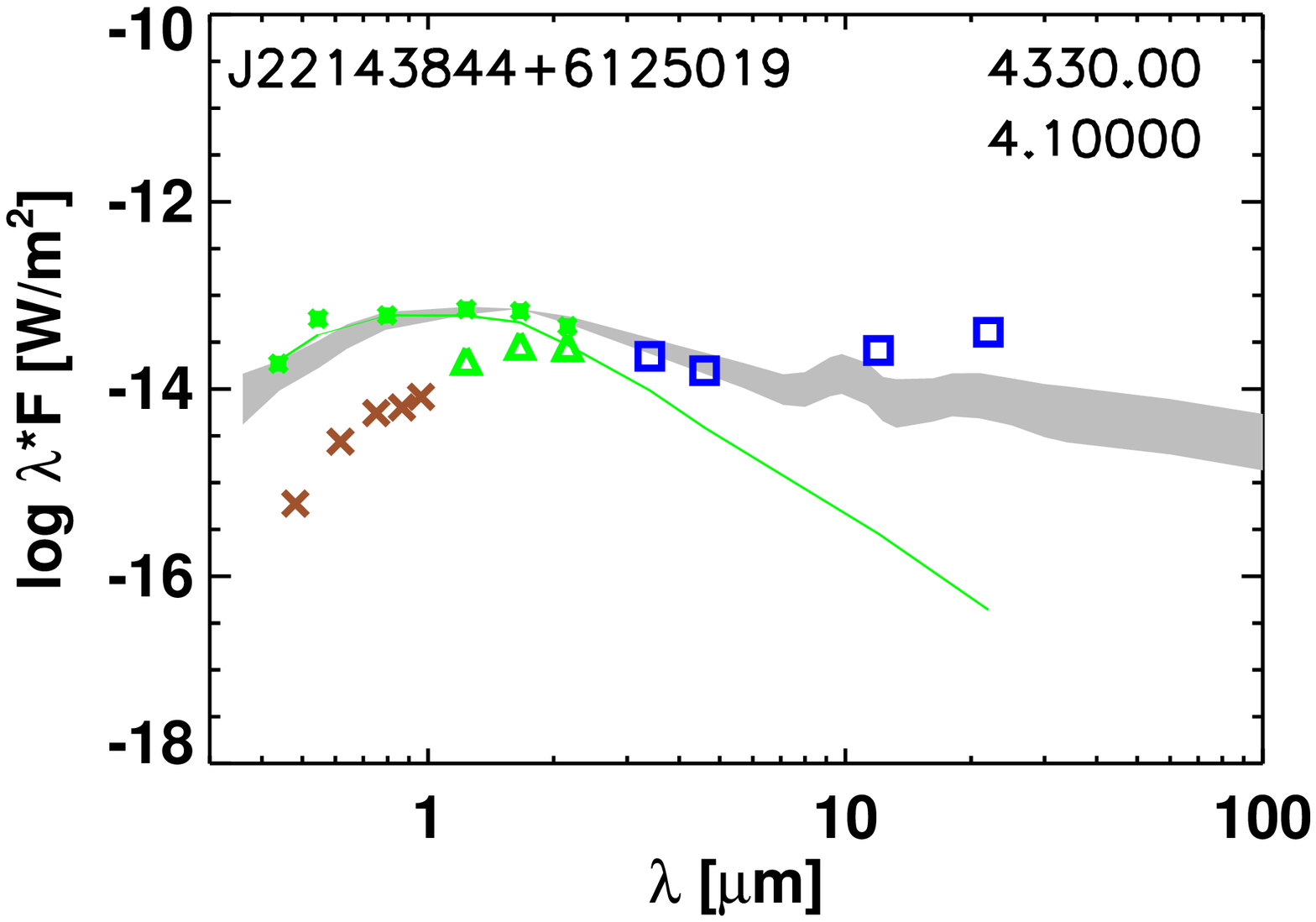}\includegraphics[scale=0.2]{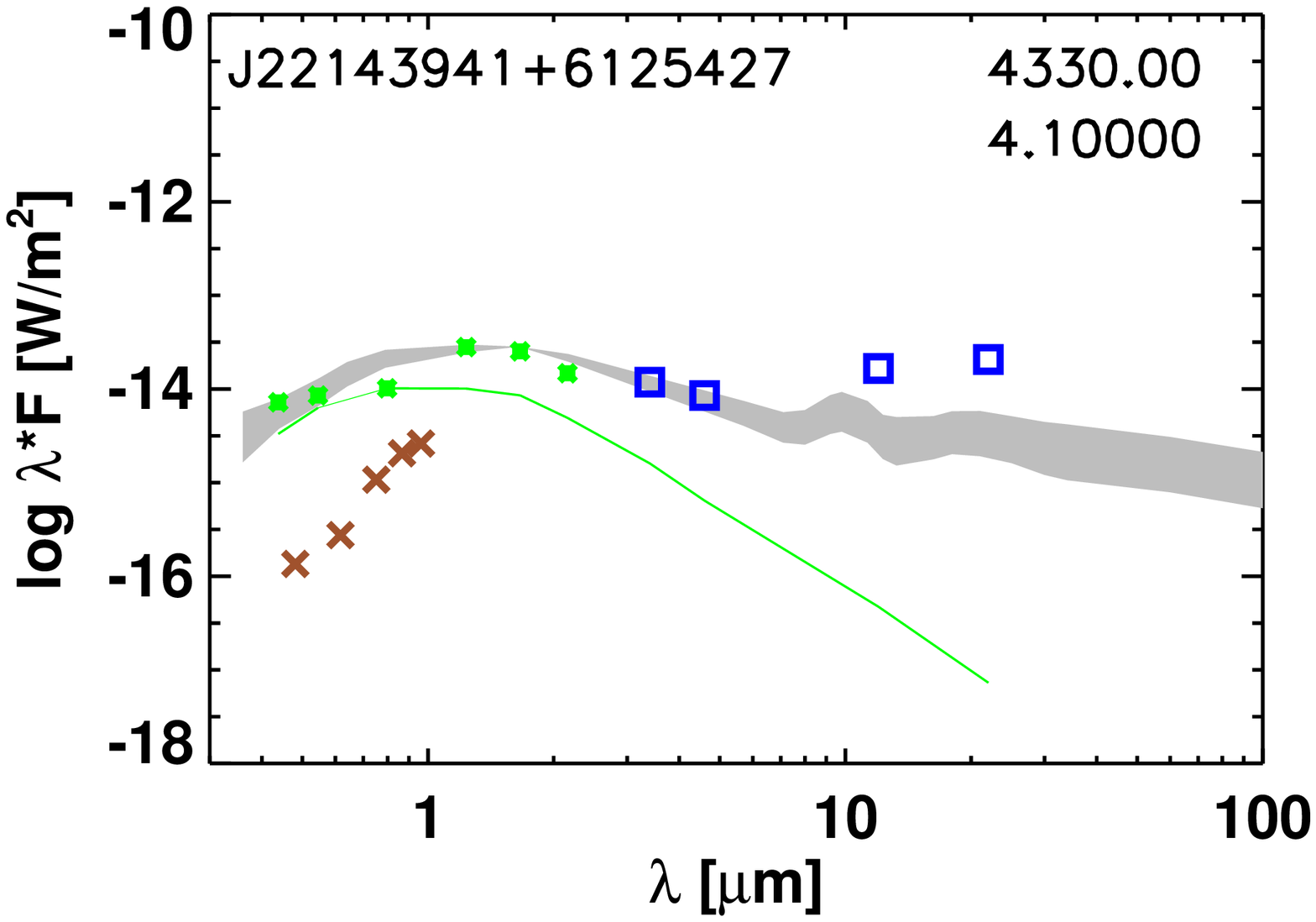}\includegraphics[scale=0.2]{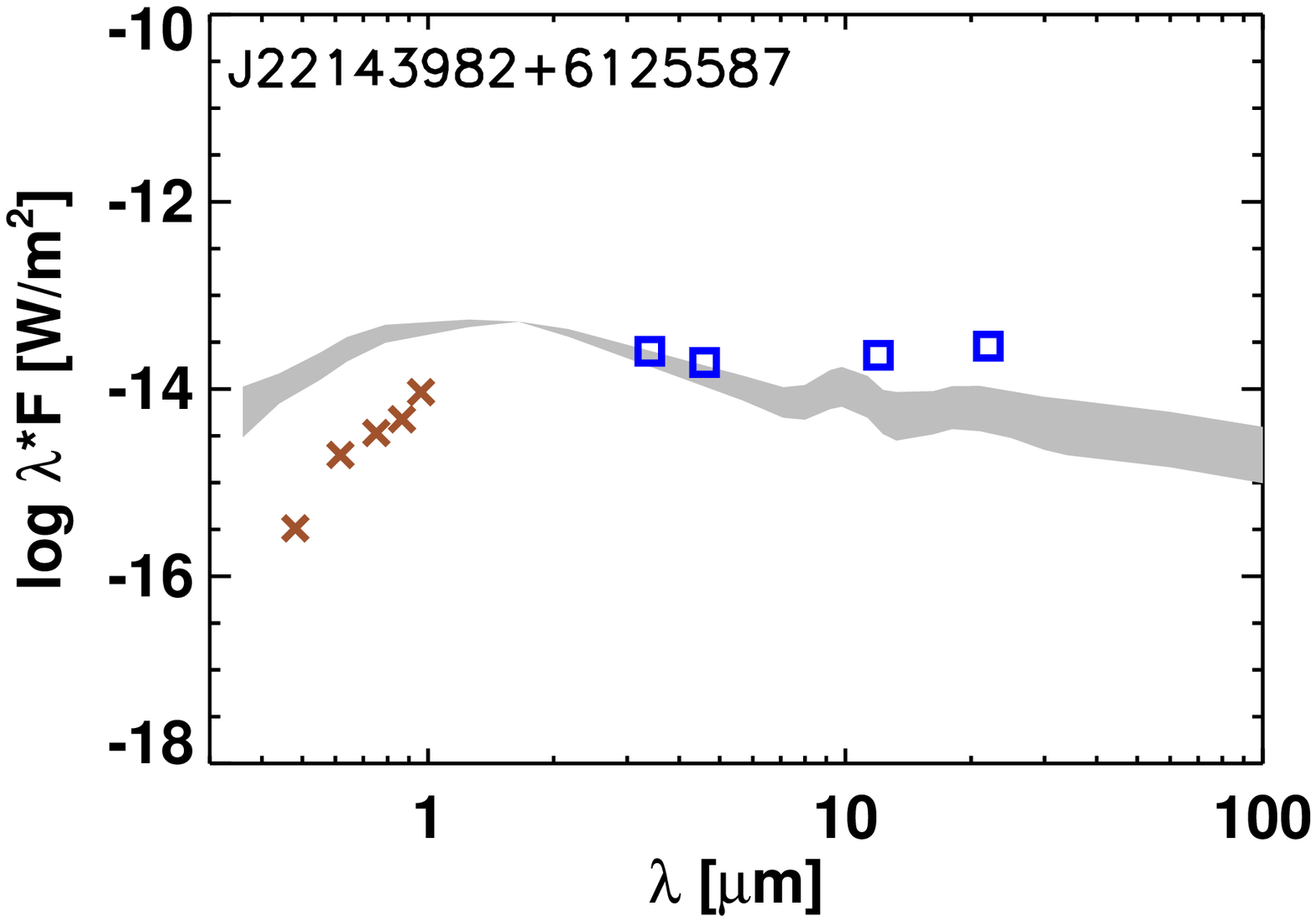}\includegraphics[scale=0.2]{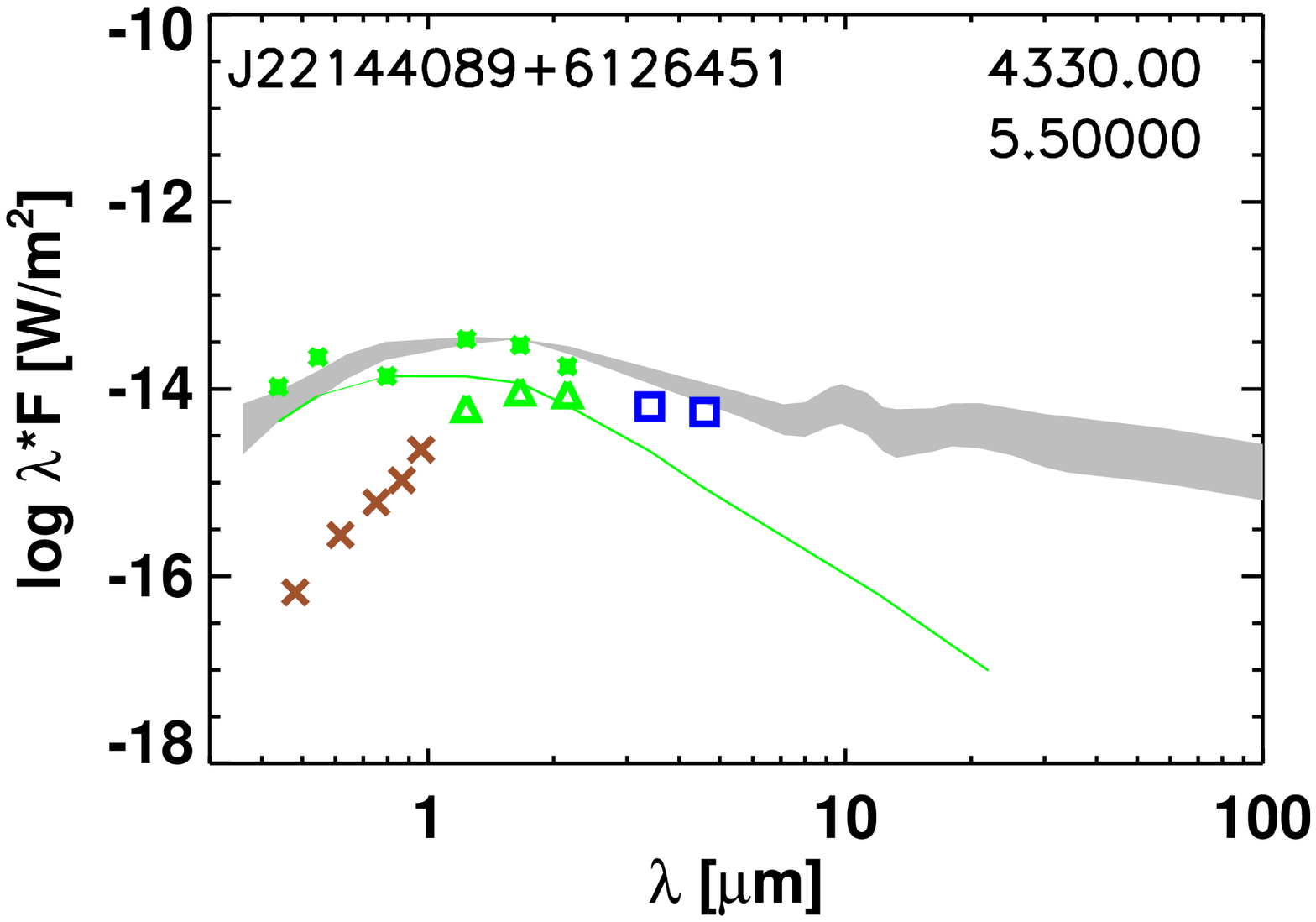}}

\centerline{\includegraphics[scale=0.2]{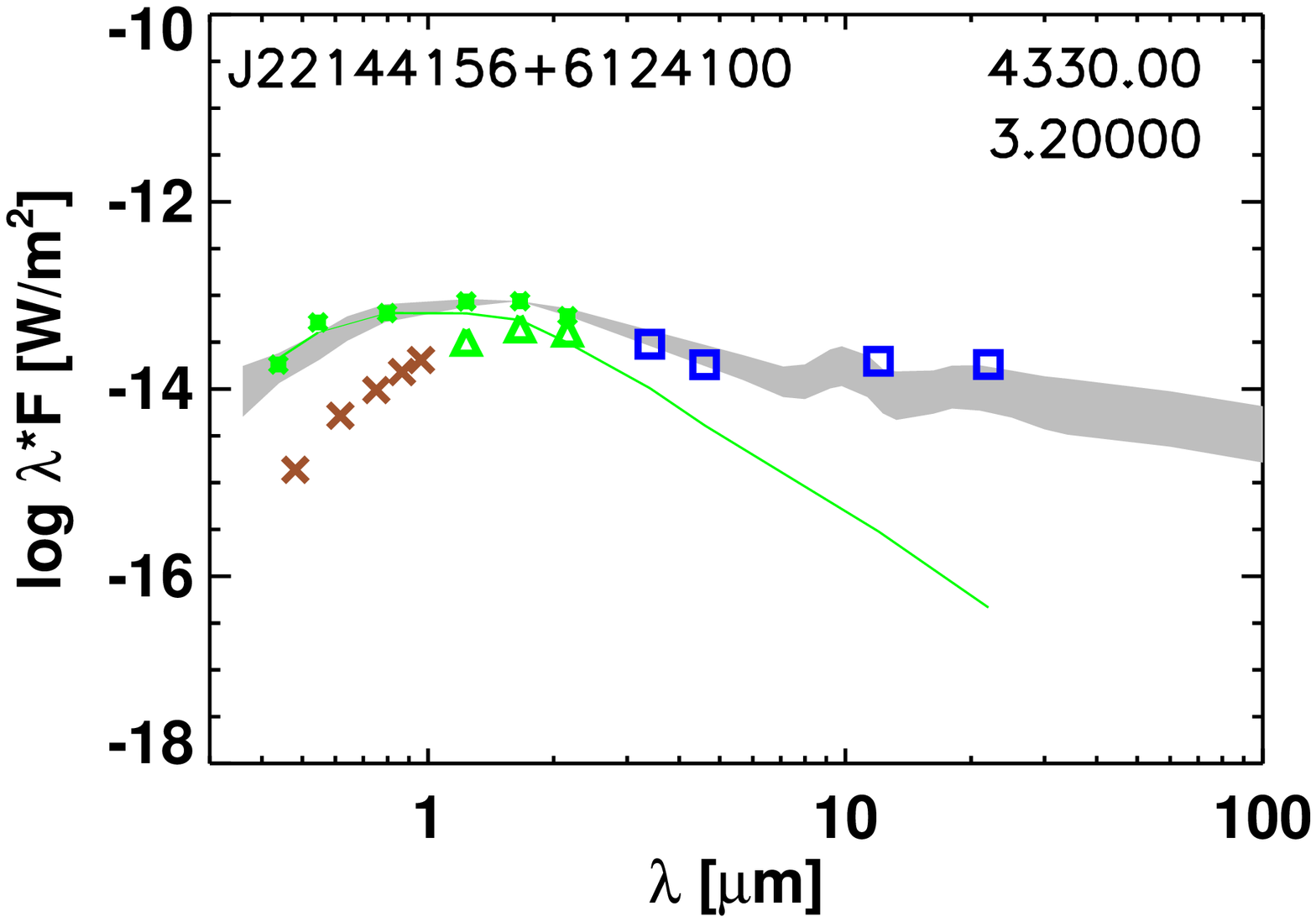}\includegraphics[scale=0.2]{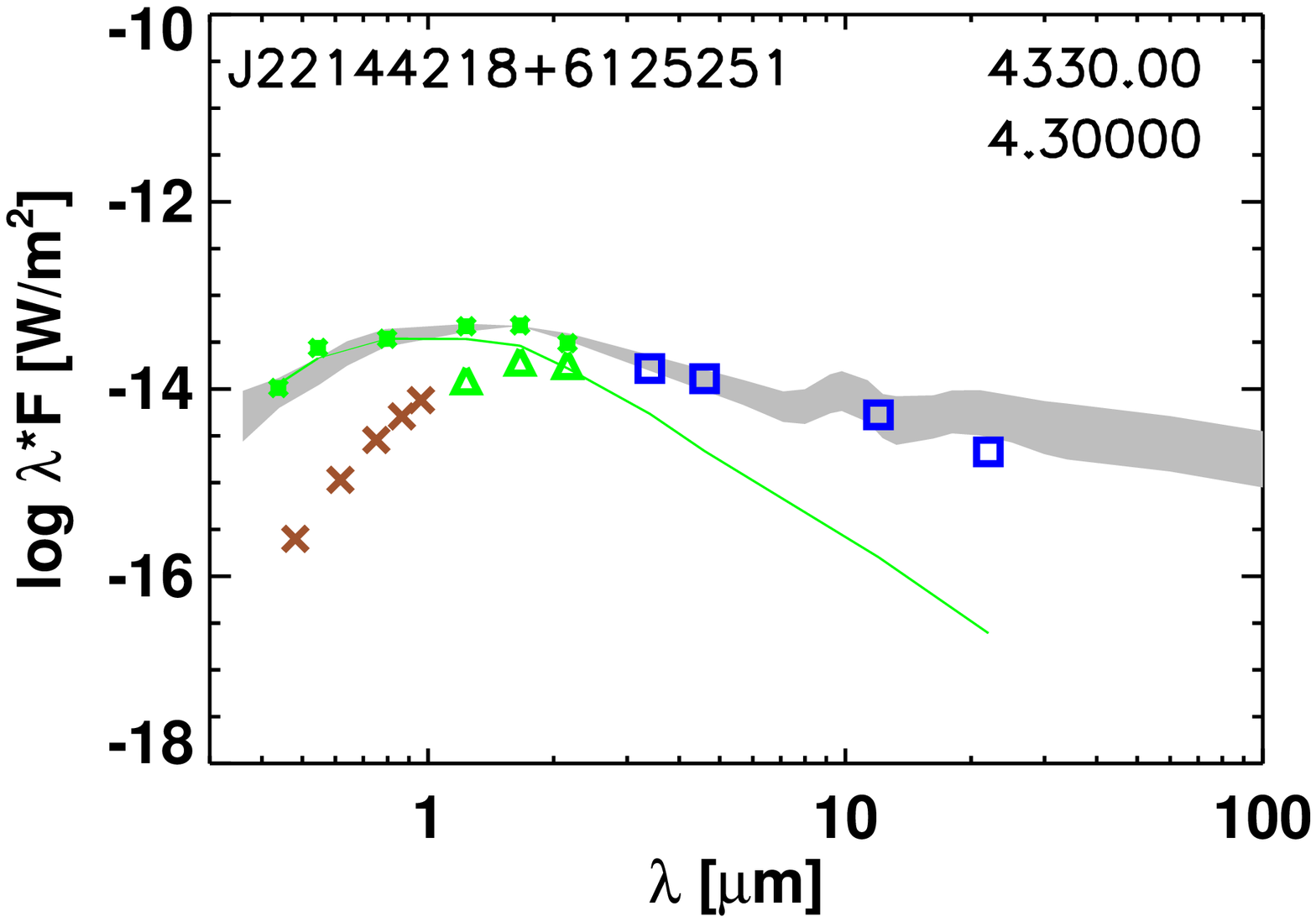}
\includegraphics[scale=0.2]{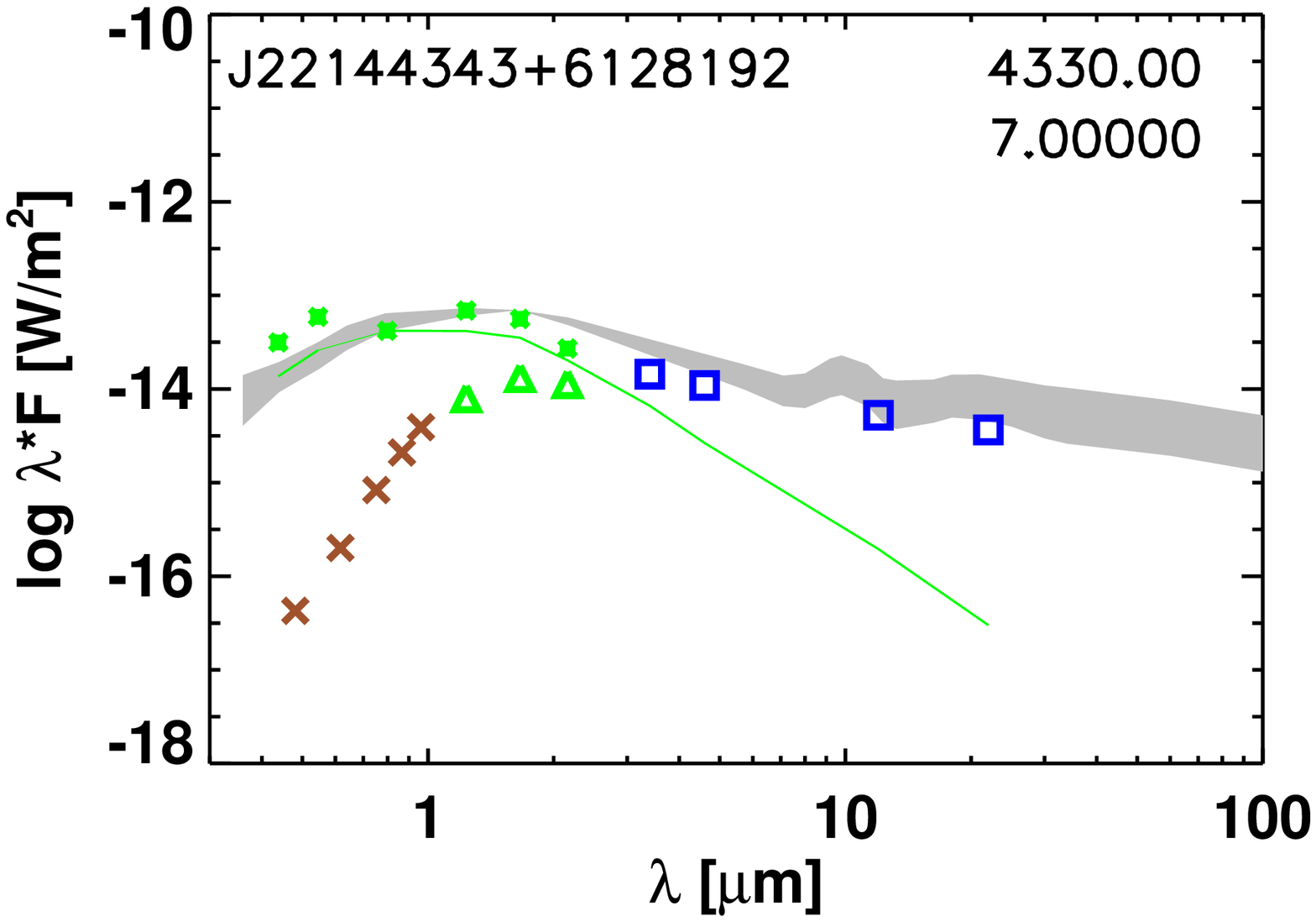}\includegraphics[scale=0.2]{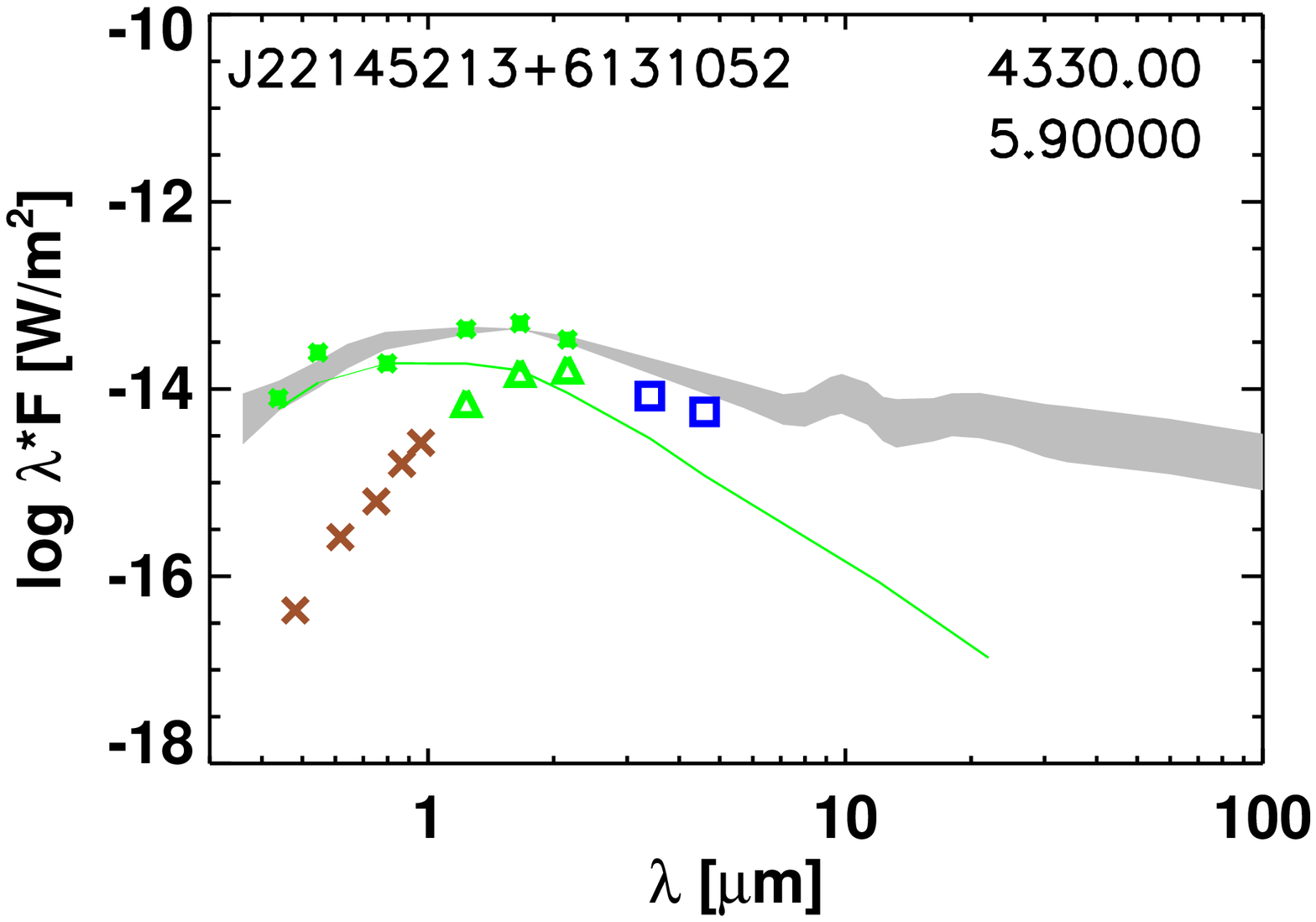}}

\centerline{\includegraphics[scale=0.2]{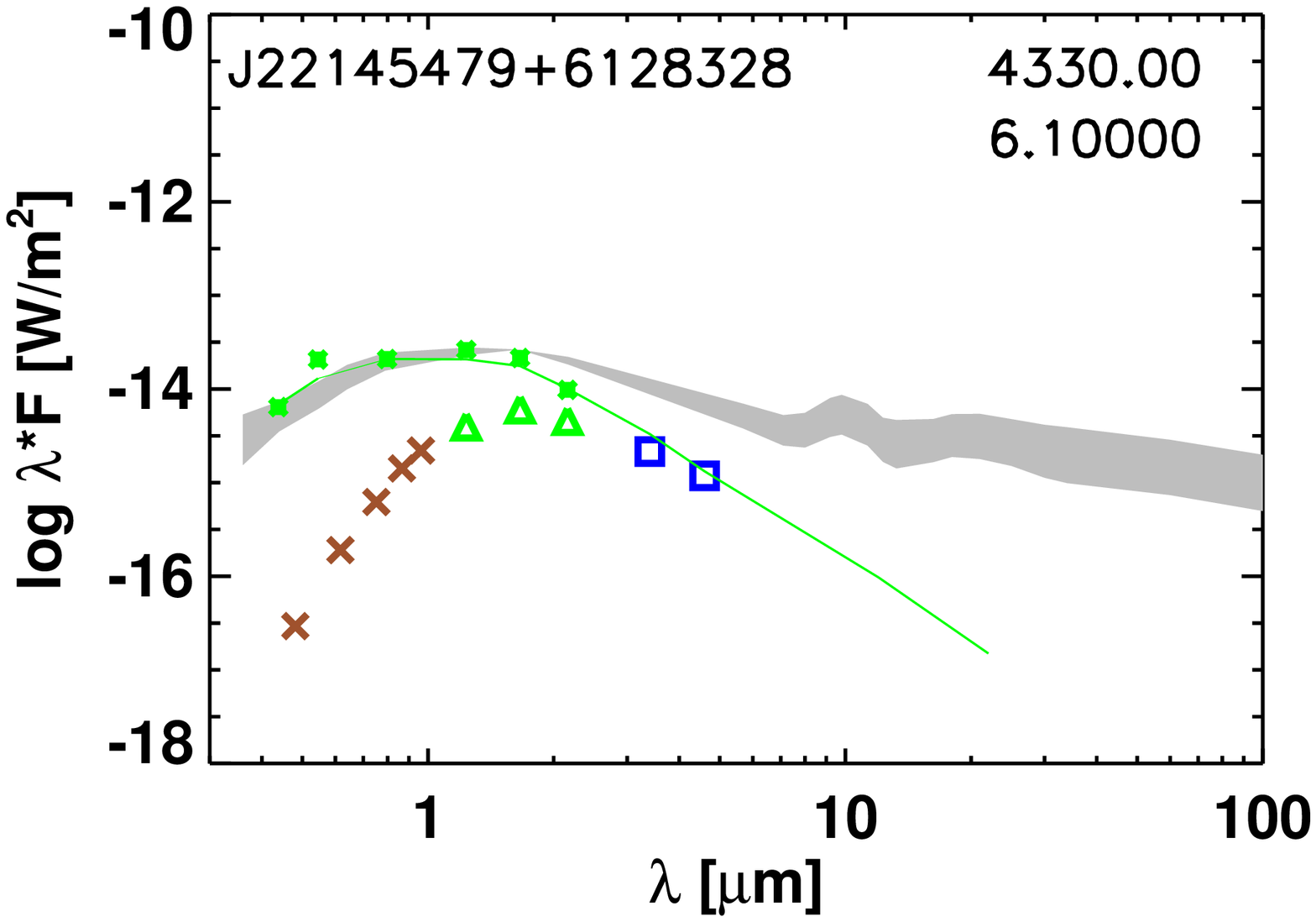}\includegraphics[scale=0.2]{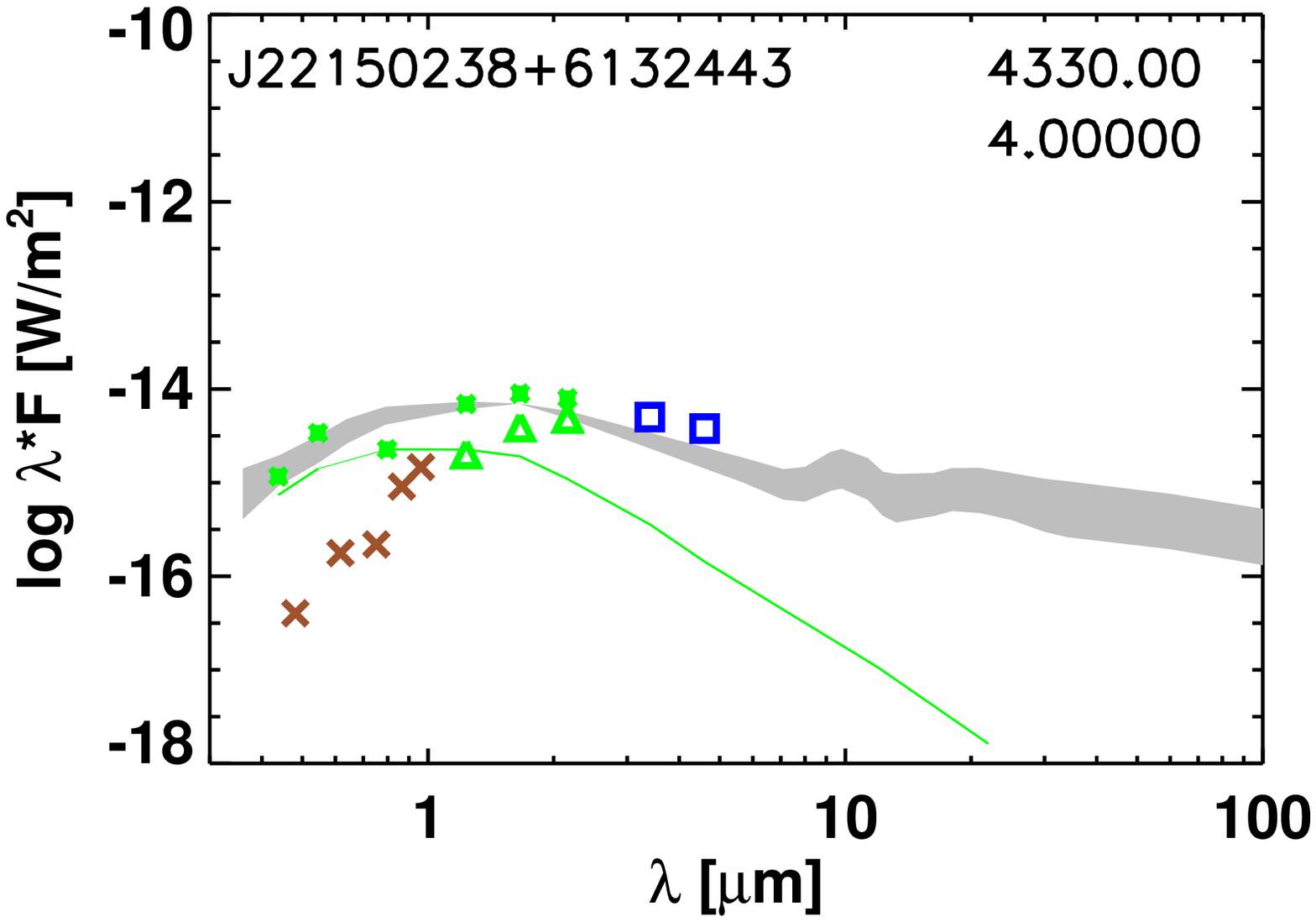}\includegraphics[scale=0.2]{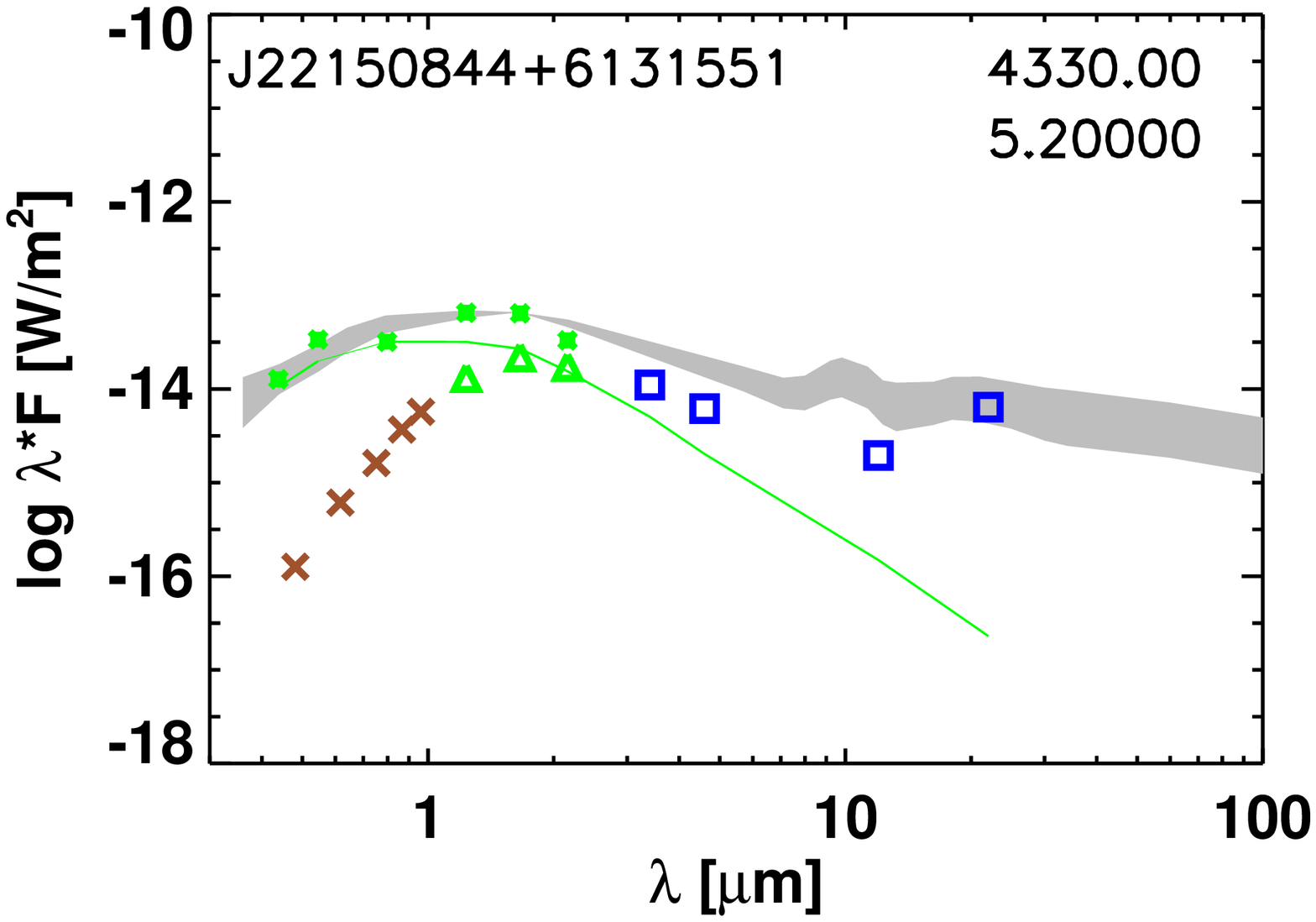}\includegraphics[scale=0.2]{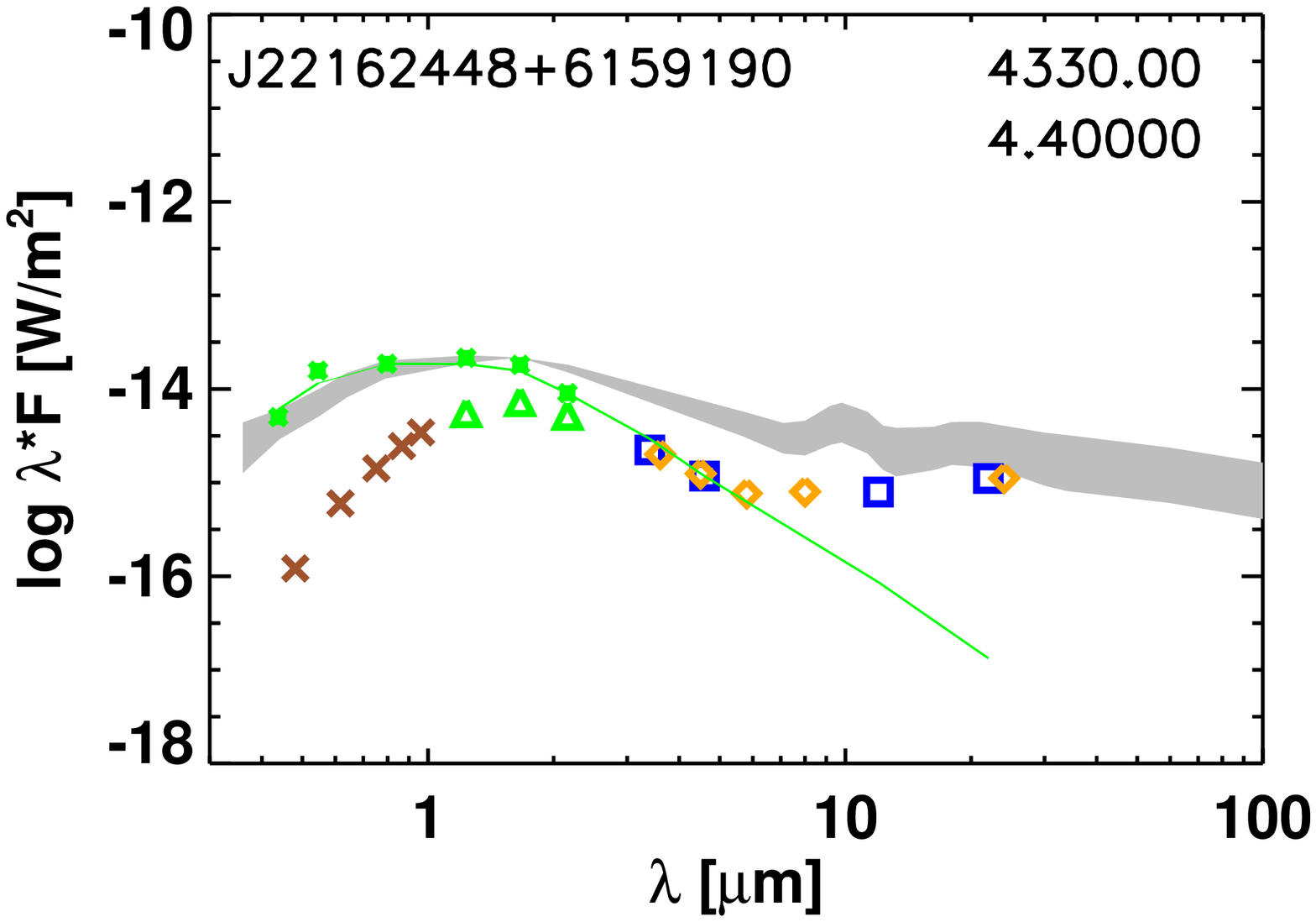}}

\centerline{\includegraphics[scale=0.2]{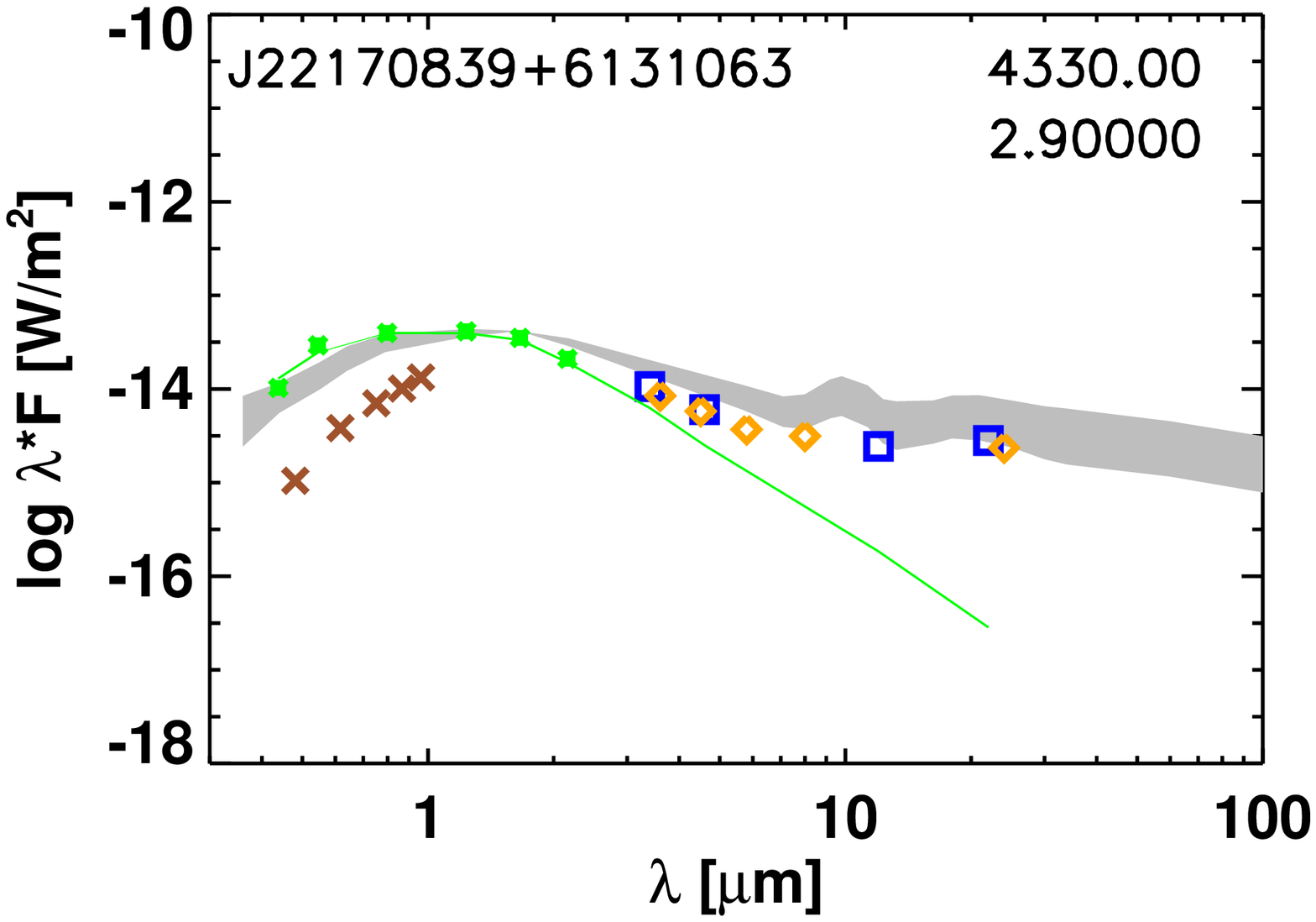}\includegraphics[scale=0.2]{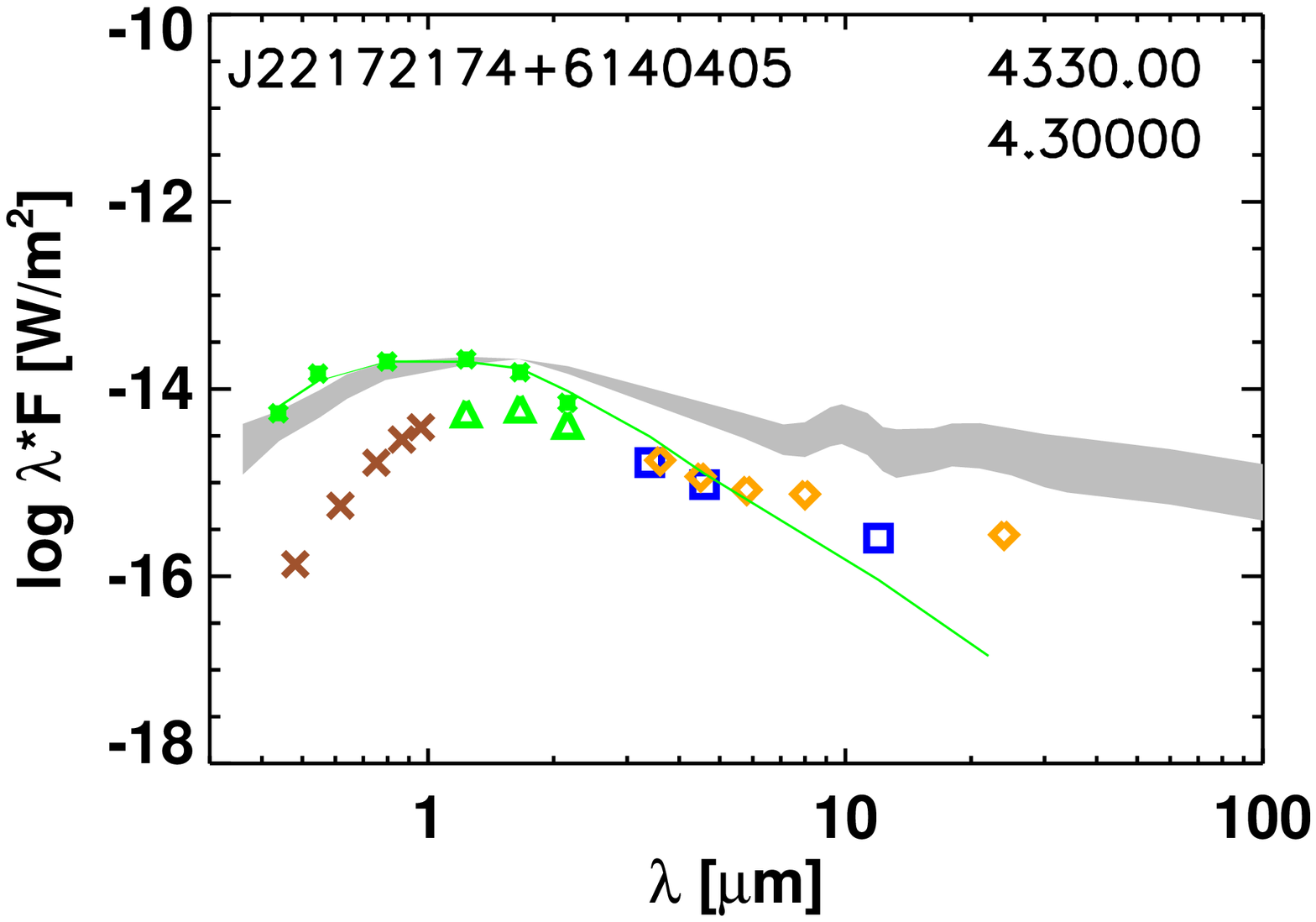}\includegraphics[scale=0.2]{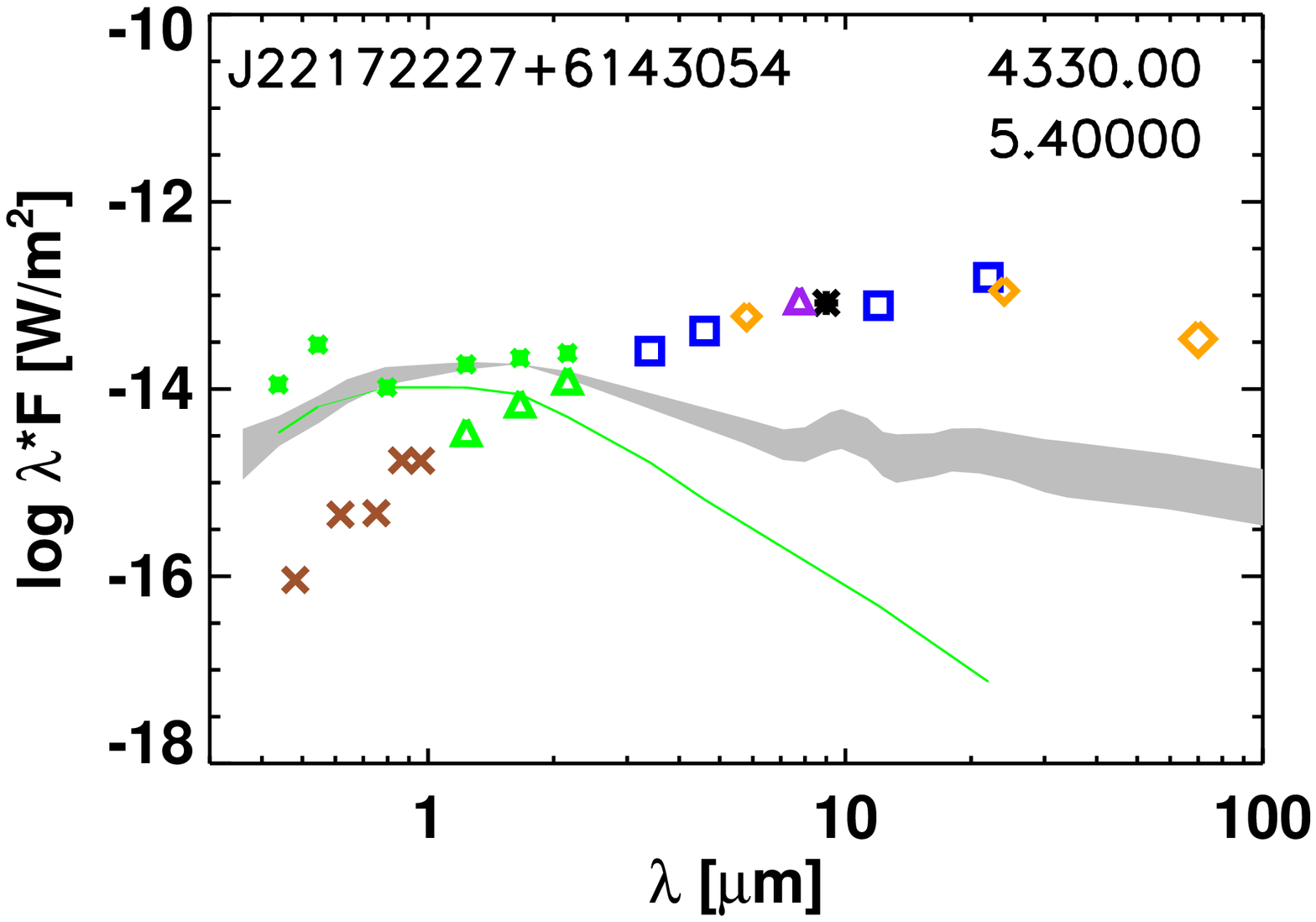}\includegraphics[scale=0.2]{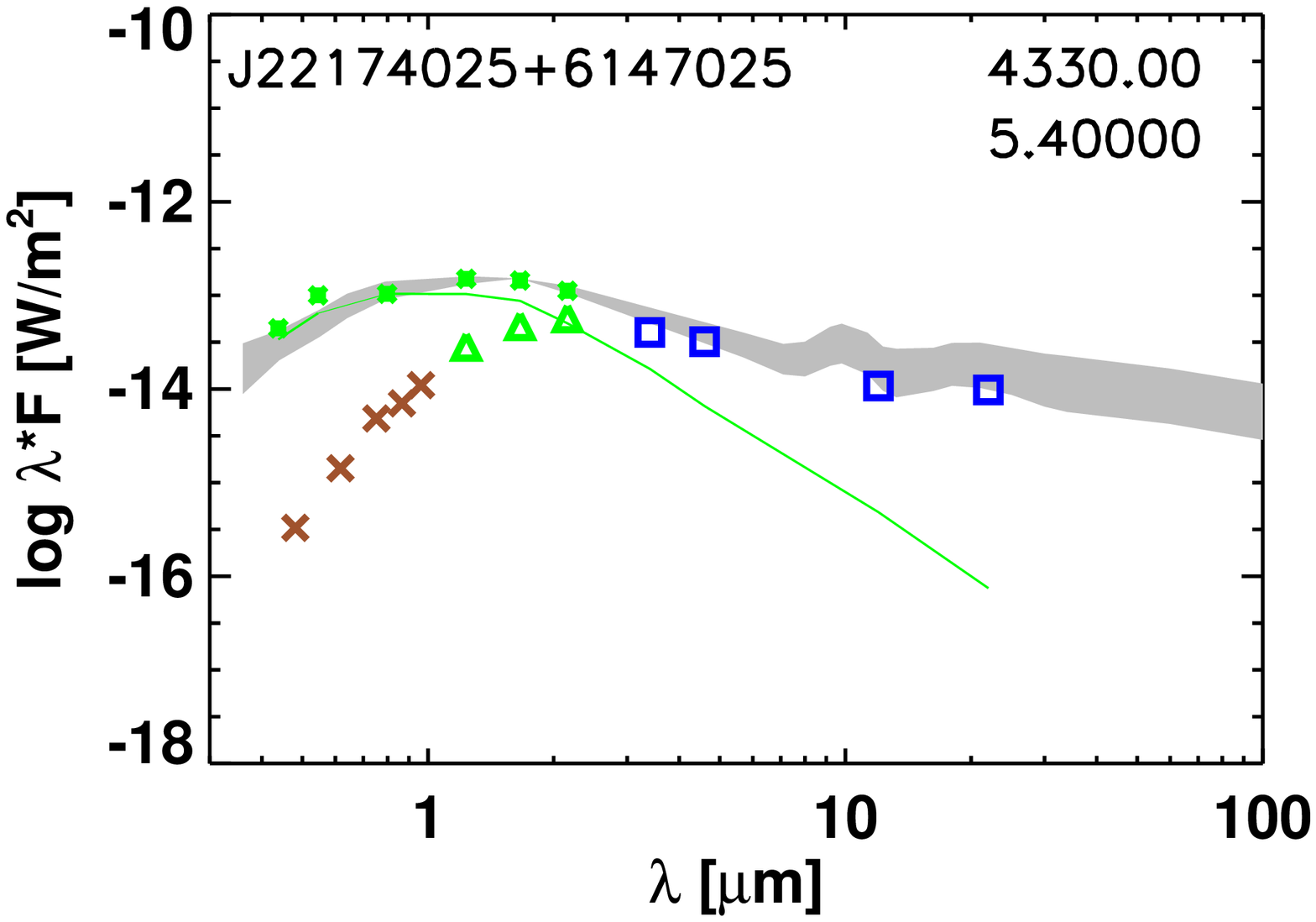}}
\caption{SEDs of the H$\alpha$ emission stars. Brown X show the Pan-STARRS data, green triangles symbol 2MASS data, blue squares indicate  {\it WISE}  data and yellow diamonds represent  {\it Spitzer} data. Green stars indicate the dereddened SED, and the solid green line shows the photospheric SED of the spectral type as a result of a fitting algorithm (see the text). The gray shaded band indicates the median SED of the T Tauri stars of Taurus star-forming region.}\label{halfa_mamajack_sed}
\end{center}                                                                                                                                                                                                                                                                                                                                                                     

\end{figure*}

\begin{figure*}
\begin{center}

\centerline{\includegraphics[scale=0.2]{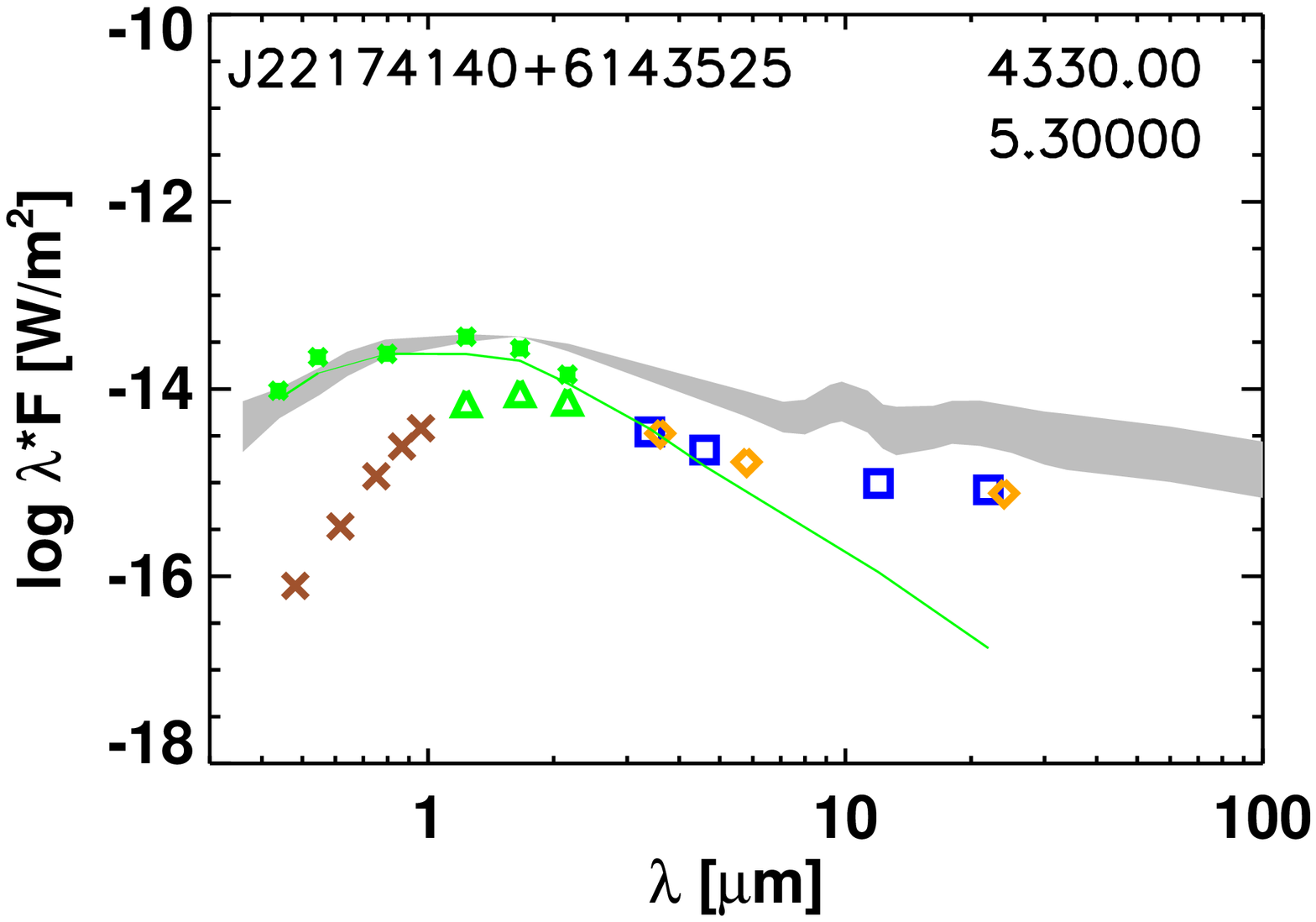}\includegraphics[scale=0.2]{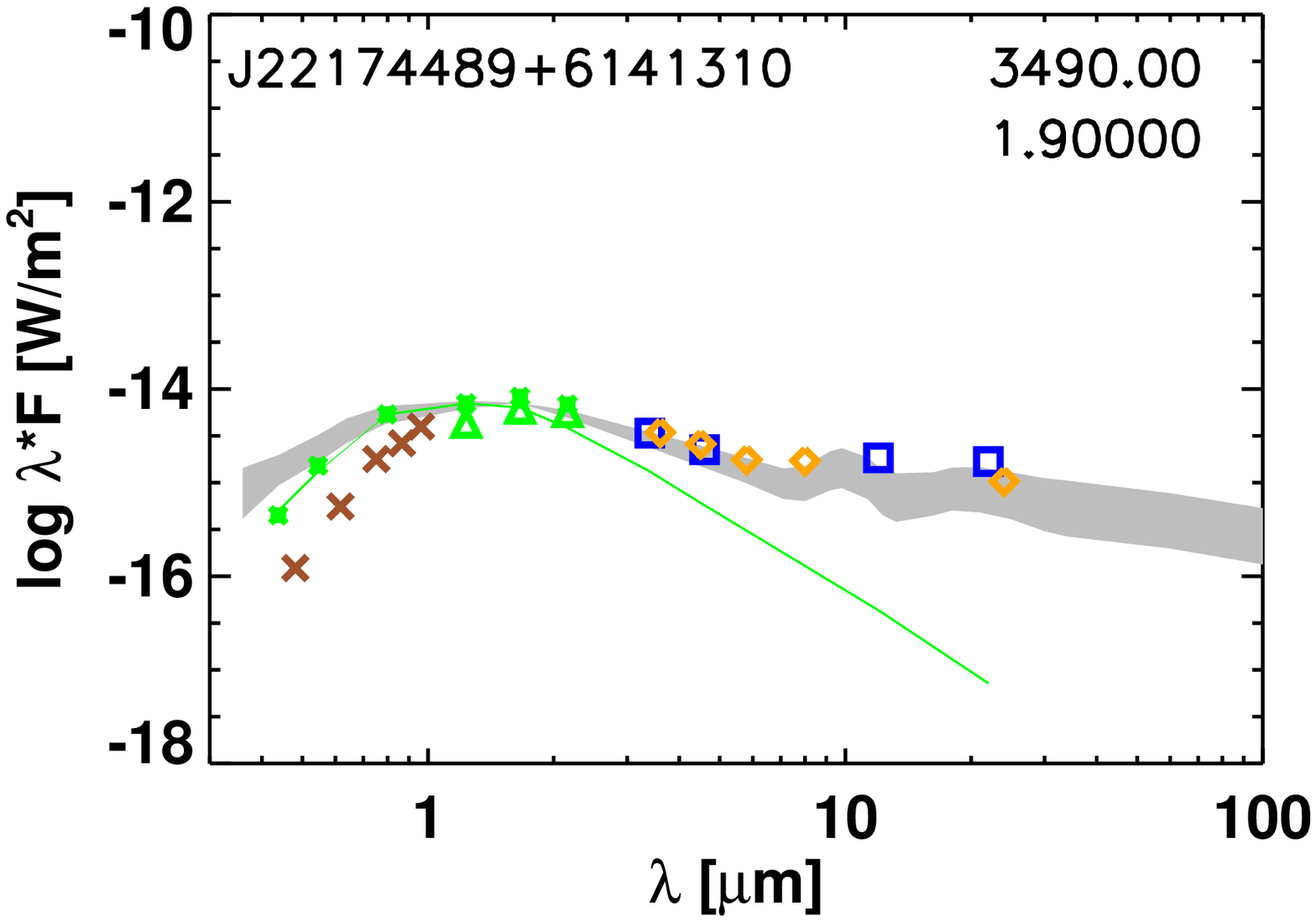}\includegraphics[scale=0.2]{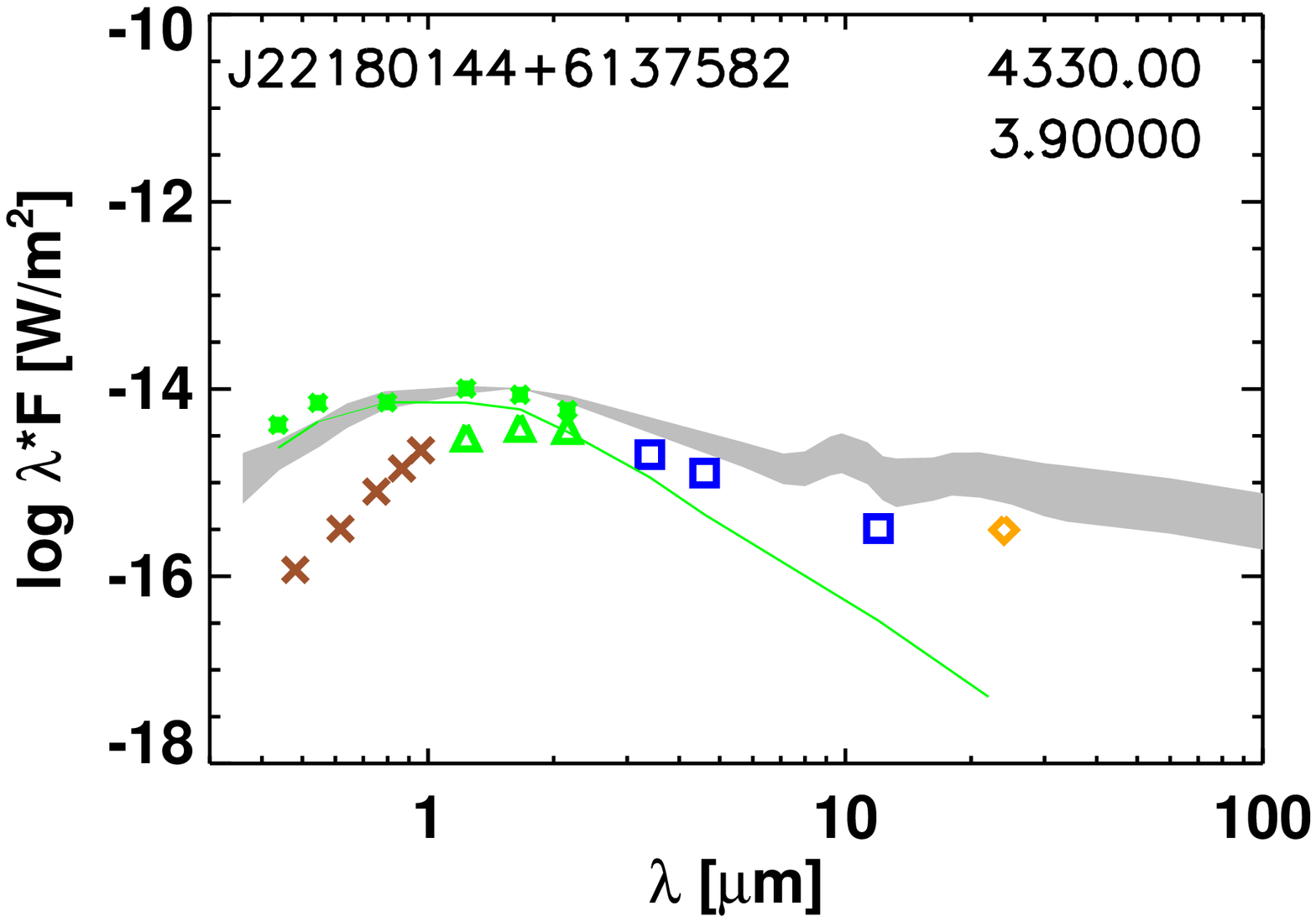}\includegraphics[scale=0.2]{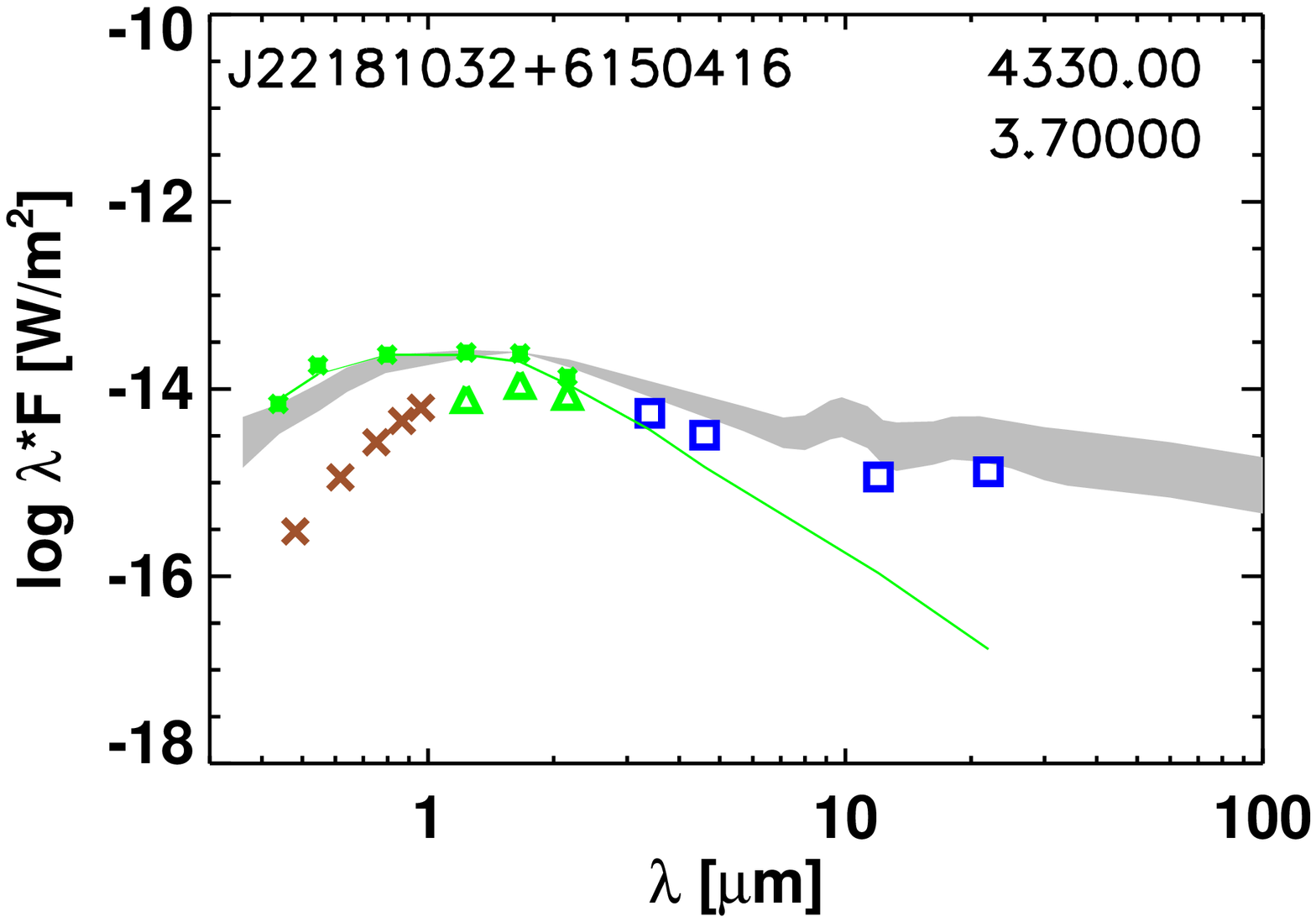}}

\centerline{\includegraphics[scale=0.2]{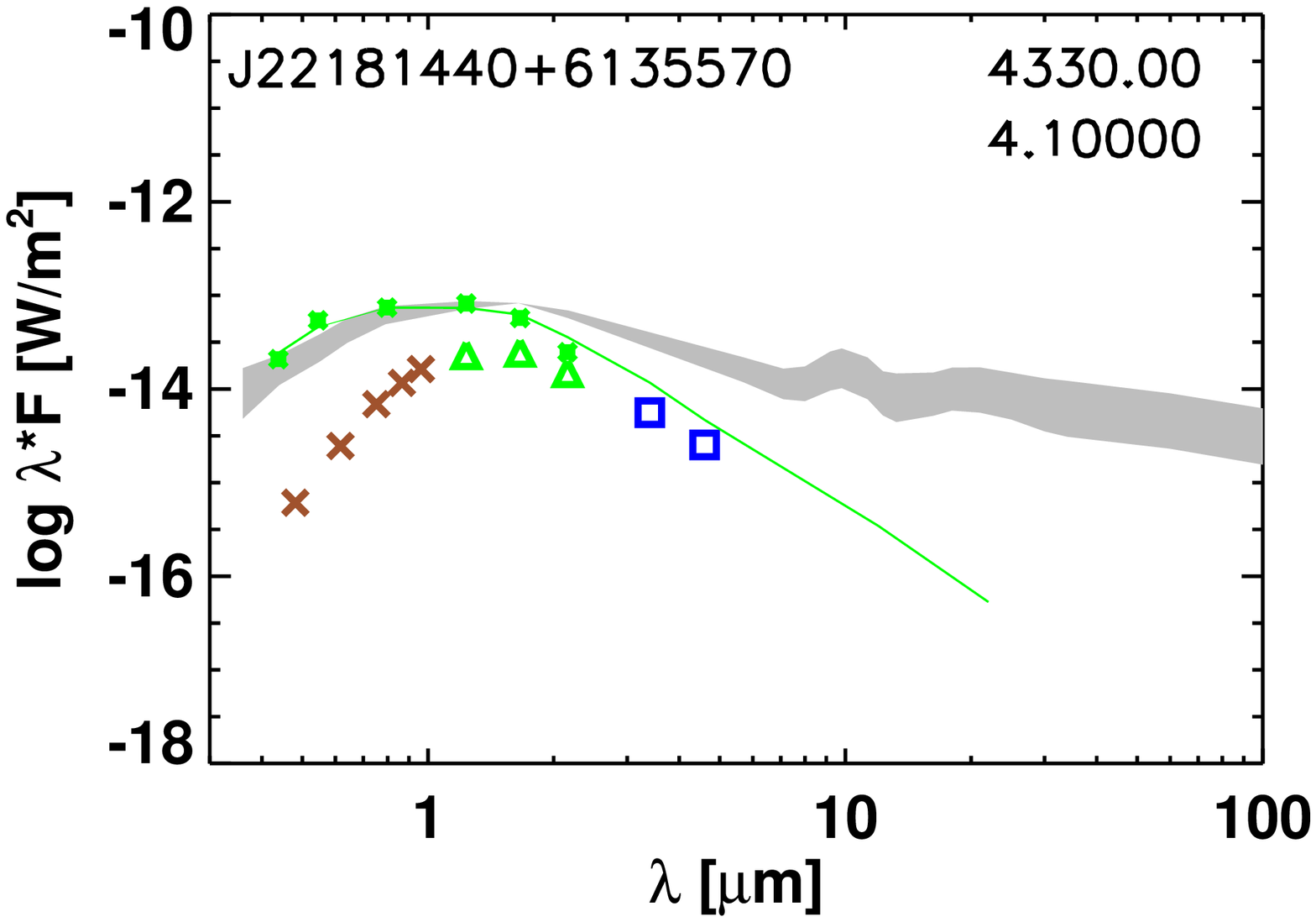}\includegraphics[scale=0.2]{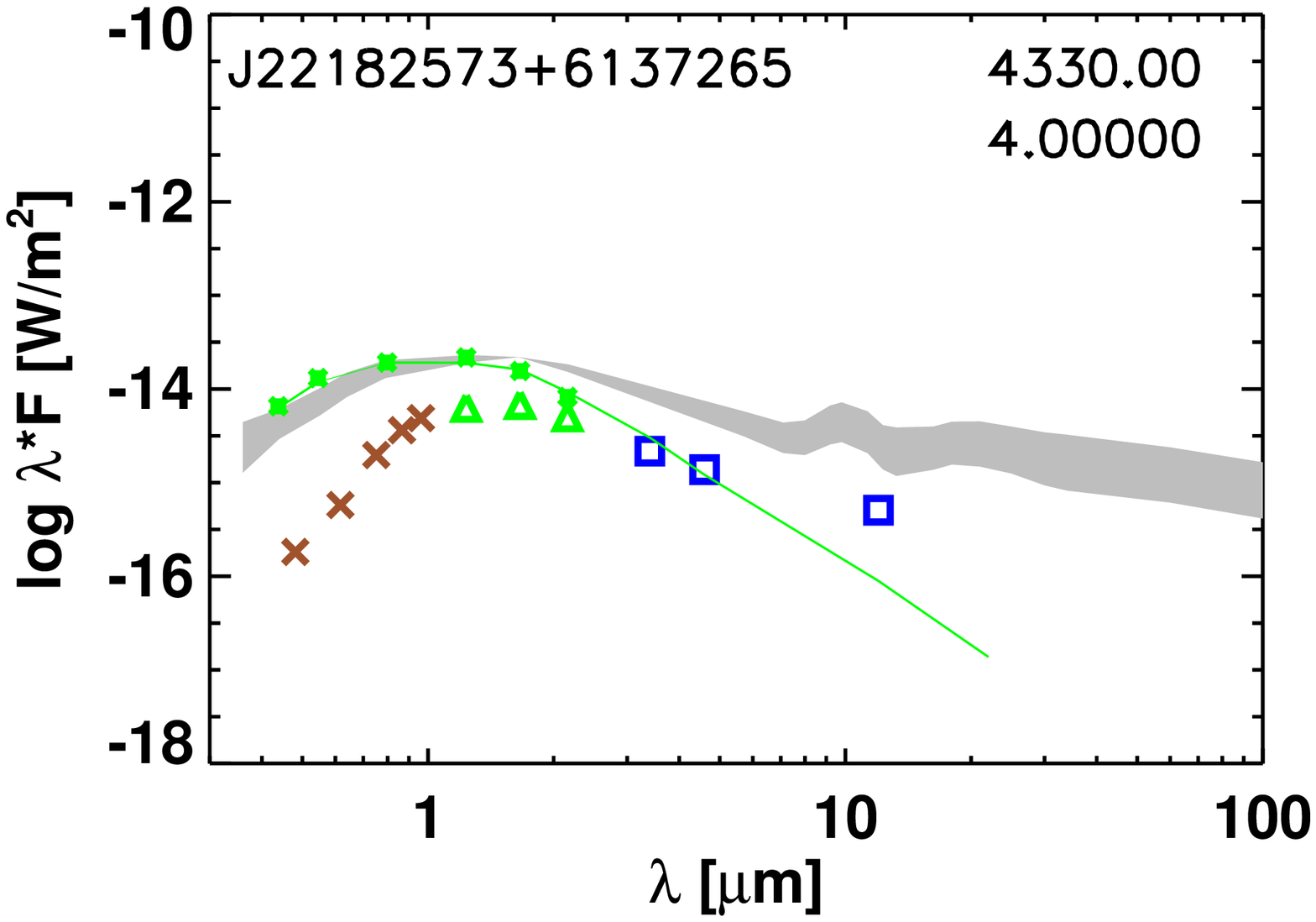}\includegraphics[scale=0.2]{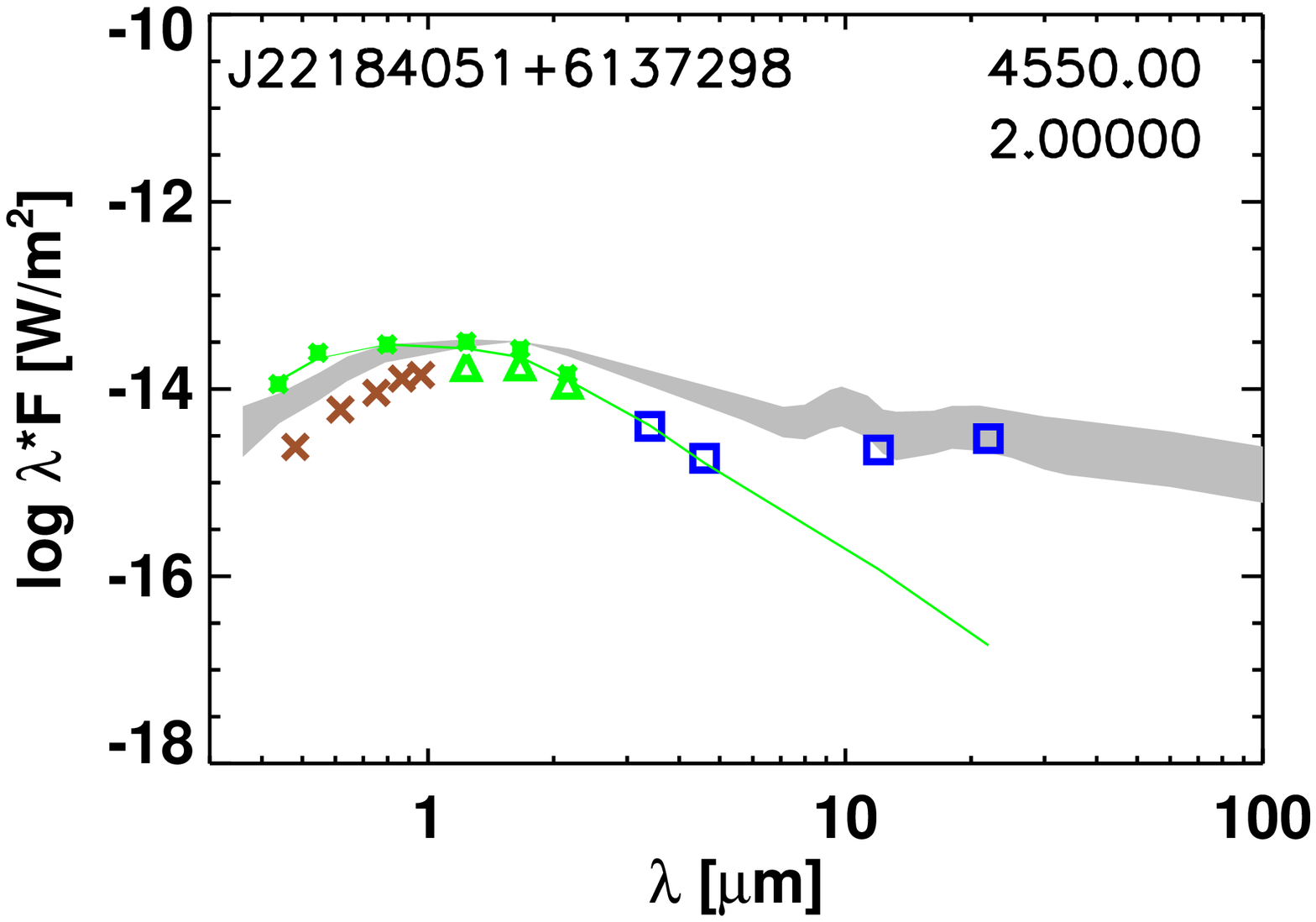}\includegraphics[scale=0.2]{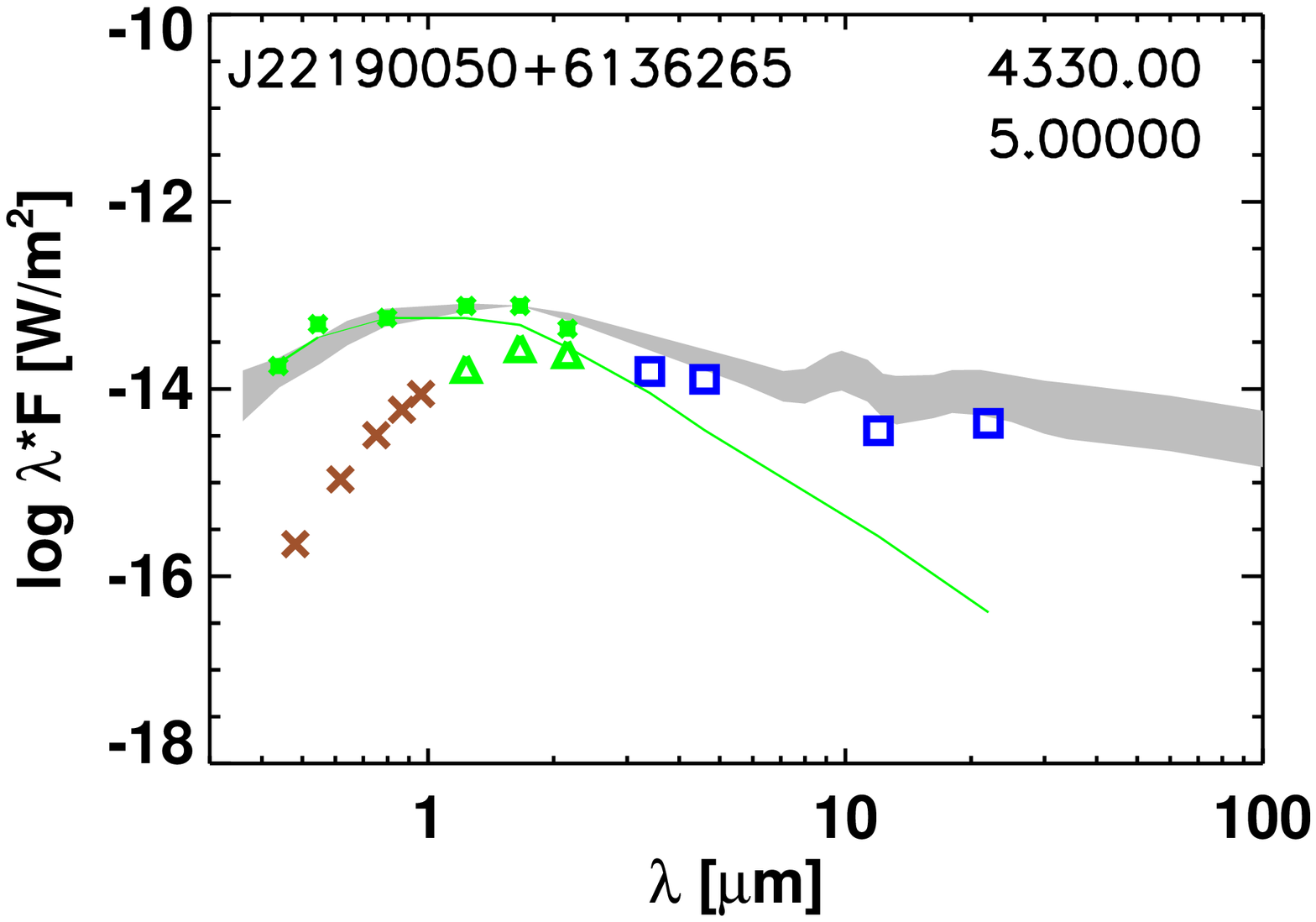}}

\centerline{\includegraphics[scale=0.2]{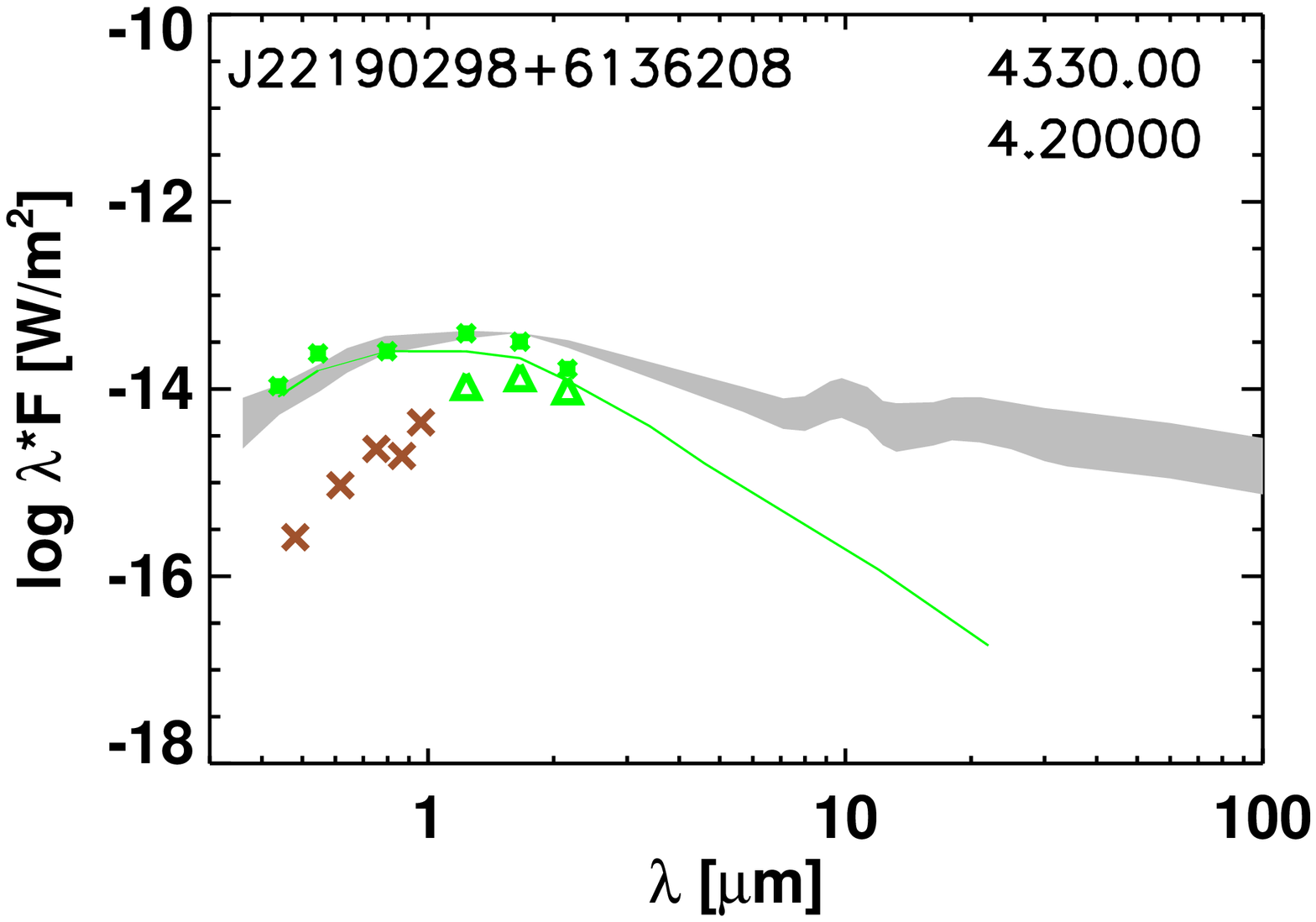}\includegraphics[scale=0.2]{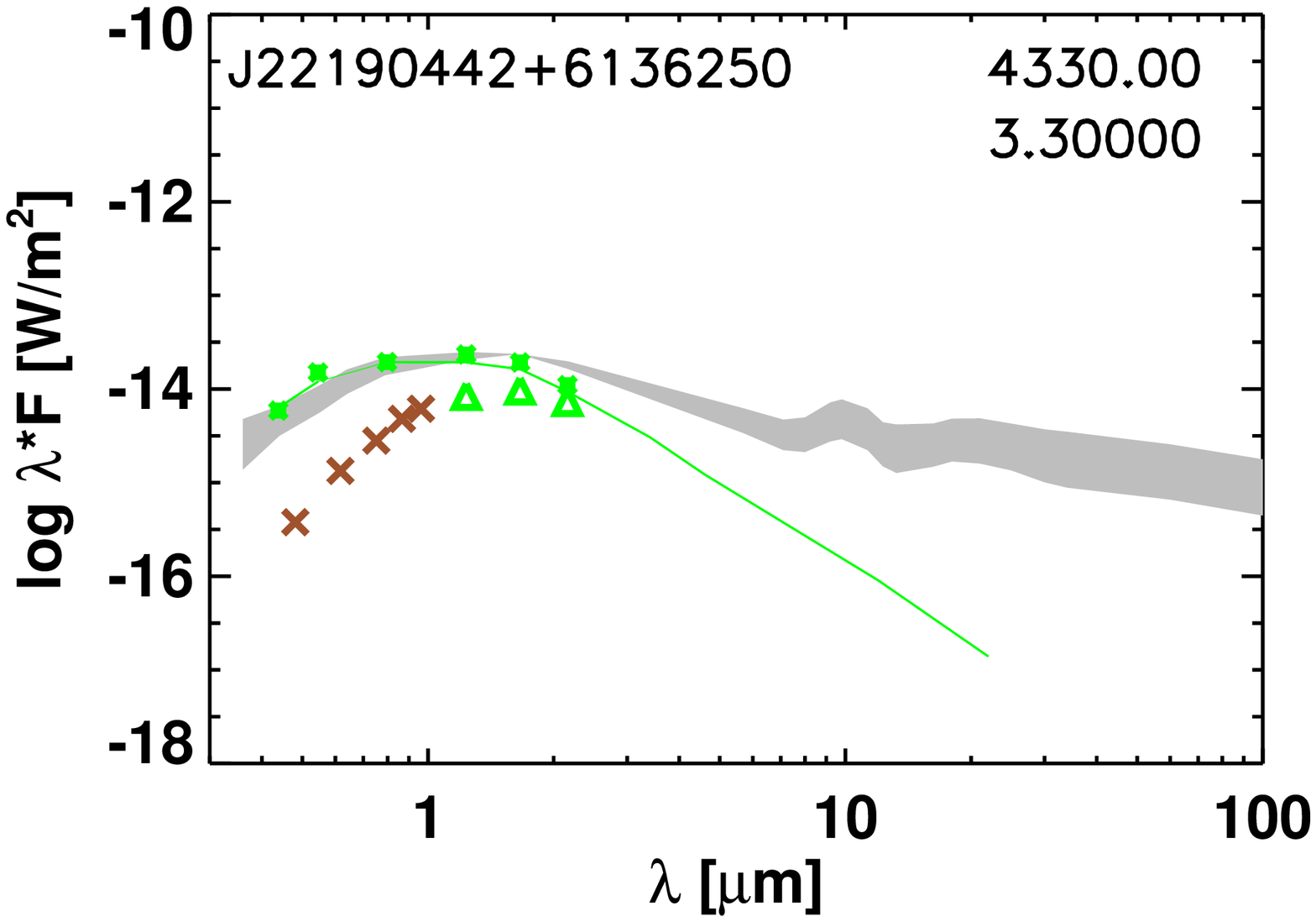}\includegraphics[scale=0.2]{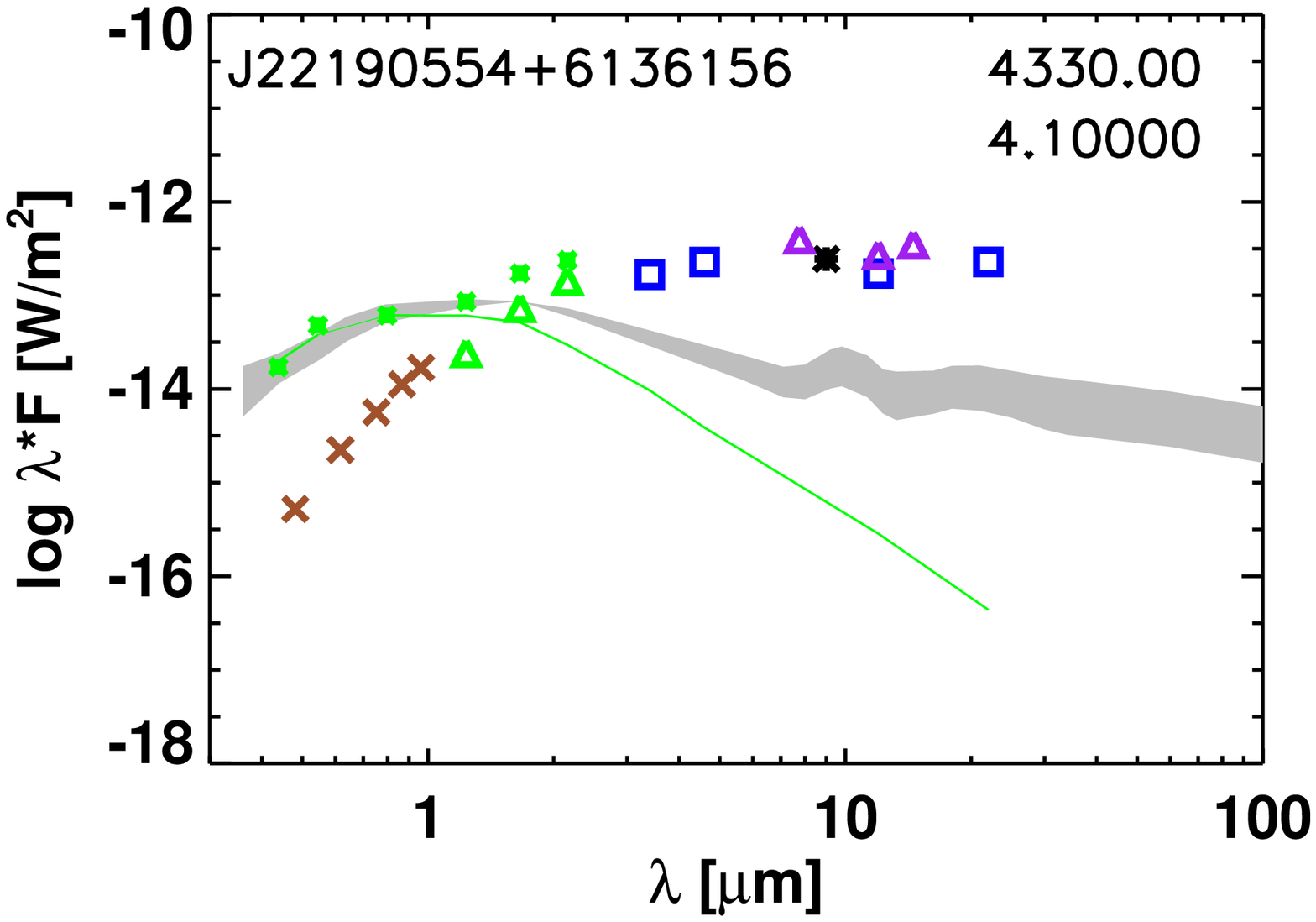}\includegraphics[scale=0.2]{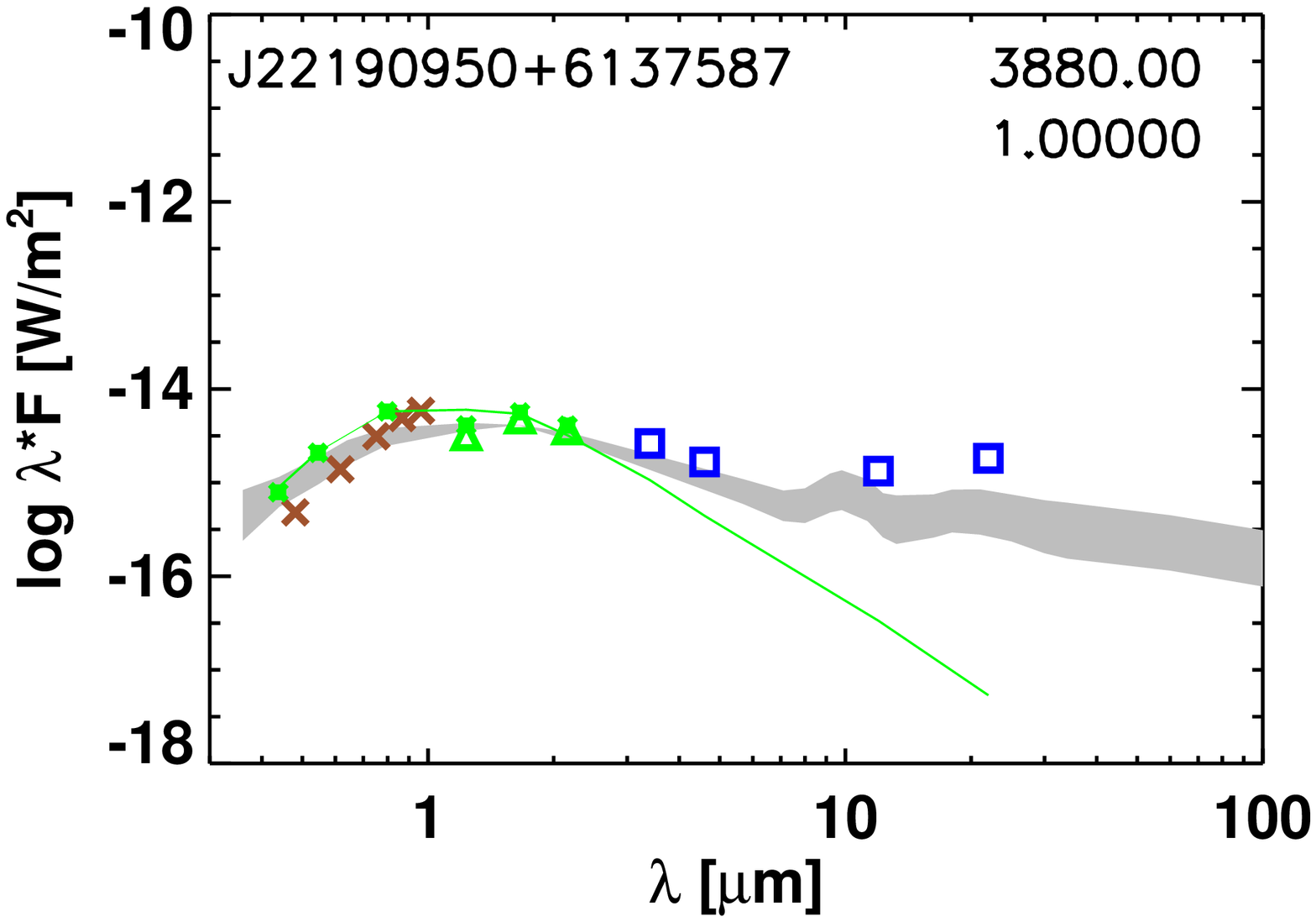}}

\centerline{\includegraphics[scale=0.2]{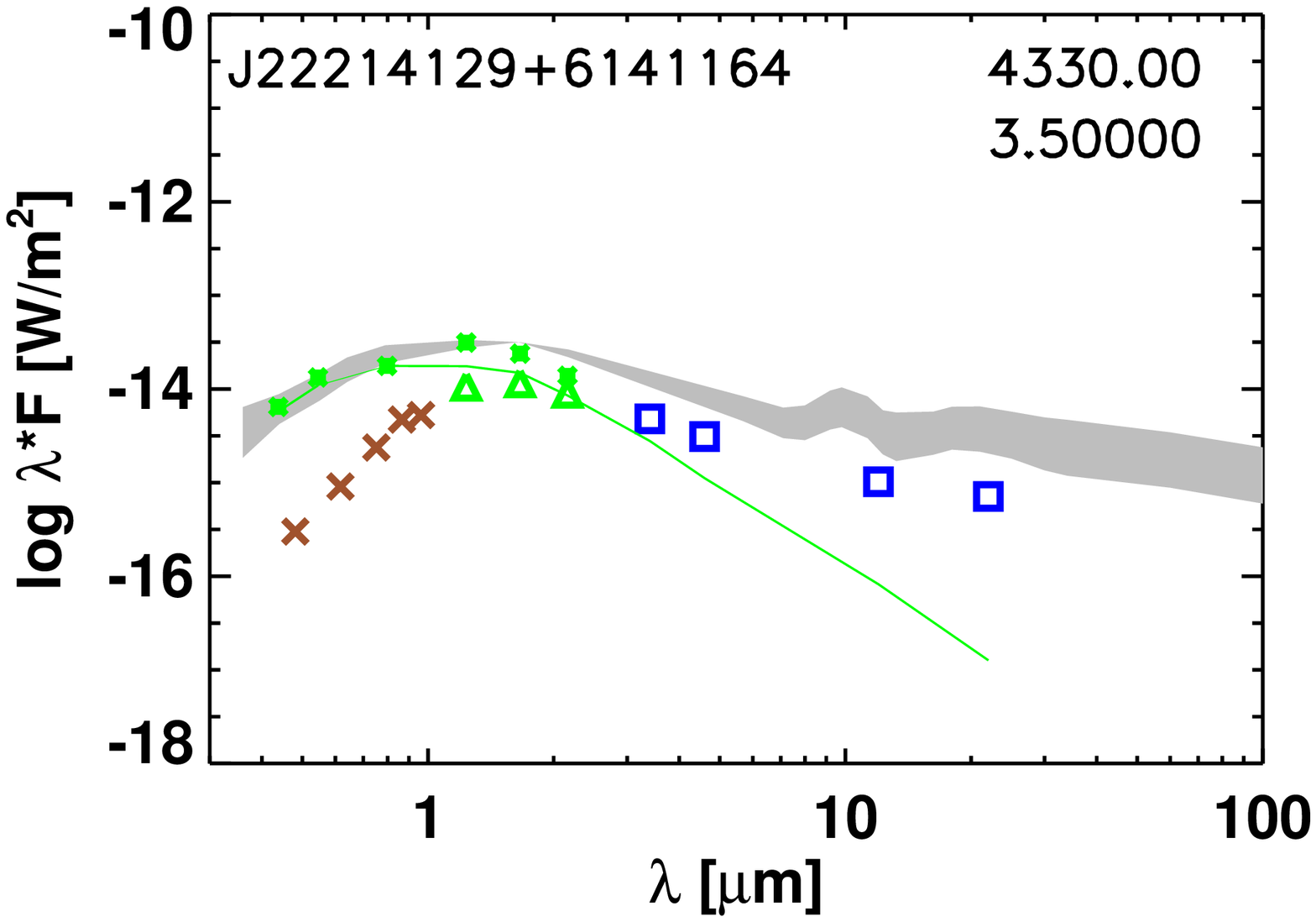}}
\contcaption{ }
\end{center}
\end{figure*}

\begin{figure*}
\begin{center}
\centerline{\includegraphics[scale=0.2]{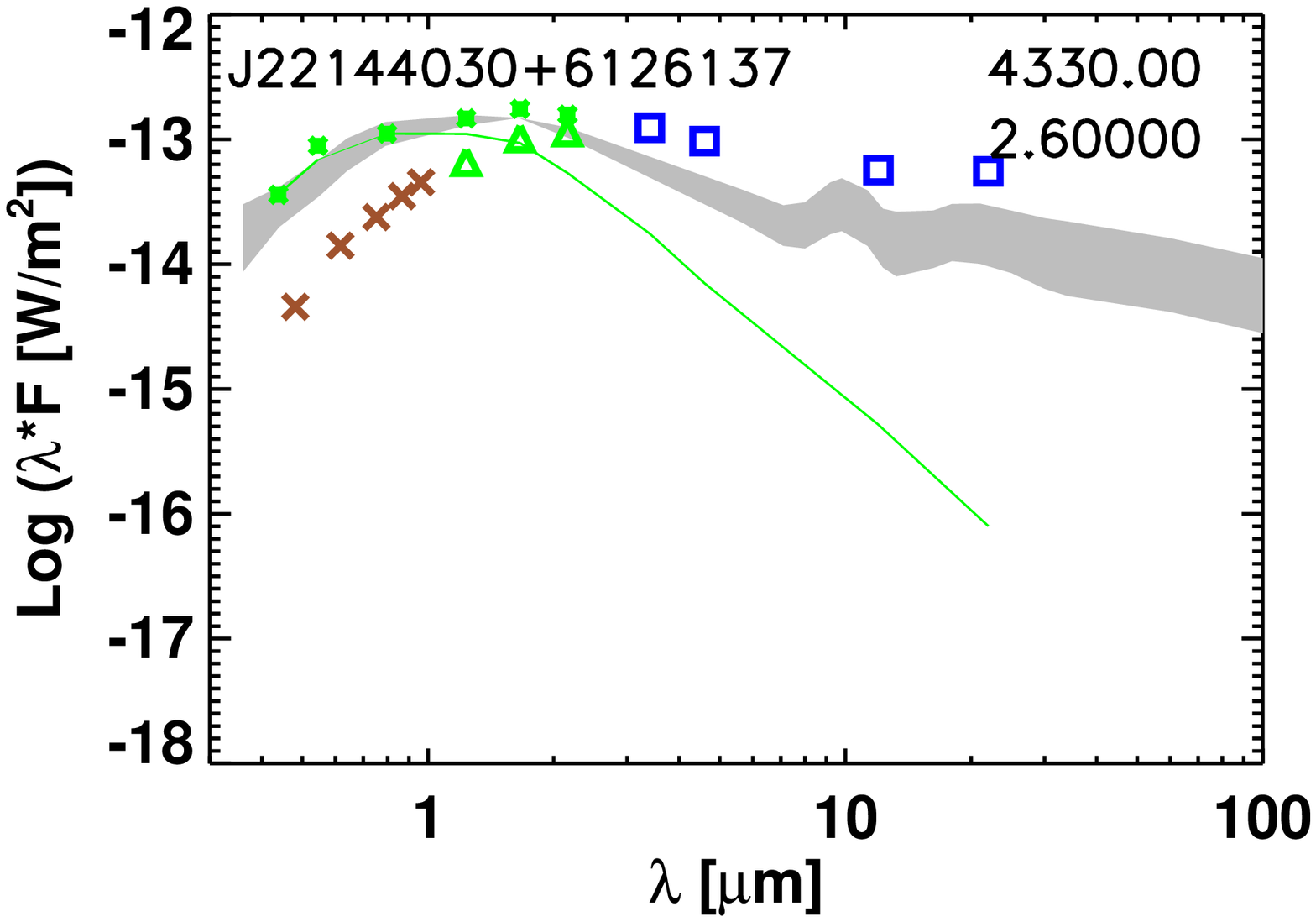}\includegraphics[scale=0.2]{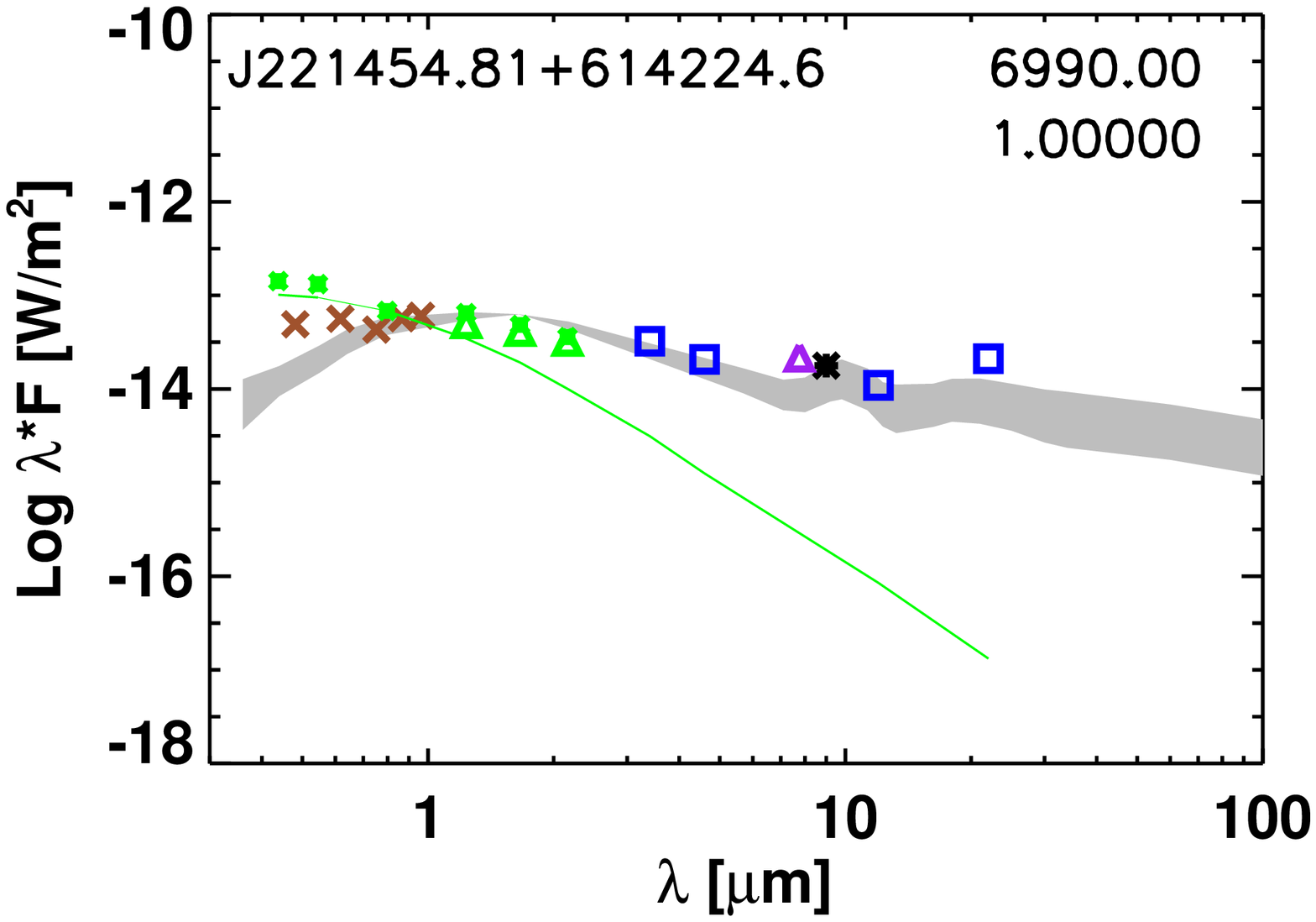}\includegraphics[scale=0.2]{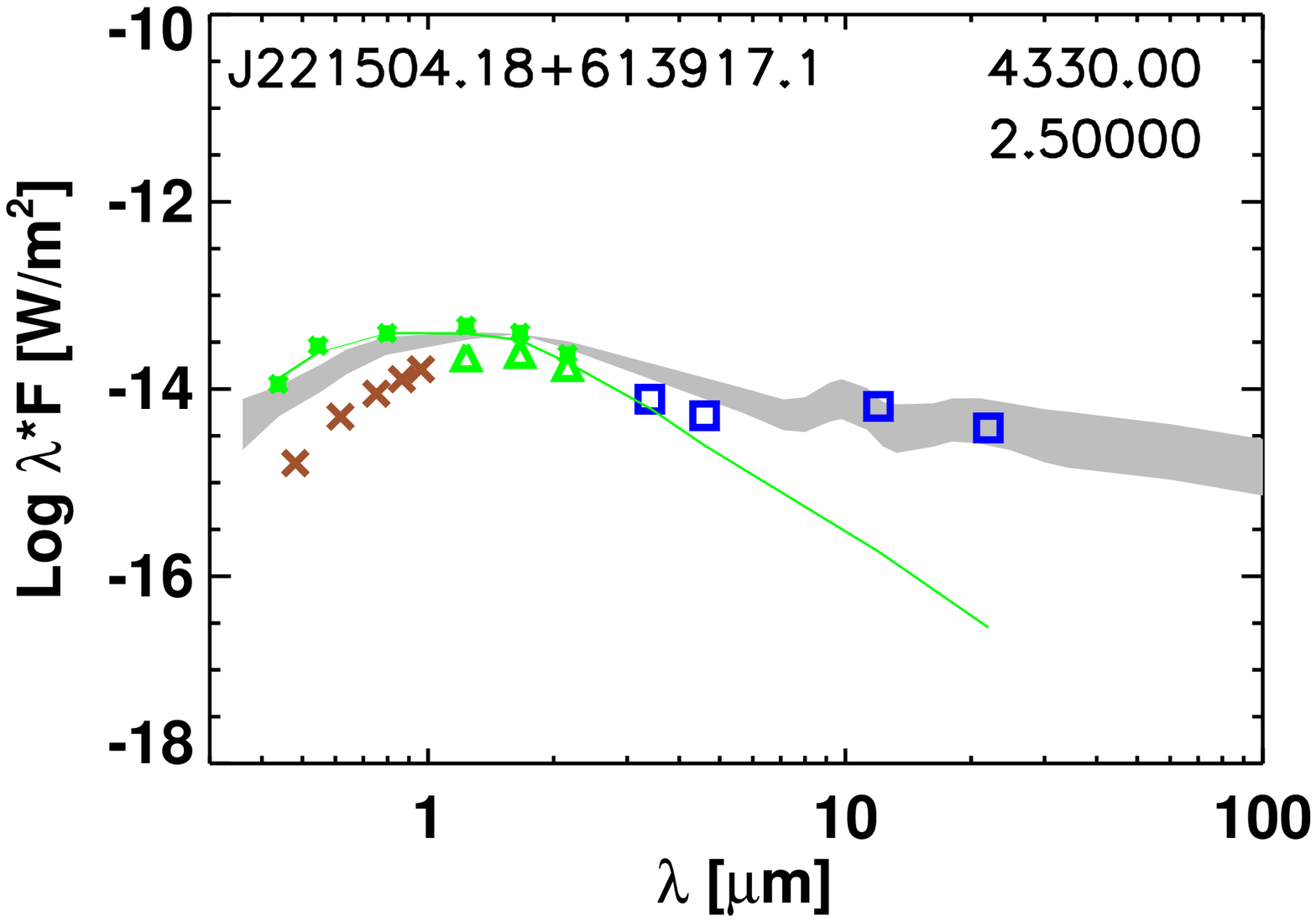}\includegraphics[scale=0.2]{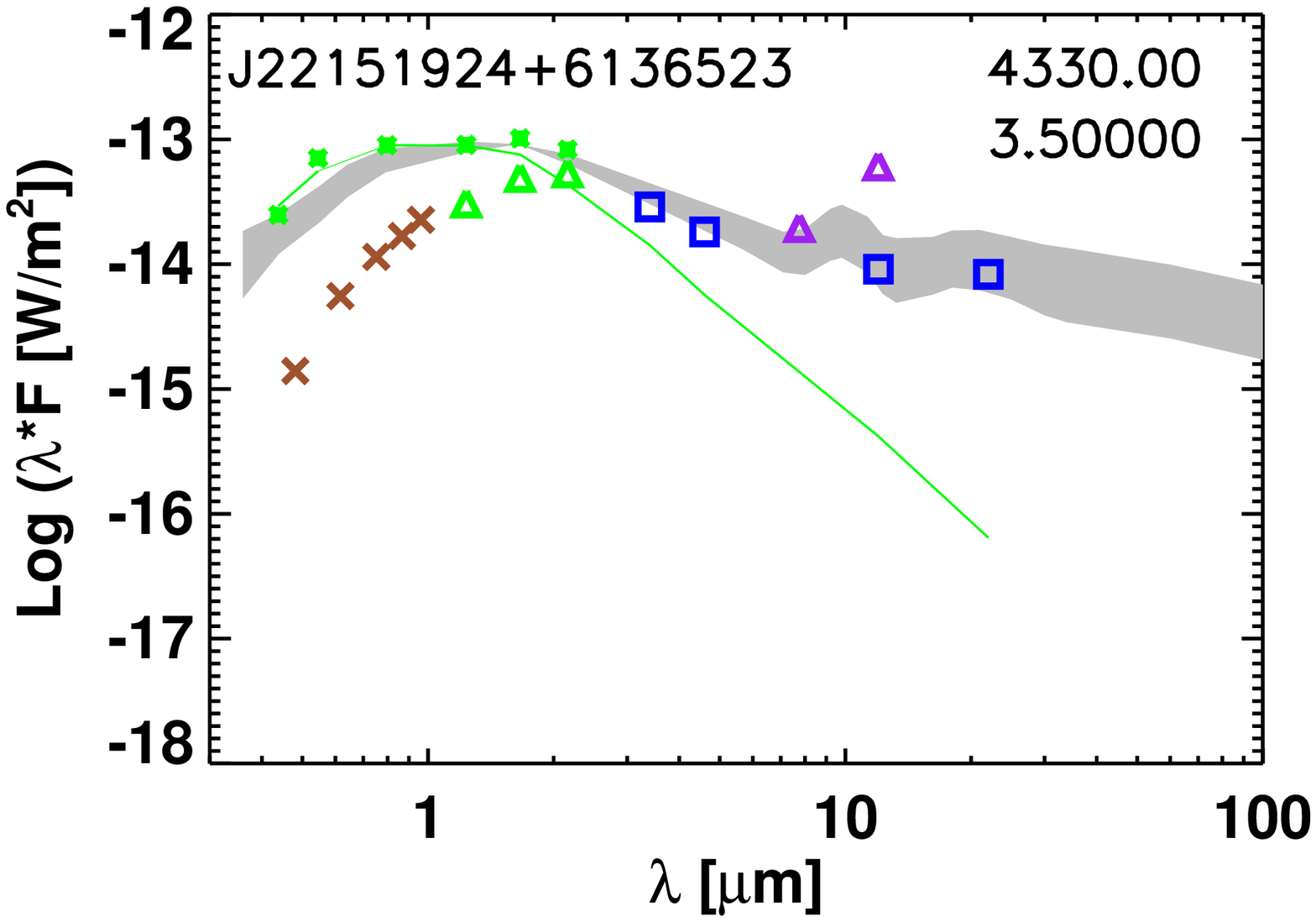}}

\centerline{\includegraphics[scale=0.2]{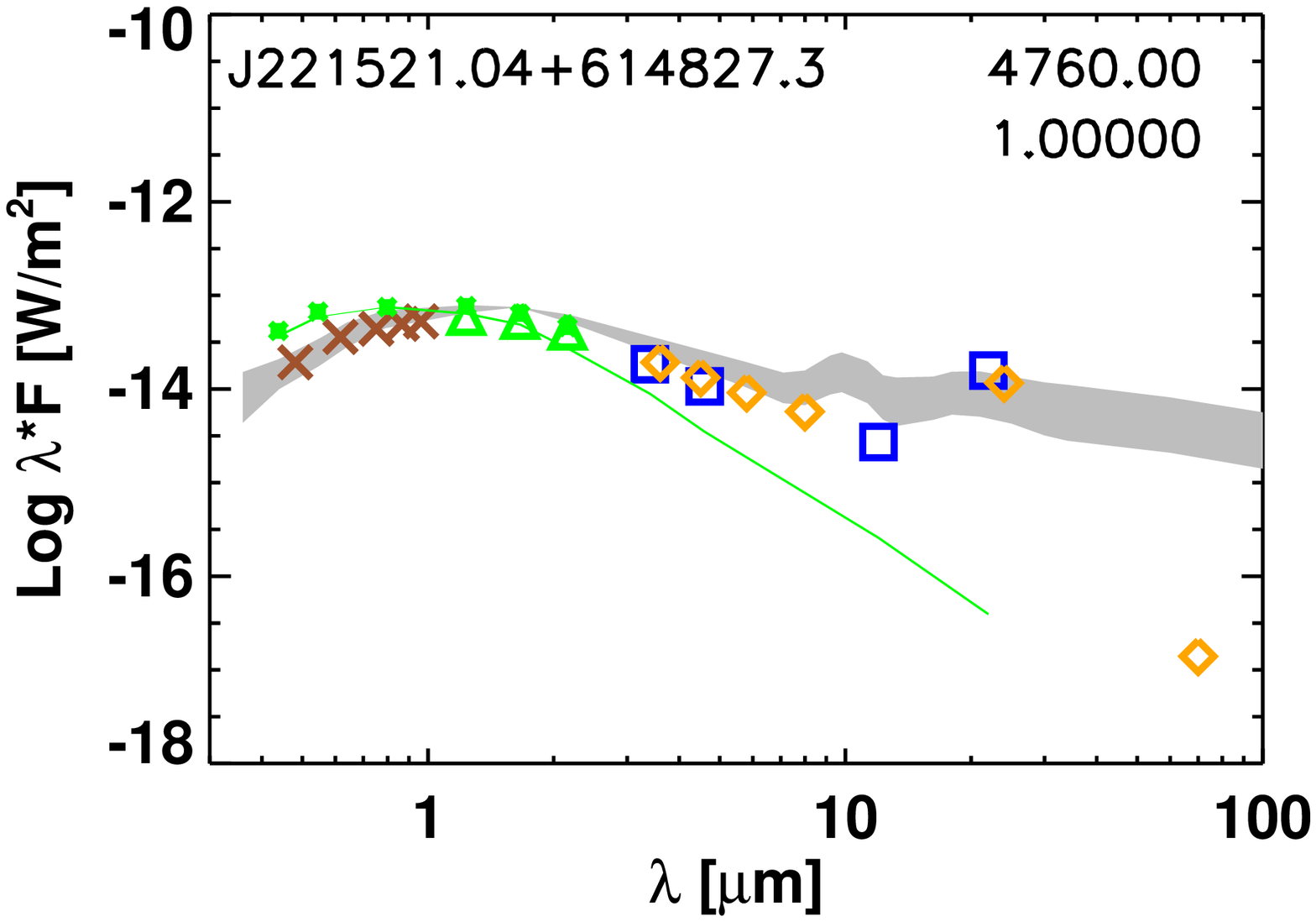}\includegraphics[scale=0.2]{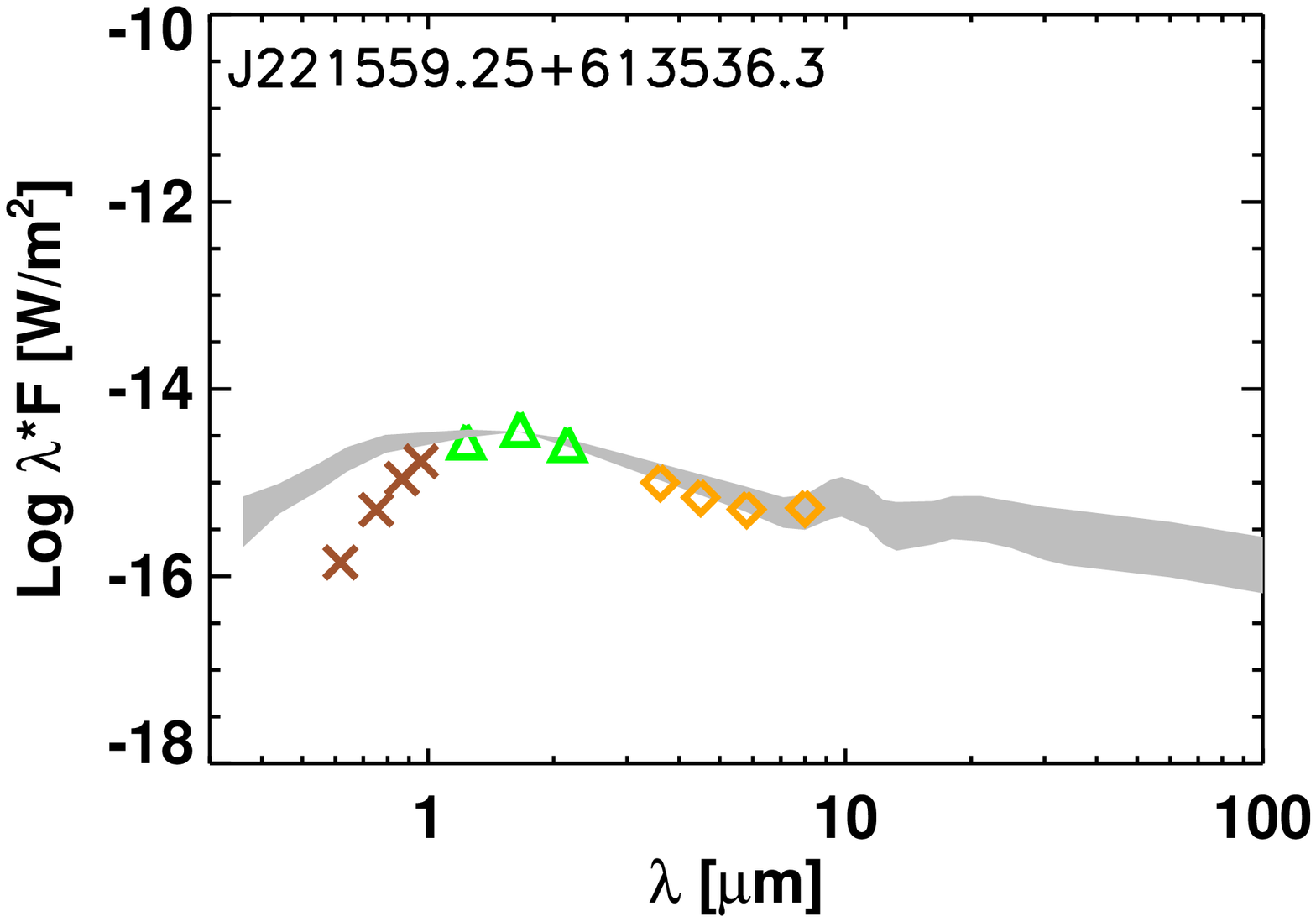}\includegraphics[scale=0.2]{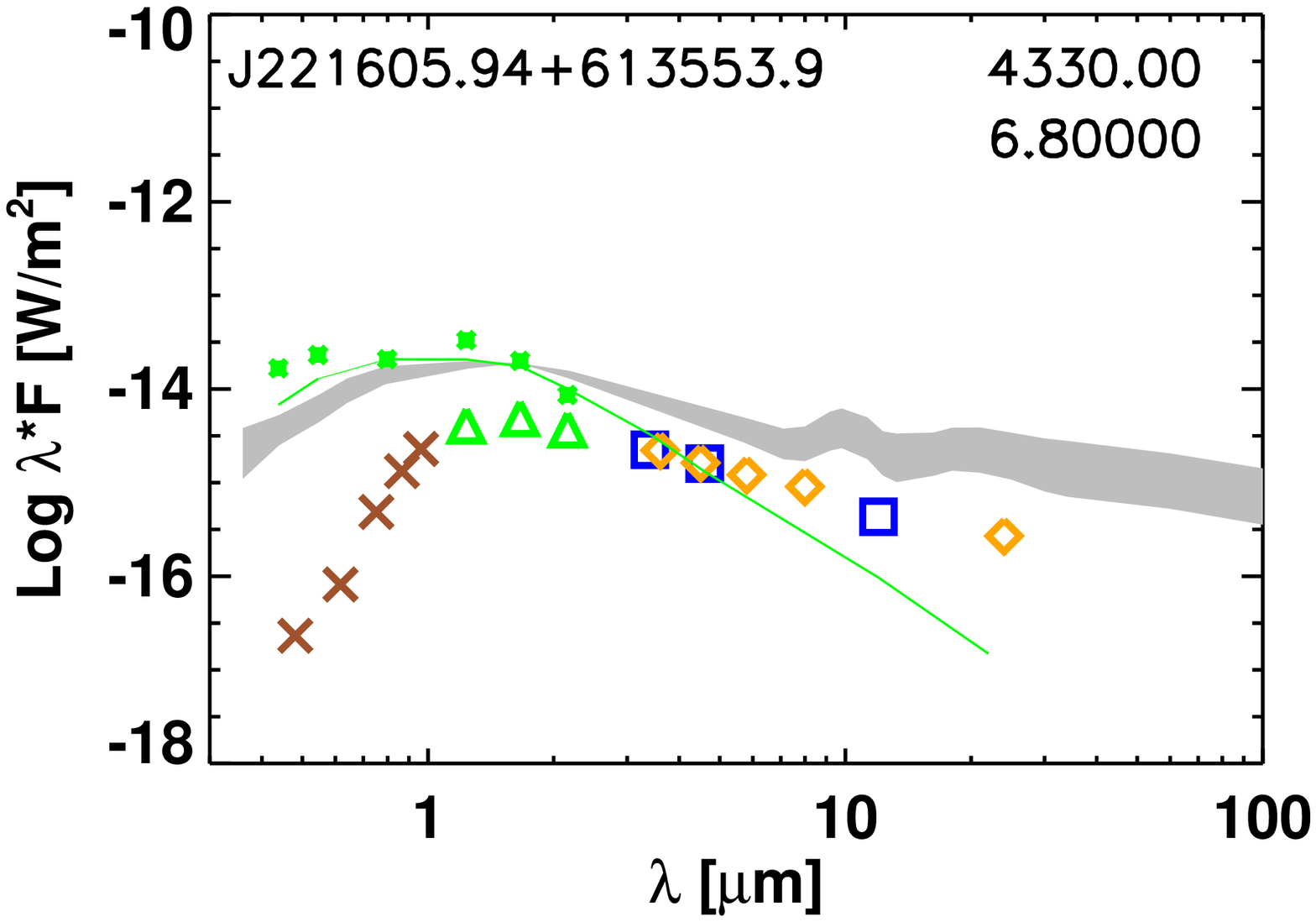}\includegraphics[scale=0.2]{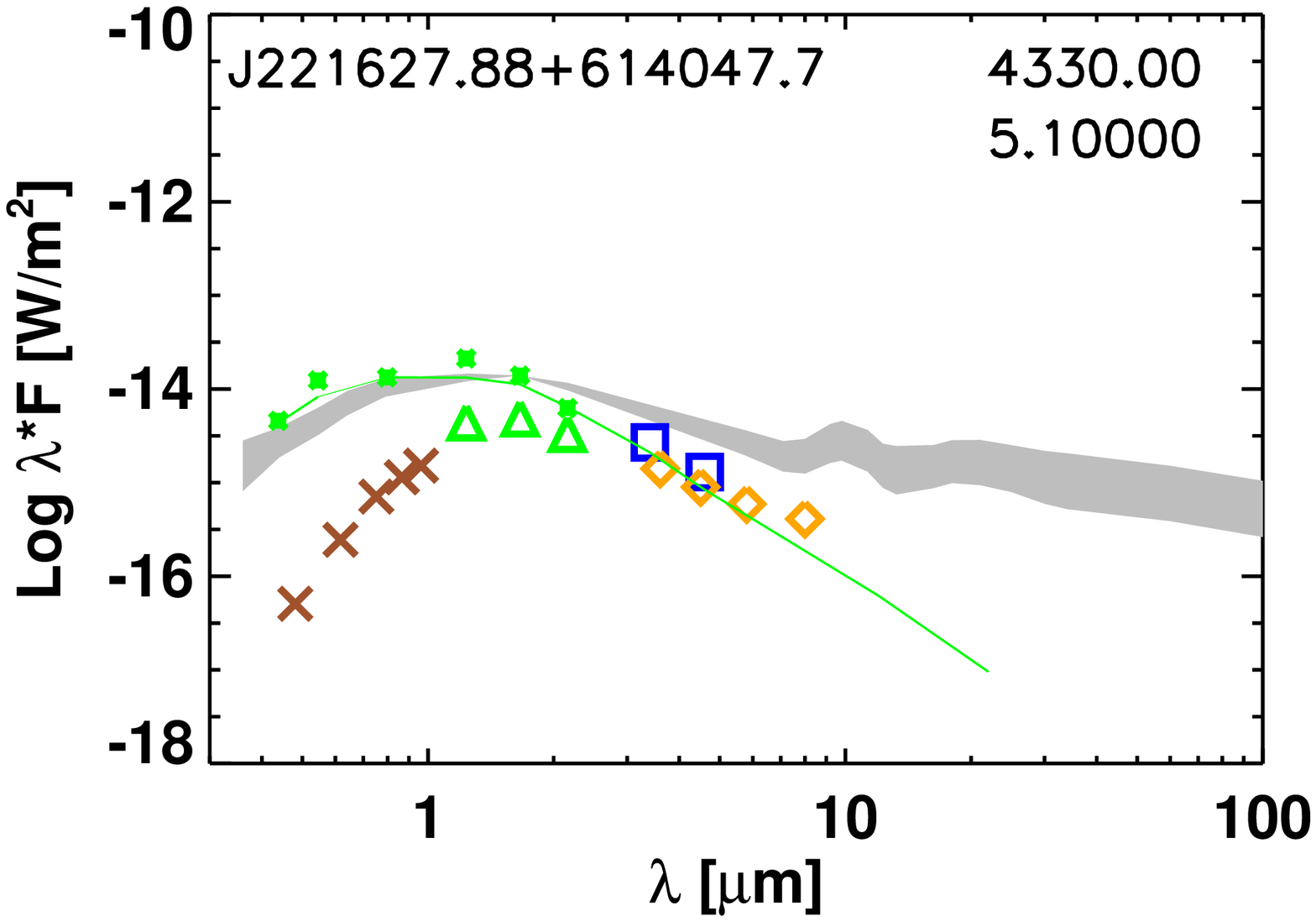}}

\centerline{\includegraphics[scale=0.2]{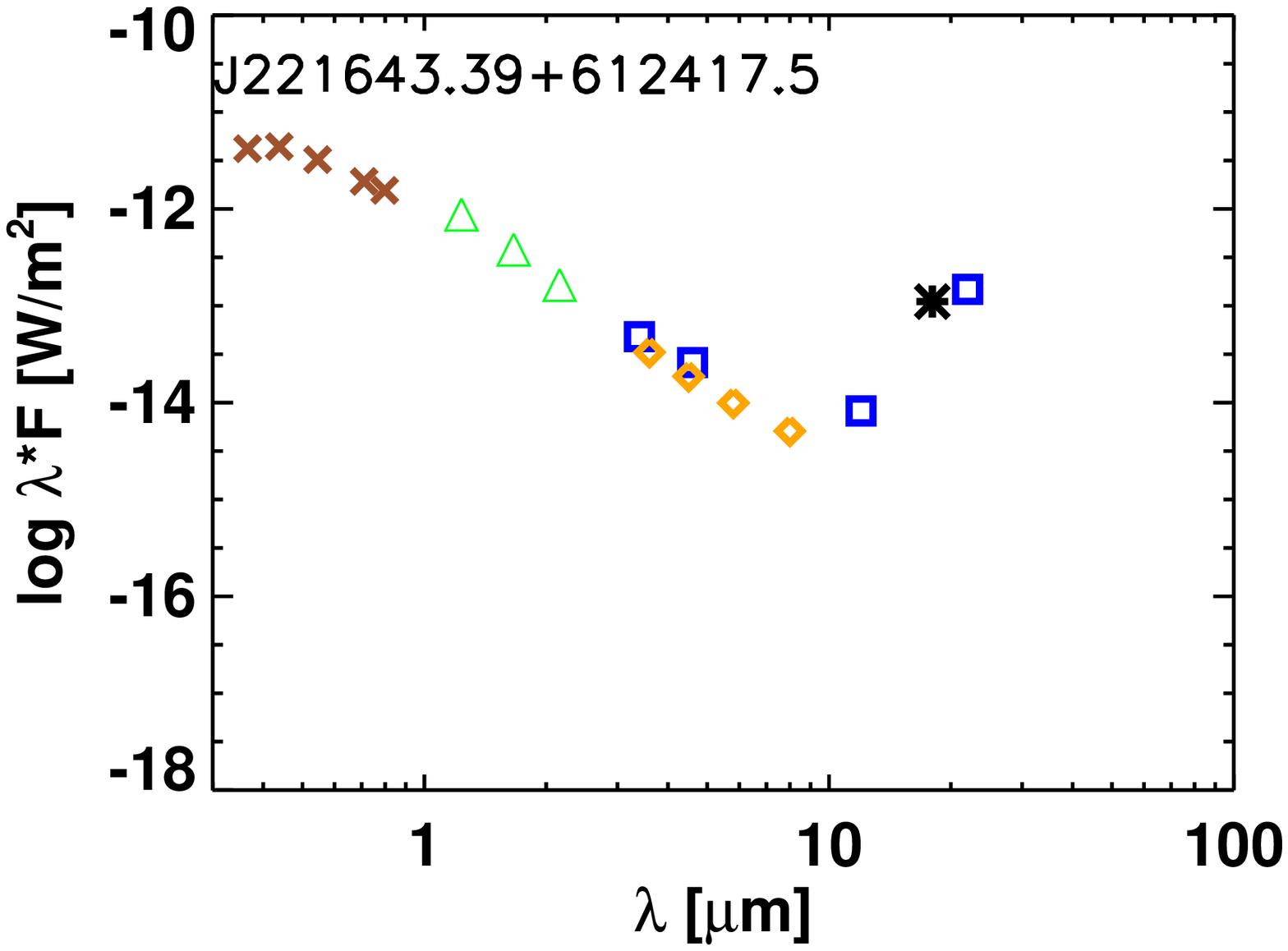}\includegraphics[scale=0.2]{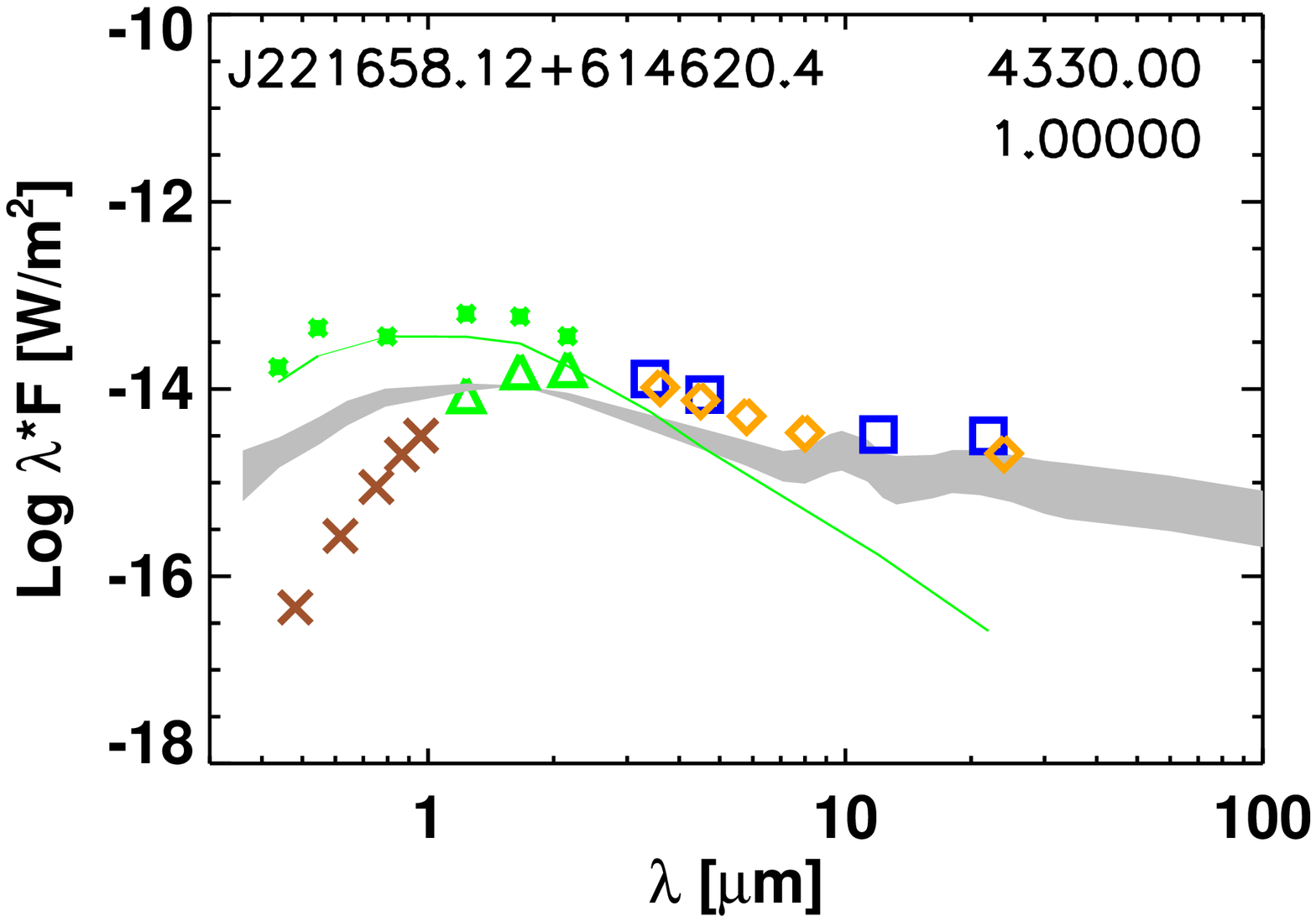}\includegraphics[scale=0.2]{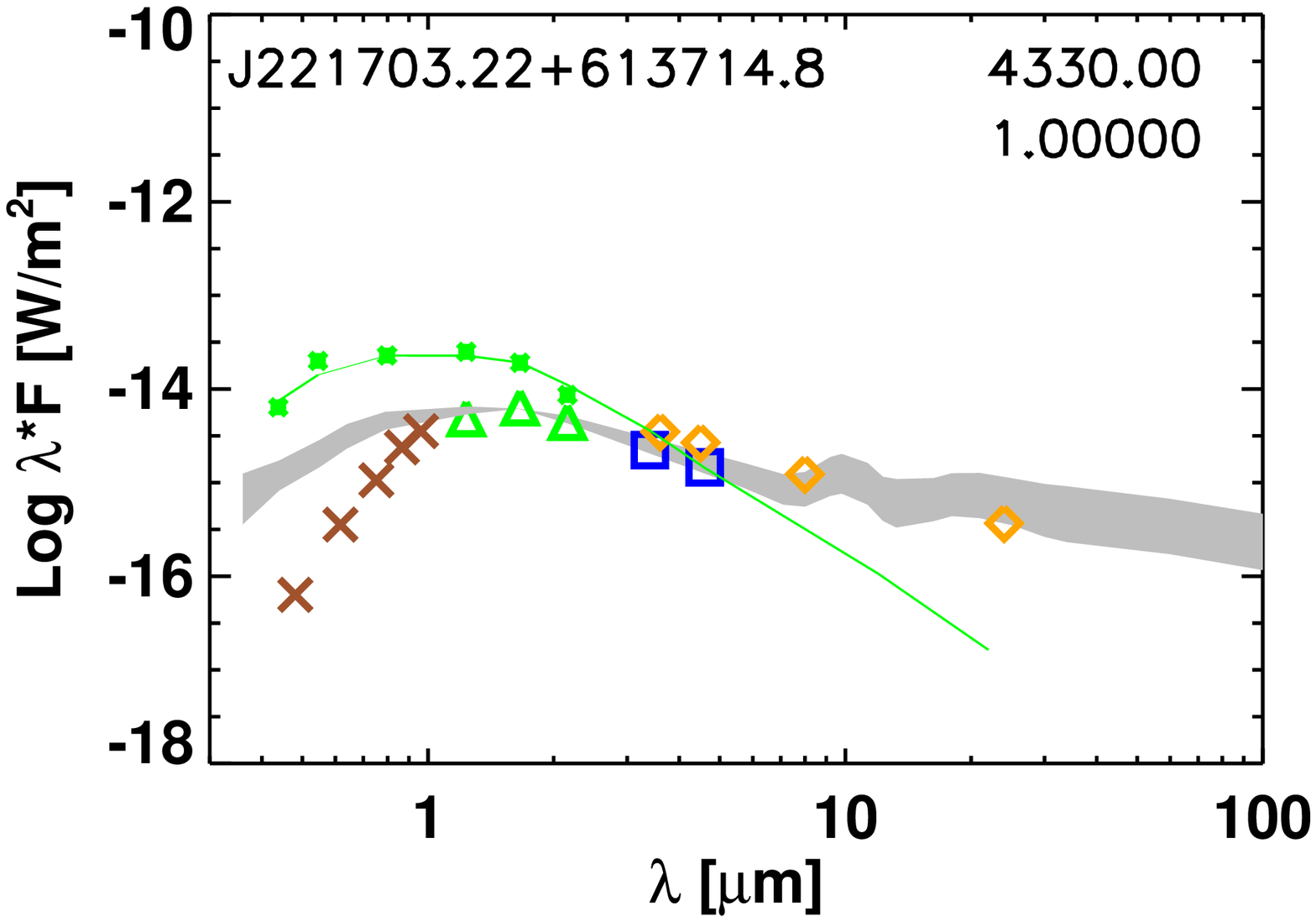}\includegraphics[scale=0.2]{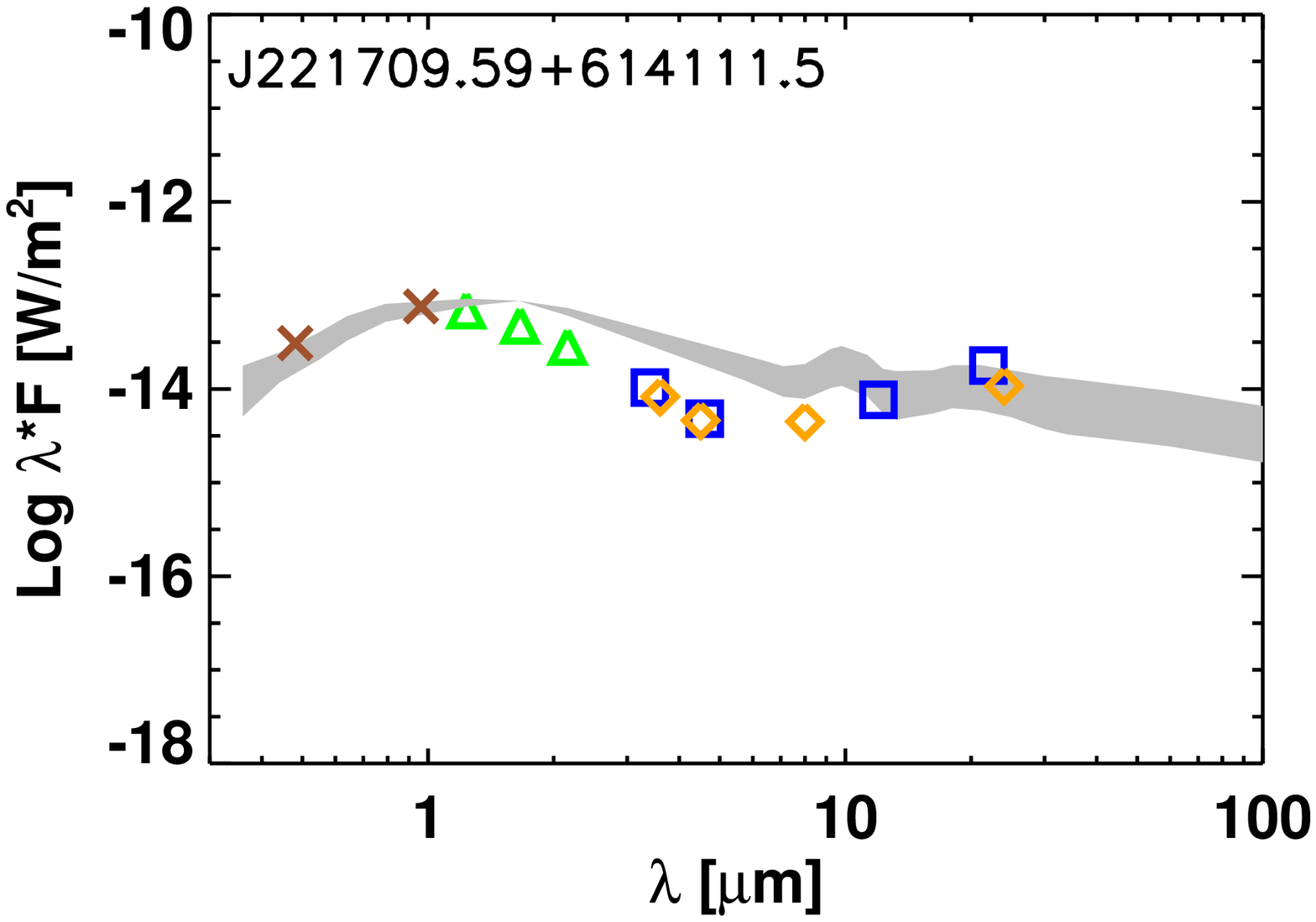}}

\centerline{\includegraphics[scale=0.2]{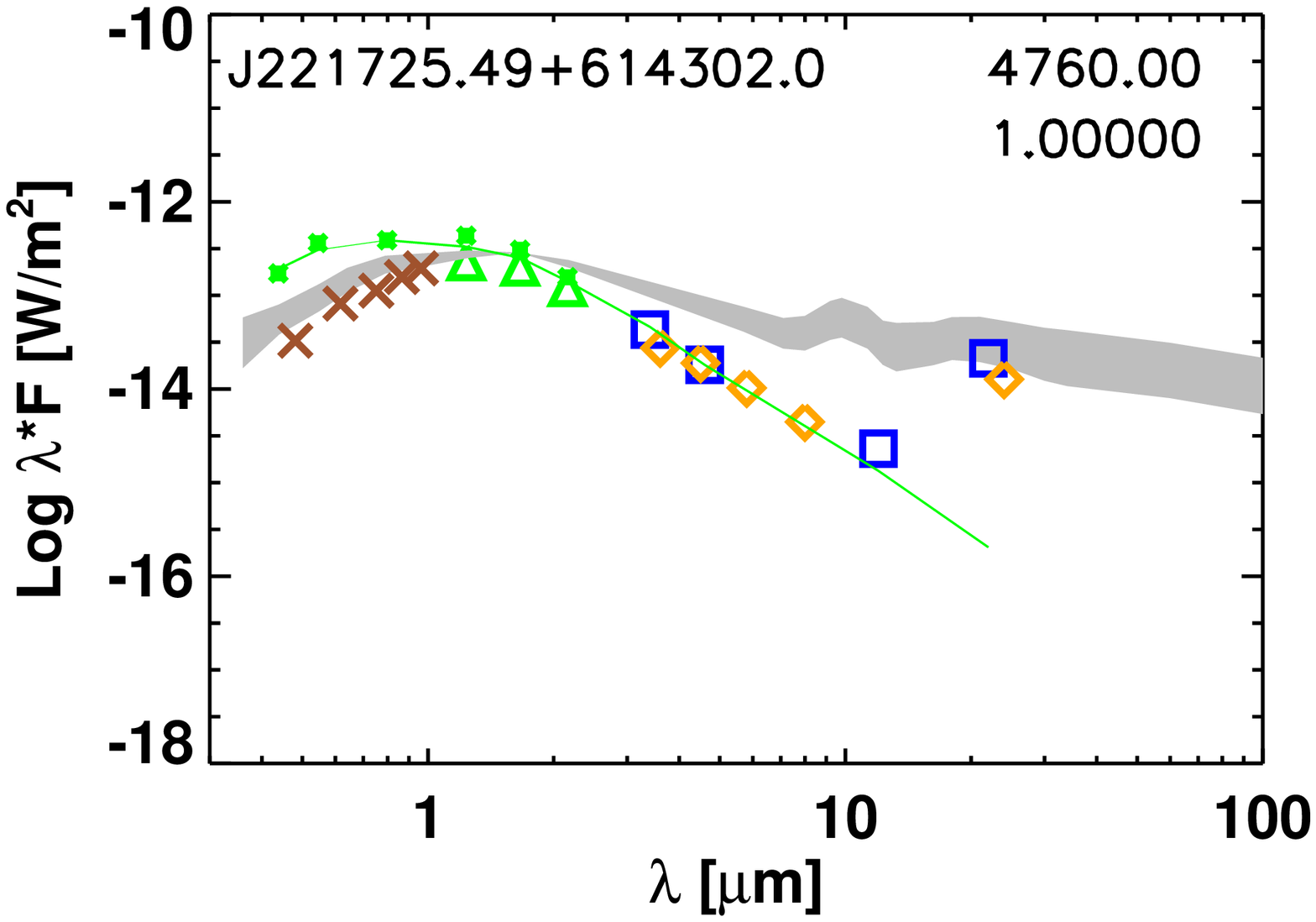}\includegraphics[scale=0.2]{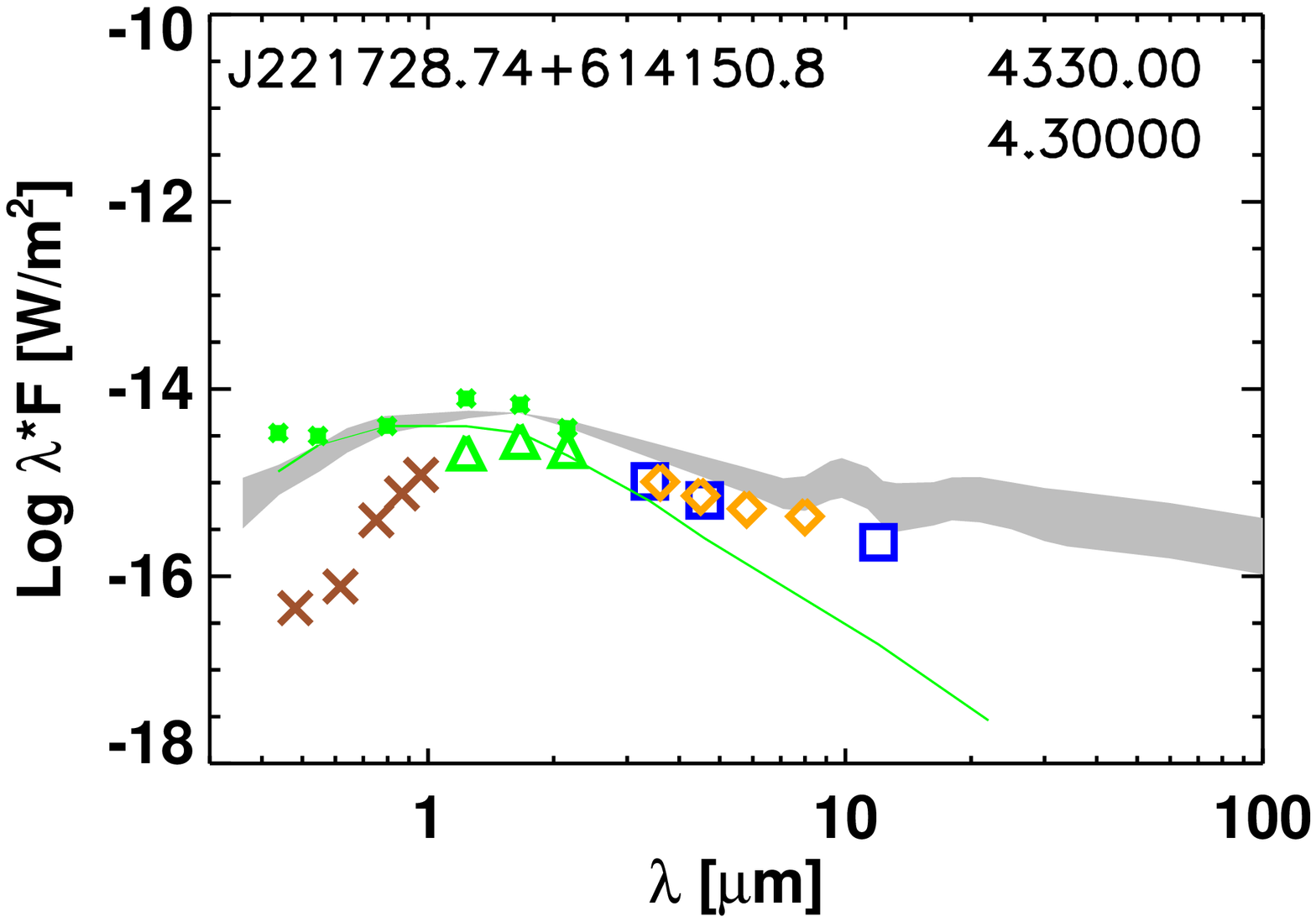}\includegraphics[scale=0.2]{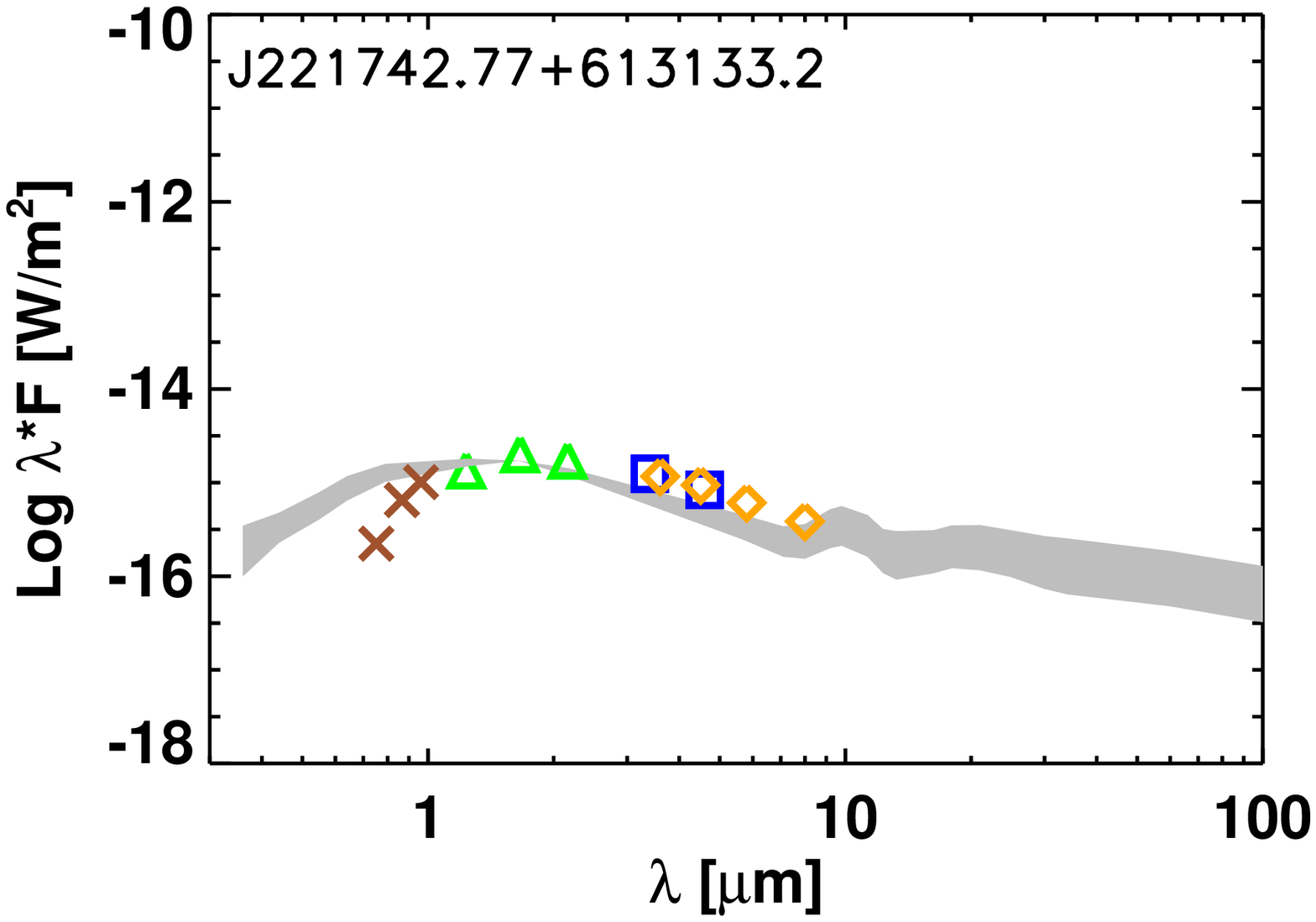}\includegraphics[scale=0.2]{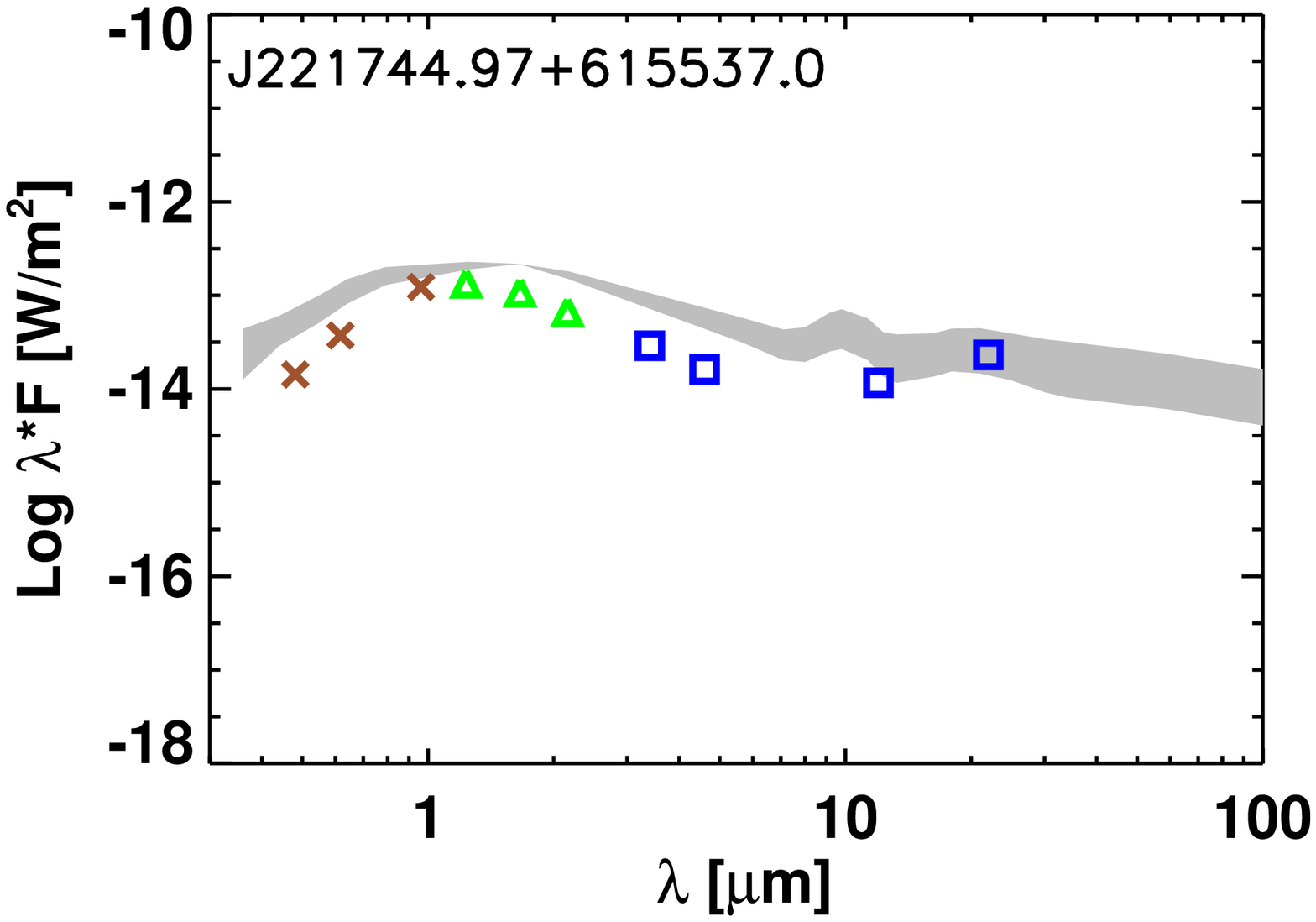}}

\centerline{\includegraphics[scale=0.2]{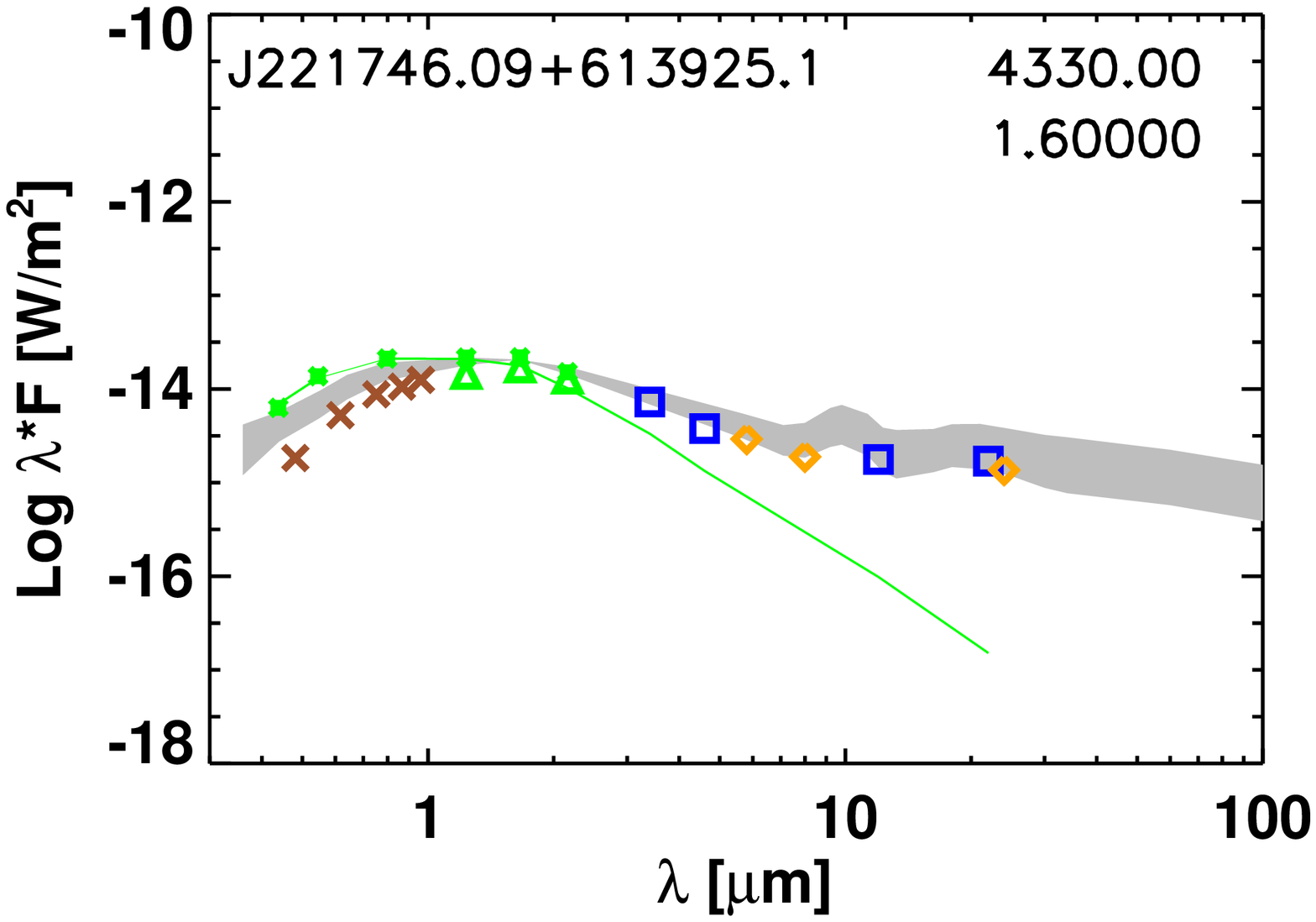}\includegraphics[scale=0.2]{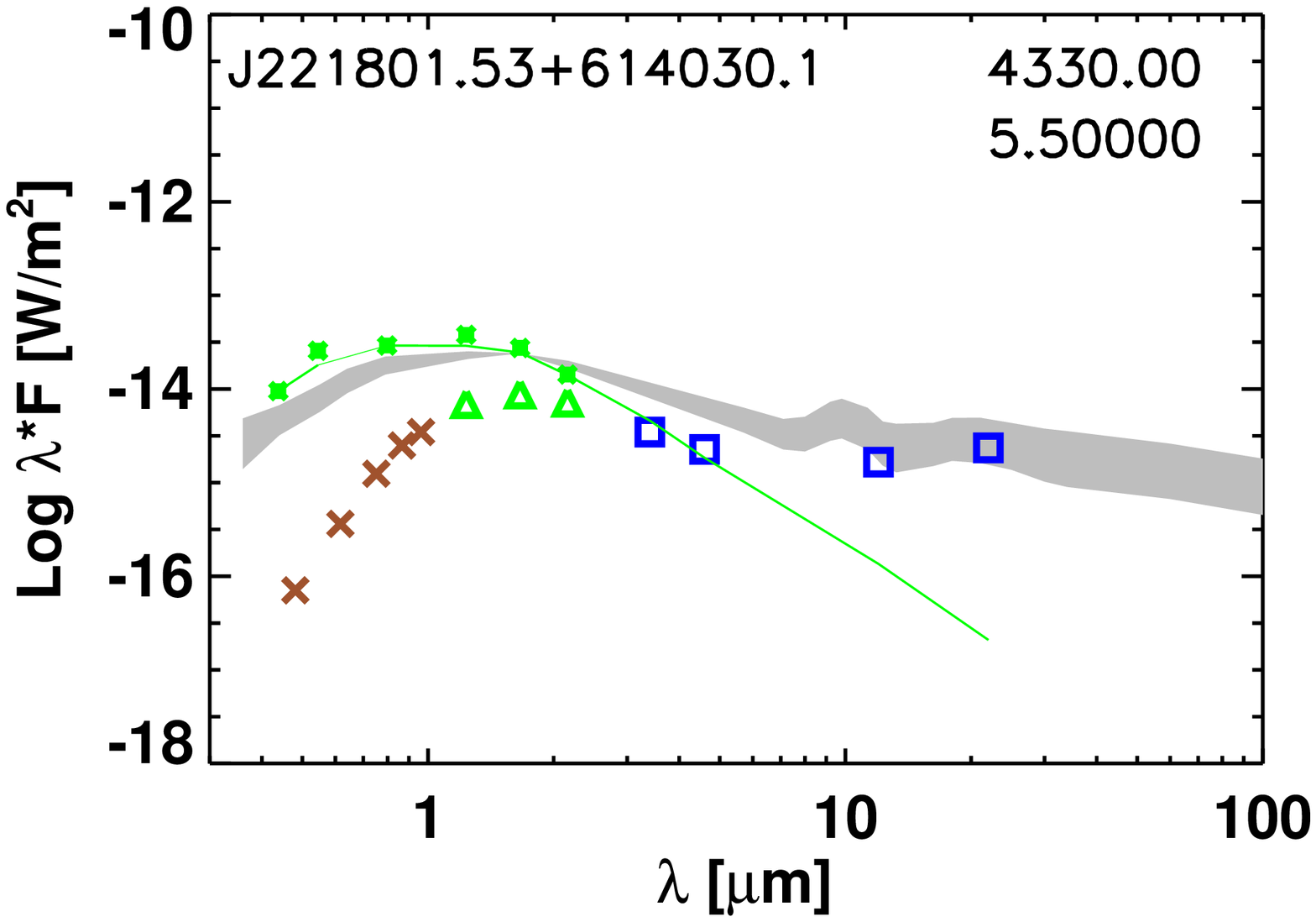}\includegraphics[scale=0.2]{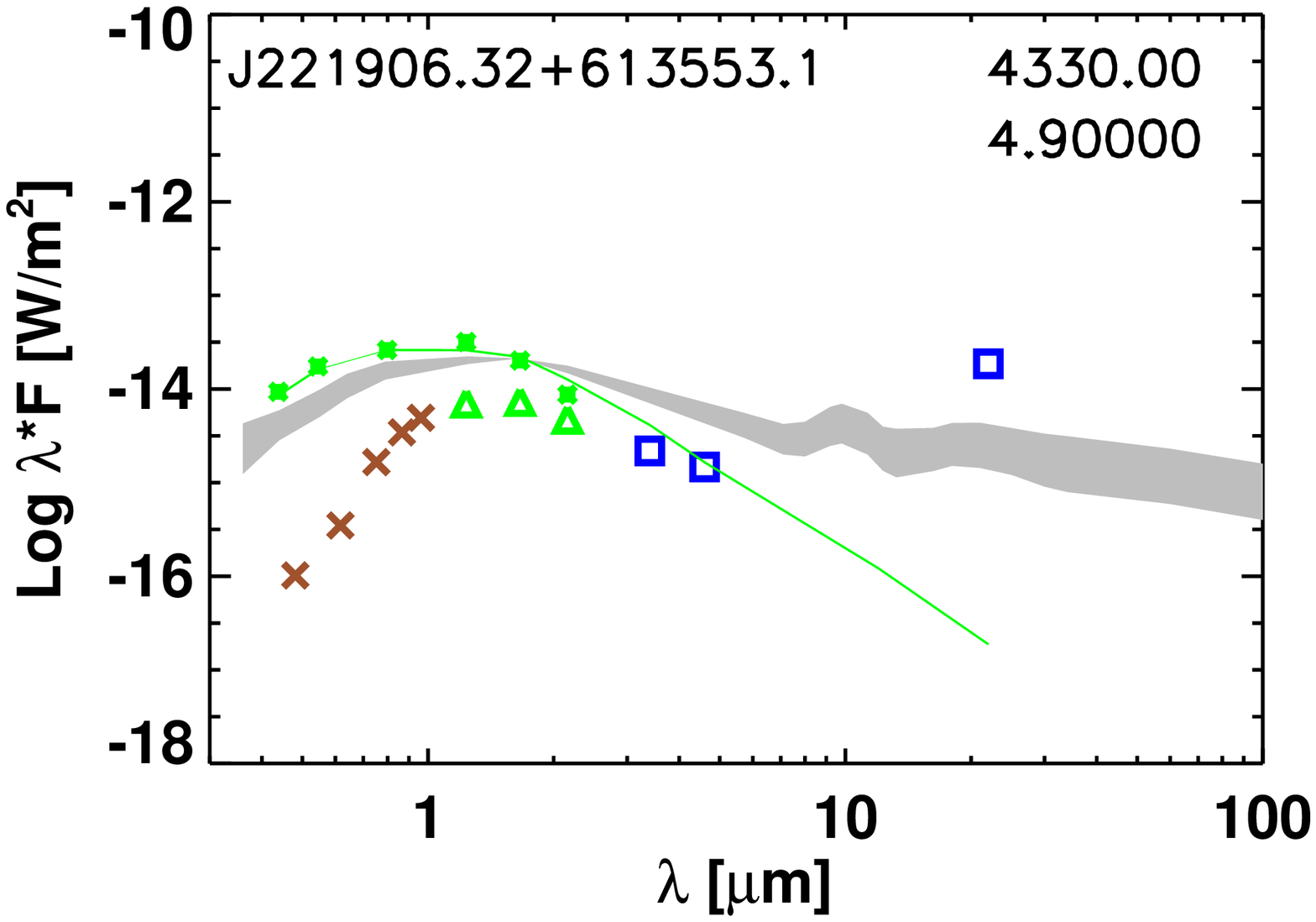}\includegraphics[scale=0.2]{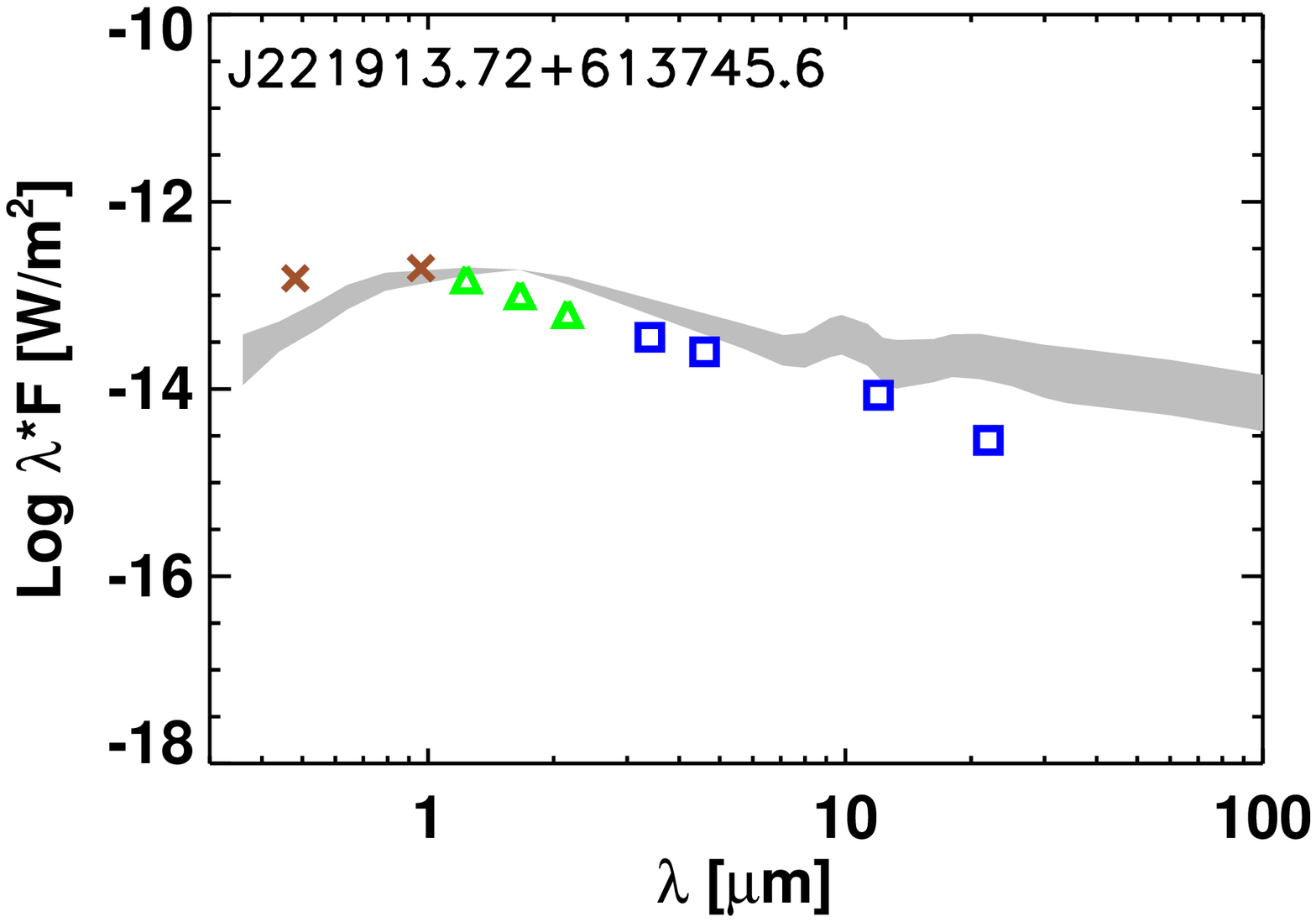}}

\centerline{\includegraphics[scale=0.2]{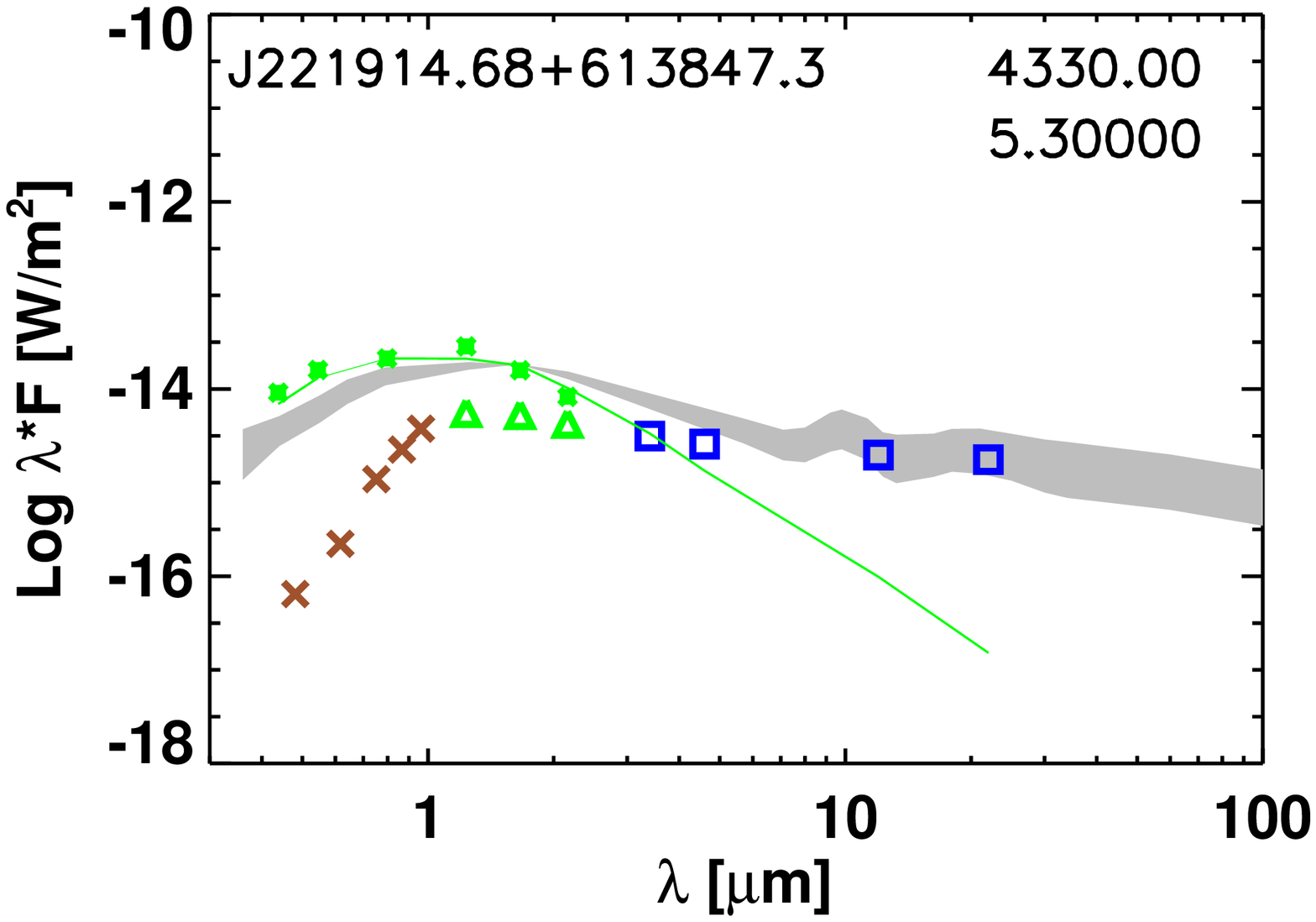}\includegraphics[scale=0.2]{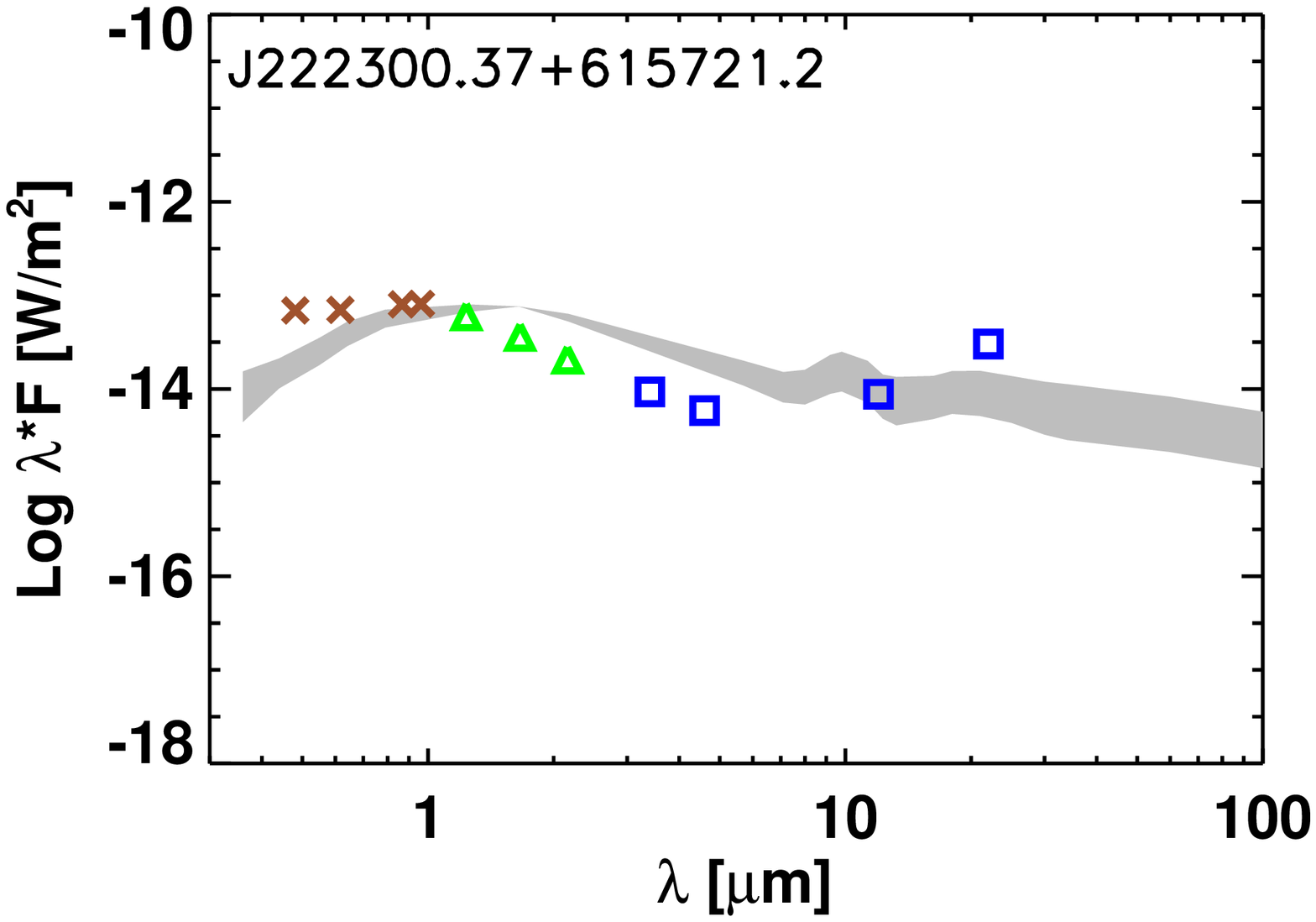}}

\caption{same as Fig.\ref{halfa_mamajack_sed}, for the IR-excess selected stars
}\label{infra_mamajack_sed}
\end{center}

\end{figure*}

\subsubsection{Accretion rates}\label{accretion}
Gas from the circumstellar disk accretes onto the star via magnetic funnel flows. This accretion process causes broad, often asymmetric H$\alpha$ emission lines. Several empirical relationships have been determined between the luminosity of the H$\alpha$ line and accretion luminosity \citep{dahm2008,barentsen}. We computed accretion rates for our H$\alpha$ emission sources and infrared excess stars using EW from the IPHAS colour-colour diagram and r$\sp{\prime}$ magnitude, applying the relationship established by \citet{barentsen} for the H$\alpha$ emission stars of IC 1396, spreading over a similar mass interval as our stars:

\begin{equation}
log(L\sb{\rm{acc}}/L\sb{\odot})=(1.13 \pm 0.07)log(L\sb{\rm{H\alpha}}/L\sb{\odot}) + (1.93 \pm 0.23)
\end{equation}
where $L\sb{\rm{H\alpha}}$ is the luminosity of the H$\alpha$ emission line. We convert the luminosity to accretion rate  $\dot{M}\rm\sb{{\rm{acc}}}$, according to the relationship
\begin{equation}
L\sb{\rm{acc}}\approx\frac{G\dot{M}\sb{\rm{acc}}M\sb{\star}}{R\sb{\star}}\left(1-\frac{R\sb{\star}}{R\sb{0}}\right). 
\end{equation}

We determined the radius $R\sb{\star}$ from the bolometric luminosity in view of the effective temperature and extinction.
\begin{equation}
L\sb{bol}=4  \pi \sigma R\sp{2}  T\sp{4}\sb{\rm{eff}} 
\end{equation}
We took the stellar masses $M\sb{\star}$ from the IPHAS colour-magnitude diagram, and adopted  $R\sb{0} \approx 5 R\sb{\star}$ for the inner radius of the gaseous disk \citep{gullbring}.
The average $\dot{M}\sb{\rm{acc}}$ is $8.96\times10\sp{-9}$ M$\sb{\odot}$ yr$\sp{-1}$, a typical value for T Tauri stars.

 We fitted the $\dot{M}\sb{\rm{acc}}$ $\propto$ $M\sp{\alpha}\sb{\star}$ power-law relationship and find a slope $\alpha$ =1.55$\pm$ 0.71, although the  spread is very large in  $\dot{M}\sb{\rm{acc}}$ for any  $M\sb{\star}$. The linear fit to the data is, 
\begin{equation}
log \dot{M}\sb{\rm{acc}}= (1.55 \pm 0.71 )log M\sb{\star} - (8.27 \pm 0.10 ).
\end{equation}
Its slope is within the range of recent published values (1.0 to 3.0) for other star-forming regions \citep[][e.g.]{barentsen,natta}.



\subsection{The surface distribution of young stars}

\begin{figure}
\begin{center}

 \includegraphics[width=\columnwidth]{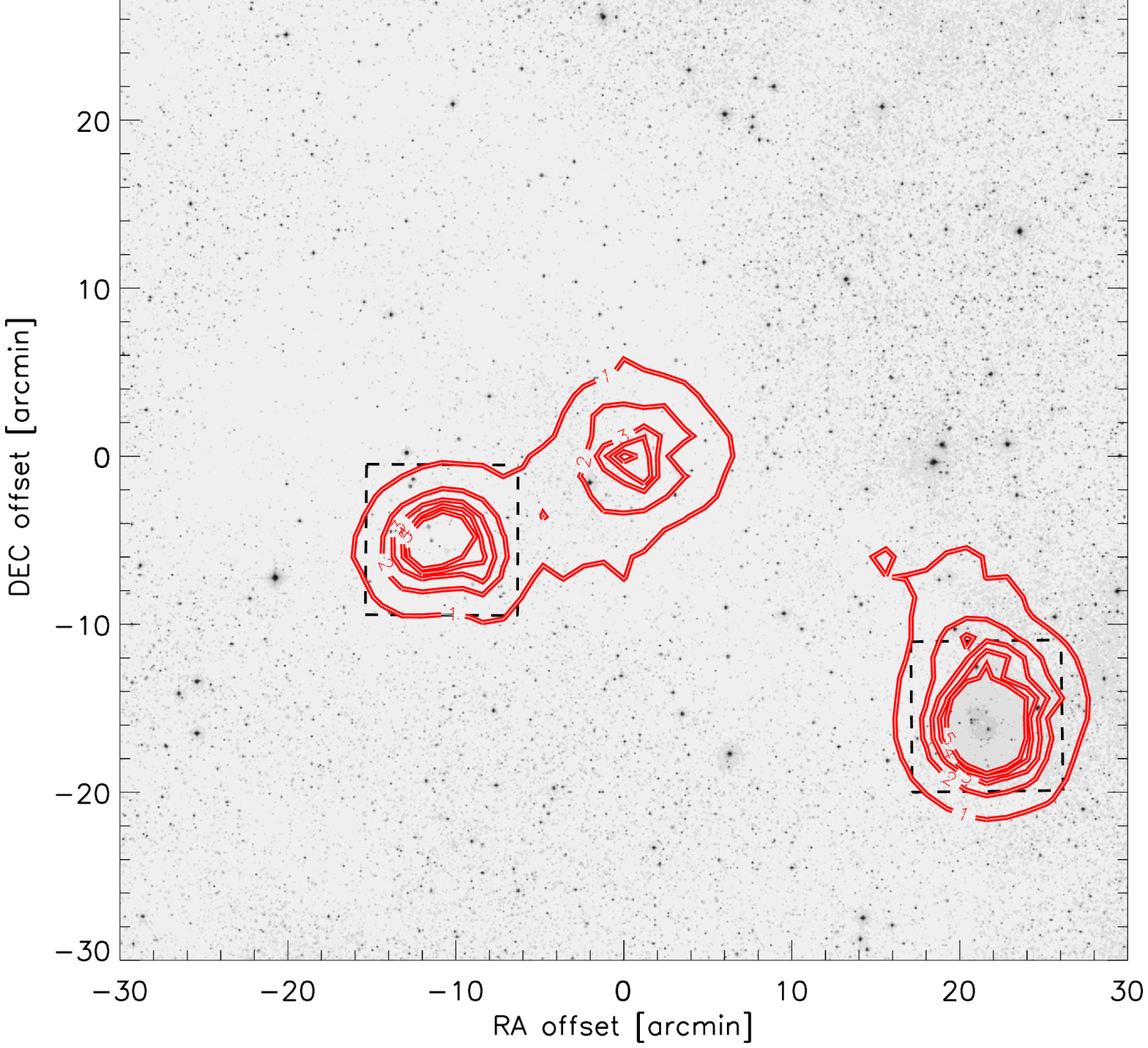}\vspace{5mm}
 \caption{Surface density distribution of the young stars, overplotted on the  DSS2 red image of the cloud, centered on R.A.(2000) = 22$\mathrm{\sp{h}}$ 17.5$\mathrm{\sp{m}}$ and Dec(2000)= 61$\degr$ 42$\arcmin$.  Red contours show the surface density of the young star candidates. The dashed squares show the regions displayed in Fig. \ref{three}. The distribution was computed from the distance of the fifth-nearest stars to the grid points.}\label{ext_surface} 
\end{center}

\end{figure}

The surface distribution of the candidate YSOs  (Figs.~\ref{distribution} and \ref{ext_map}) reveals two  compact groups of young stars. One of them is associated with the westernmost $^{13}$CO clump L1188~A, \citep{abraham}, and the other one is projected on the $^{13}$CO clump L1188~C. A loose aggregate of YSOs is projected on the clump L1188~B. We constructed a surface density map of the YSOs following the method described by \citet{gutermuth2005}. We determined the r$\sb{N}$(i, j) distance of the ${N}$th nearest star at each (i,j) position of a uniform grid and obtained the local surface density of YSOs at the grid point as $\rho$ (i, j) = $N/\pi r\sb{N}\sp{2} $(i, j). 
The surface density contour plot  (shown  on Figure \ref{ext_surface}) was constructed using a 30$\arcsec$ grid and N = 5. 




\subsubsection{Young clusters}

To find and characterize clusters in the YSO population of L1188, we used the Minimal Spanning Tree (MST) method presented by \citet{mst} and described by \citet{gutermuth2009} to identify groups. Within the MST structure, clusters can be isolated from the background, when one adopts a critical length longer than the typical separation of cluster members, and shorter than the intermediate spacings between the cluster and background. This critical length was calculated from the cumulative dist\-ribution function of the MST branch length.
We used a branch length of 150~arcsec  to identify the unique clusters within the MST network.
In this manner we were able to identify two remarkable clusters. The number of YSO candidates located in these groups is 27 (41.5\% of all of the young star candidates). This result is accordance with the surface density map presented in Fig.~\ref{ext_surface}.

One of the two clusters was also identified by \citet{bica} during a hunt for new infrared star clusters based on 2MASS data. This cluster is associated with the DG~180 reflection nebula and consists of 19 young star candidates selected mainly based on H$\alpha$ emission. The [ADM95]~IRAS-1 source is also associated with this cluster. 
The other cluster contains 8 young star candidates, inc\-luding [ADM95]~IRAS-6.
The extended, loose aggregate of some 12--15 YSOs, associated with the clump L1188~B and RNO~140 contains the only Class~I source of our sample 2MASS~J22172227+6143054 and [ADM95]~IRAS-4. A compact group of some 10 faint \textit{Spitzer\/} sources can be seen in the composite IRAC image of this region, presented in \citet{alap}. \citet{bica} identify this group as an embedded cluster.



\begin{figure*}
\begin{center}
 \includegraphics[scale=0.3]{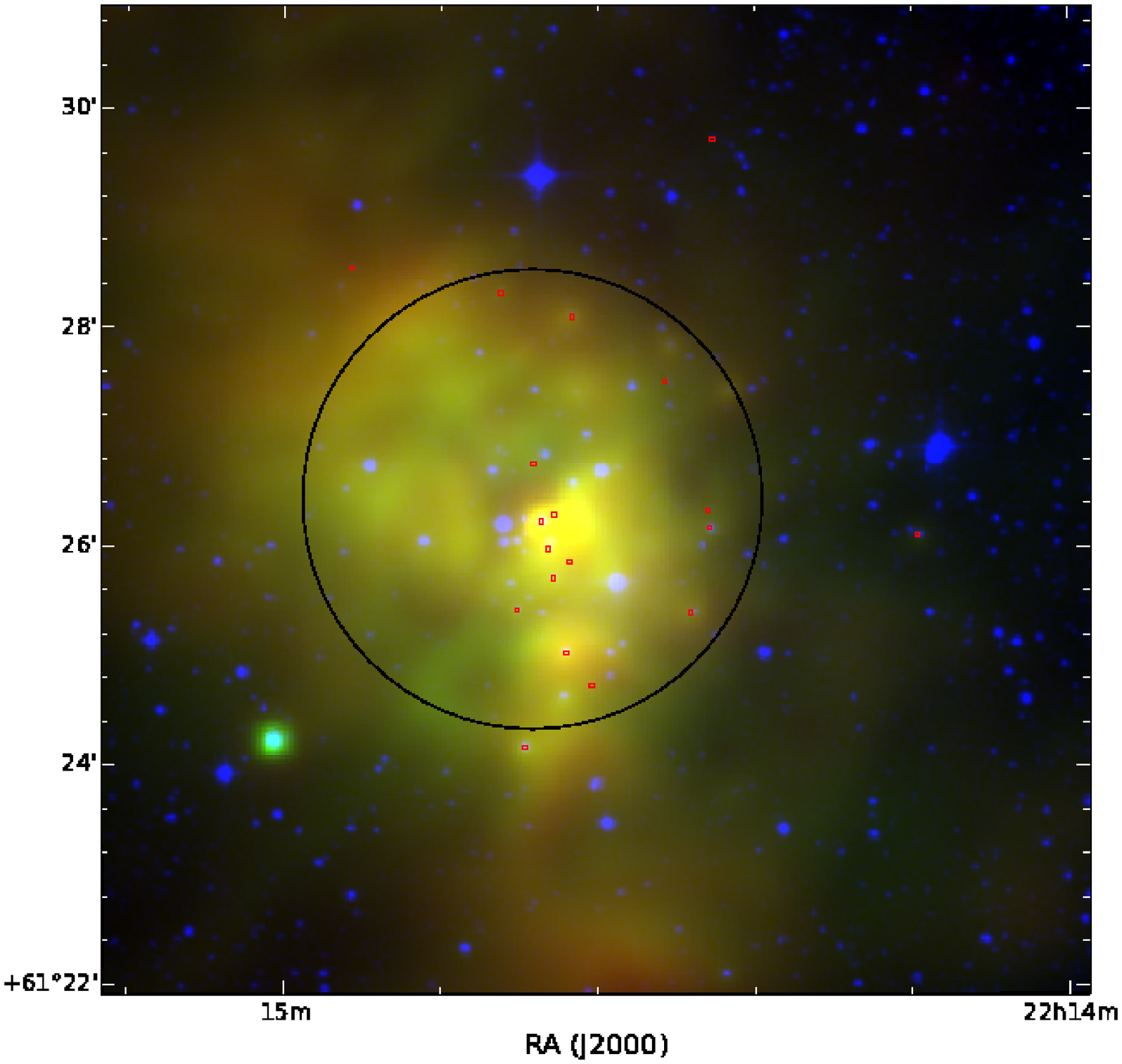}\includegraphics[scale=0.3]{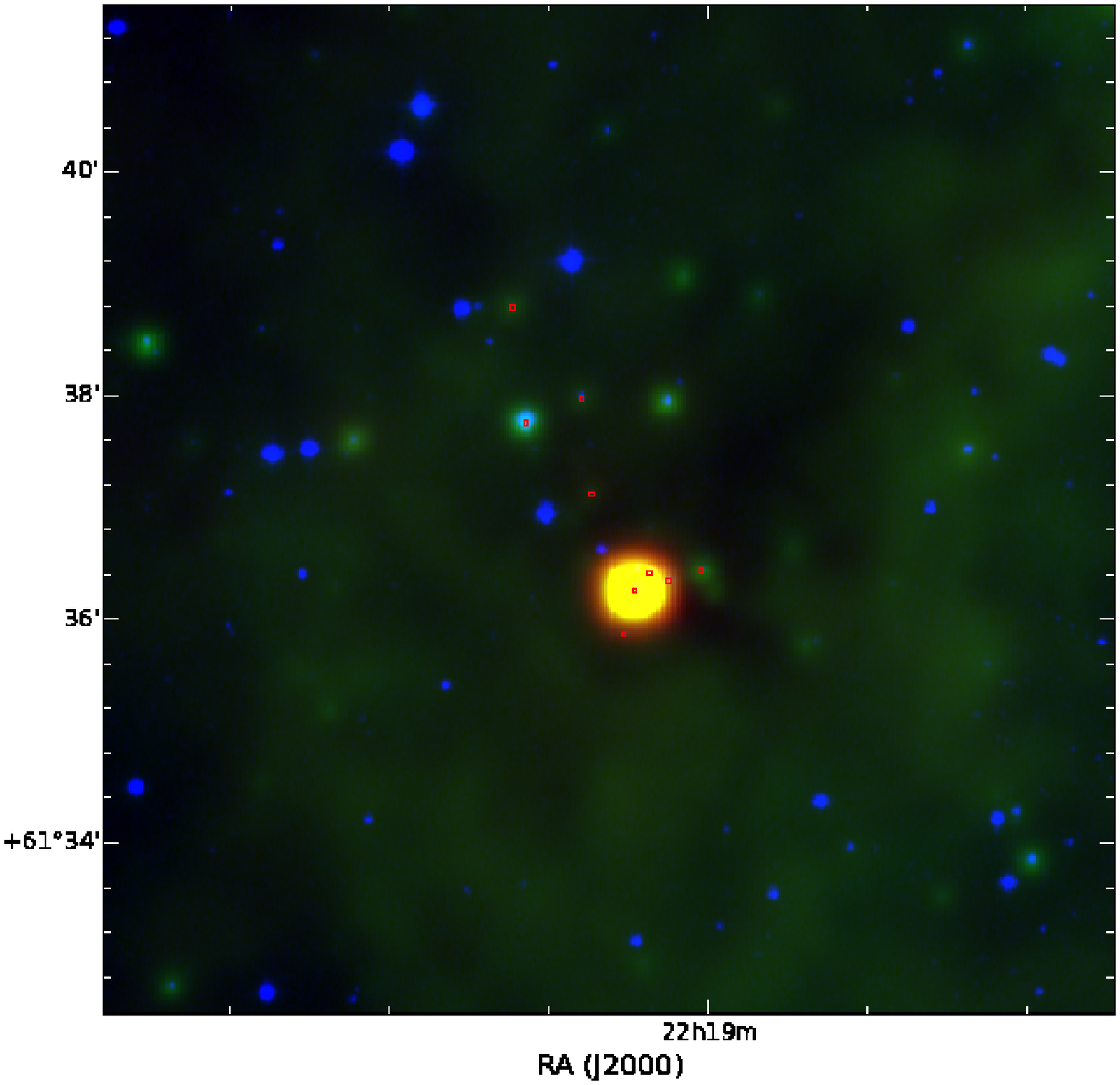}
 \caption{Composite images of the two identified clusters using {\it WISE} 3.6, 12 and 22 $\micron$ images. Left: the cluster around DG~180, red squares indicate the young star candidates, while solid black circle shows the region identified by \citet{bica} as an infrared cluster. Right: the cluster around the [ADM95]~IRAS-6 source.}\label{three}
\end{center}

\end{figure*}

\subsubsection{ L1188 and the Cepheus Bubble}

\begin{figure*}
\begin{center}
\includegraphics[width=\columnwidth]{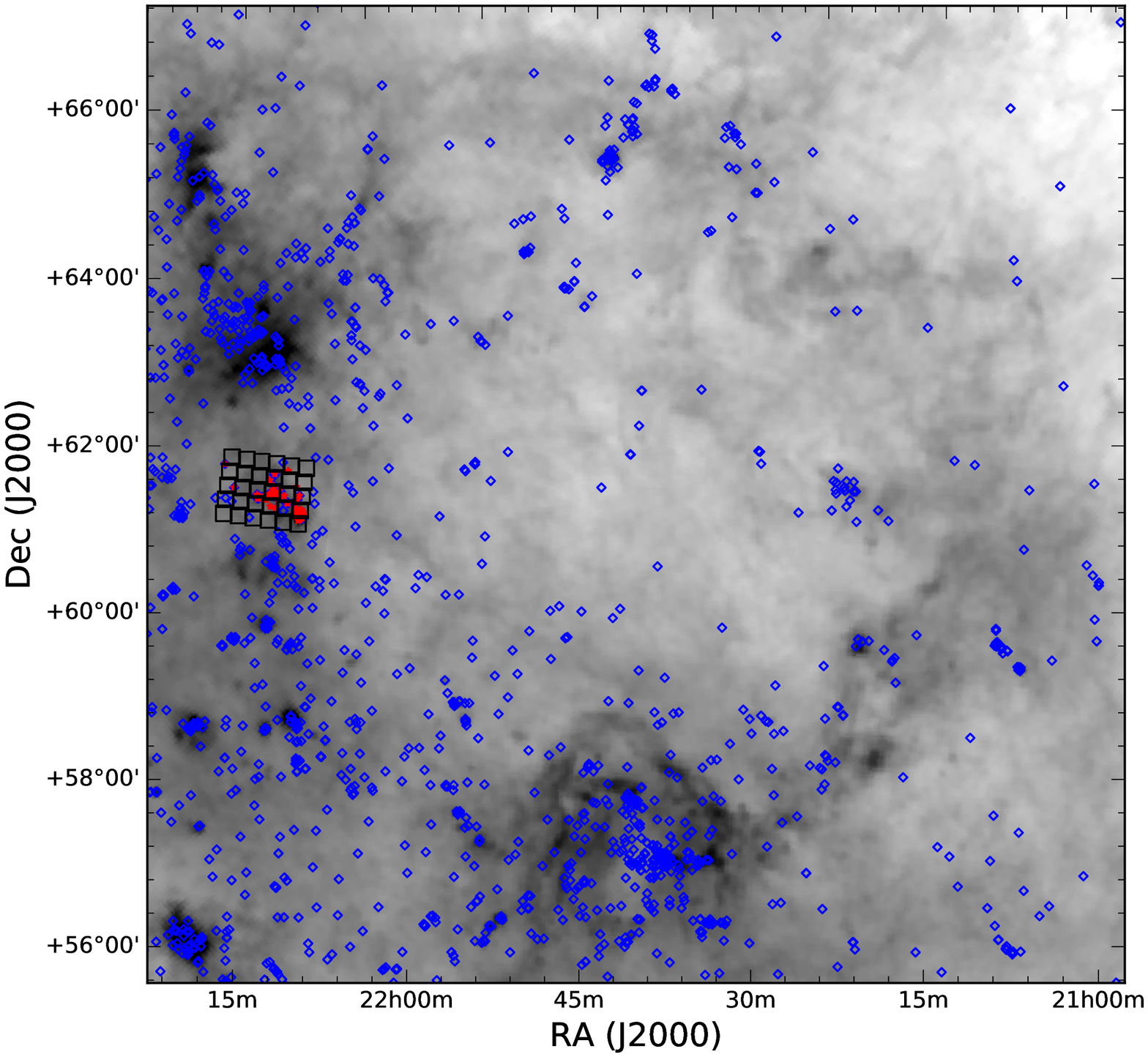}
 \caption{IRAS  100 \micron  (IRIS) image of the Cepheus Bubble. Blue diamonds are {\it WISE} YSO candidates presented by \citet{marton2016}, while red crosses represent young star candidates belonging to L1188. The area bordered by the black squares was the target of our H$\alpha$ survey.}
\label{bubble}
\end{center}

\end{figure*}

L1188 is located at the periphery of the Cepheus Bubble. Figure \ref{bubble} shows the 100\,$\micron$ {\it IRAS\/} (IRIS) image of the Cepheus Bubble. Black squares mark the observed regions of L1188, blue diamonds symbol candidate YSOs selected based on {\it WISE} data by \citet{marton2016}, and red dots are YSO candidates associated with L1188, identified in the present work. The image suggests that an enhanced star-formation is ongoing in the Cepheus Bubble. 
Figure~\ref{diagram} suggests that star formation in L1188 started at least 5 million years ago, i.e. the age of this star-forming region is comparable with that of Tr~37, the richest subgroup of Cep~OB2 \citep{aurora}. Our selection excluded embedded sources fainter that the {\it Gaia\/} limit at optical wavelengths, thus the Class~I/Class~II number ratio is uncertain. The large number of candidate YSOs at far larger distances than L1188 indicates that abundance of Class~I protostars, suggested by \citet{alap} and \citet{gong}, may contain sources at various distances. On-going and possible future star formation in L1188 is indicated by the ammonia cores \citep{verebelyi}. These cores are located to the east from the  clusters of Class~II YSOs. This structure suggests that star formation proceeds from the west to the east, according to the triggered star formation scenario. Another hint of the same age gradient is that the westernmost cluster around [ADM95]~IRAS-1 lies at the outskirts of the L1188-A CO clump, whereas the [ADM95]~IRAS-6 cluster and the aggregate associated with [ADM95]~IRAS-4 are closely associated with molecular clumps and also with \textit{Planck\/} cold clumps.  


\subsection{ Comparison with Other Nearby Cluster-forming Regions}

\citet{gutermuth2009} studied 36 young clusters within the 1~kpc of the Sun.  Similarly to L1188, most of these star-forming regions are parts of molecular cloud complexes and their most massive stars are around 5$-$6 M$\sb{\odot}$ (e.g. IC~348, NGC~7129, BD+40$\degr$4124). Although the  known YSO po\-pulation of L1188 is not complete, the number of the YSO candidates is  of the same order of magnitude as of the po\-pulation of NGC~7129 or L1211. The MST branch length of L1188 is also within the range found for this sample.


\section{Summary}
\label{summary}

We studied the little studied young stellar po\-pulation towards the dark nebulae in  L1188. During our slitless grism H$\alpha$ survey we found 76 H$\alpha$ emission line stars,  across an area of $60\arcmin \times 50\arcmin$. Based on archival infrared data we identified further 61 young sources. Finally, we studied the distance of the sources based on {\it Gaia} DR2 data and selected 63 sources related to L1188, nine young star candidates are foreground objects (236$-$718 pc), 45 young star candidates are background objects (1014$-$6507 pc) and the distance of 15 sources are unknown. We constructed SEDs of our target stars associated with L1188 based on Pan-STARRS, 2MASS, {\it WISE} and {\it Spitzer} photometric data, derived their spectral types, extinctions, estimated masses by pre-main sequence evolutionary models, and examined the disc shapes utilizing the 2$-$24 $\micron$ interval of the SED. We measured the equivalent width of the H$\alpha$ line  for a subsample of our targets and derived accretion rates. The average $\dot{M}\sb{\rm{acc}}$ is $8.96\times10\sp{-9}$ M$\sb{\odot}$ yr$\sp{-1}$. The new extinction map of L1188, based on Pan-STARSS data, reveals a filamentary  structure, with very compact, high density regions. 
 We found that star formation in L1188 started about 5 million years ago. The age of the newly identified pre-main-sequence po\-pulation is comparable to that of the nearby Trumpler~37. The L1188 star-forming region was probably created during the fragmentation of the shell, compressed by the expan\-ding ionization front and stellar wind of the Cep~OB2a association \citep{patel}. Collision of clouds, proposed by \citet{gong}, might have played a role in star formation. The presence of young clusters outside the proposed collision region, however, suggests further agents. The apparent age gradient of star formation signposts across the cloud supports the role of the expansion of the Cepheus Bubble.  

\section*{Acknowledgements}
Our results are based on observations with the 2.2-m telescope of the University of Hawaii and we thank Colin Aspin and Mark Willman for their interest and support. Special thanks to L\'aszl\'o  Szabados for the careful reading of the manuscript and for the suggestions to improve it.
We would like to thank  P\'eter \'Abrah\'am for the $\sp{13}$CO data for.
 This work makes use of observations made with the {\it Spitzer Space Telescope}, which is operated by the Jet Propulsion Laboratory, California Institute of Technology under a contract with NASA. 
This research has made use of the NASA/IPAC Infrared Science Archive, which is operated by the Jet Propulsion Laboratory, California Institute of Technology, under contract with the National Aeronautics and Space Administration. This research has made use of the VizieR catalog access tool, CDS, Strasbourg, France. The original description of the VizieR service was published in A\&AS 143, 23. 
This work has made use of data from the European Space Agency (ESA) mission {Gaia} (\url{https://www.cosmos.esa.int/gaia}), processed by the { Gaia}
Data Processing and Analysis Consortium (DPAC,
\url{https://www.cosmos.esa.int/web/gaia/dpac/consortium}). Funding for the DPAC
has been provided by national institutions, in particular the institutions
participating in the {\it Gaia} Multilateral Agreement.
We thank the anonymous referee for helpful suggestions.

\bibliographystyle{mnras}
\bibliography{l1188} 

\begin{table*}
\caption{H$\alpha$ emission line stars associated with L1188}
\label{halfa_tab}
\begin{tabular}{cccccc}
\hline
Name &2MASS ID&\multicolumn{2}{c}{EW(H$\alpha$)(\AA)} &r'$\sb{\mathrm{IPHAS}}$&SED \\
&&WFGS &IPHAS&(mag)&\\
\hline
L1188	H$\alpha$\,2		&		J22140809+6130163		&		10.56	$\pm$	4.51		&		$\cdots$	&	17.18  $\pm$  0.01	&		Class II			\\
L1188	H$\alpha$\,3		&		J22141156+6126063		&		27.08	$\pm$	2.77		&		70	&	17.56  $\pm$  0.01	&		Class II			\\
L1188	H$\alpha$\,5		&		J22142723+6129433		&		$\cdots$		&		60	&	19.39 $\pm$   0.06	&							\\
L1188	H$\alpha$\,6		&		J22142746+6126103		&		22.29	$\pm$	5.13		&		$\cdots$	&	17.53  $\pm$  0.01	&		Flat			\\
L1188	H$\alpha$\,7		&		J22142761+6126200		&		$\cdots$		&	$\cdots$		&	20.98  $\pm$  0.23	&							\\
L1188	H$\alpha$\,8		&		J22142887+6125239		&		$\cdots$		&	60		&	17.68  $\pm$  0.01	&		Flat					\\
L1188	H$\alpha$\,10		&		J22143014+6132132		&		17.86	$\pm$	9.36		&	$\cdots$		&	18.47 $\pm$   0.03	&					\\
L1188	H$\alpha$\,11		&		J22143083+6127304		&		$\cdots$		&	$\cdots$		&	19.72 $\pm$   0.08	&							\\
L1188	H$\alpha$\,13		&		J22143197+6132216		&		$\cdots$		&	$\cdots$		&	17.48  $\pm$  0.01	&							\\
L1188	H$\alpha$\,15		&		J22143646+6124439		&		$\cdots$		&	60		&	19.05  $\pm$  0.04	&		Flat					\\
L1188	H$\alpha$\,18		&		J22143799+6128060		&		$\cdots$		&		$\cdots$	&	19.66  $\pm$  0.07	&		Class II			\\
L1188	H$\alpha$\,19		&		J22143816+6125514		&		$\cdots$		&		70	&	20.03   $\pm$ 0.10	&							\\
L1188	H$\alpha$\,20		&		J22143844+6125019		&		33.22	$\pm$	4.37		&		45	&	16.91   $\pm$ 0.01	&		Flat			\\
L1188	H$\alpha$\,23		&		J22143941+6125427		&		76.98	$\pm$	30.22		&	$\cdots$		&	19.97  $\pm$  0.09	&	Flat	\\
L1188	H$\alpha$\,25		&		J22143982+6125587		&		$\cdots$		&	40		&	17.68  $\pm$  0.01	& Flat	\\	
L1188	H$\alpha$\,26		&		J22144089+6126451		&		$\cdots$		&		$\cdots$	&	20.27   $\pm$ 0.12	&  Class II	\\
L1188	H$\alpha$\,27		&		J22144156+6124100		&		16.27	$\pm$	1.94		&	40		&	16.28 $\pm$   0.01	&		Class II		\\
L1188	H$\alpha$\,28		&		J22144218+6125251		&		$\cdots$		&		20	&	18.18   $\pm$ 0.02	&		Class II \\
L1188	H$\alpha$\,29		&		J22144343+6128192		&		85.92	$\pm$	85.48		&	$\cdots$		&	20.17   $\pm$ 0.11	&		Class II	\\
L1188	H$\alpha$\,31		&		J22145213+6131052		&		$\cdots$		&		20	&	19.46  $\pm$  0.06	&	Class II		\\
L1188	H$\alpha$\,32		&		J22145479+6128328		&		$\cdots$		&		45	&	19.93  $\pm$  0.09	&							\\
L1188	H$\alpha$\,33		&		J22150238+6132443		&		97.37	$\pm$	25.84		&	$\cdots$		&	20.70 $\pm$   0.18	& Class II	\\
L1188	H$\alpha$\,36		&		J22150844+6131551		&		$\cdots$		&		20	&	18.50 $\pm$   0.03	&		Class II	\\
L1188	H$\alpha$\,43		&		J22162448+6159190		&		$\cdots$		&		60	&	18.57  $\pm$  0.03	&		Class II					\\
L1188	H$\alpha$\,44		&		J22170839+6131063		&		38.71	$\pm$	6.02		&	$\cdots$		&	16.54  $\pm$  0.01	&		Class II		\\
L1188	H$\alpha$\,47		&		J22172174+6140405		&		$\cdots$		&		20	&	18.65  $\pm$  0.03	&		Class II				\\
L1188	H$\alpha$\,48		&		J22172227+6143054		&		$\cdots$		&	$\cdots$		&	18.86  $\pm$  0.03	&		Class I					\\
L1188	H$\alpha$\,51		&		J22174025+6147025		&		122.69	$\pm$	15.36		&	$\cdots$		&	17.34 $\pm$   0.01	&		Class II		\\
L1188	H$\alpha$\,52		&		J22174140+6143525		&		$\cdots$		&	$\cdots$		&	19.05 $\pm$   0.04	&		Class II				\\
L1188	H$\alpha$\,53		&		J22174489+6141310		&		56.99	$\pm$	31.12		&	40		&	18.46  $\pm$  0.03	&		Class II		\\
L1188	H$\alpha$\,54		&		J22180144+6137582		&		$\cdots$		&	$\cdots$		&	18.54  $\pm$  0.03	&		Class II				\\
L1188	H$\alpha$\,55		&		J22181032+6150416		&		14.37	$\pm$	3.85		&	$\cdots$		&	17.95   $\pm$ 0.02	&		Class II		\\
L1188	H$\alpha$\,56		&		J22181440+6135570		&		$\cdots$		&		10	&	17.18  $\pm$  0.01	&							\\
L1188	H$\alpha$\,57		&		J22182573+6137265		&		$\cdots$		&		60	&	18.45  $\pm$  0.03	& 			Class II				\\
L1188	H$\alpha$\,58		&		J22184051+6137298		&		29.39	$\pm$	4.41		&		40	&	16.07  $\pm$  0.01	&		Class II		\\
L1188	H$\alpha$\,60		&		J22190050+6136265		&		58.57	$\pm$	15.13		&		70	&	17.97  $\pm$  0.02	&		Class II		\\
L1188	H$\alpha$\,62		&		J22190298+6136208		&		$\cdots$		&		20	&	18.66   $\pm$ 0.03	&							\\
L1188	H$\alpha$\,64		&		J22190442+6136250		&		84.54	$\pm$	18.07		&		170	&	17.68  $\pm$  0.02	&					\\
L1188	H$\alpha$\,66		&		J22190554+6136156		&		$\cdots$		&	$\cdots$		&	18.03  $\pm$  0.02	&		Flat					\\
L1188	H$\alpha$\,69		&		J22190950+6137587		&		40.46	$\pm$	7.13		&	$\cdots$		&	17.66  $\pm$  0.02	&		Class II			\\
L1188	H$\alpha$\,75		&		J22214129+6141164		&		$\cdots$		&		$\cdots$	&	18.12  $\pm$  0.02	&	Class II					\\

\hline
\end{tabular}
\end{table*}

\begin{table*}
\caption{Young stars identified based on infrared excess}
\label{infra_table}
\begin{tabular}{ll}
\hline
Name &SED\\
&\\
\hline
2MASS J22144030+6126137 	&			Class II \\
AllWISE J221454.81+614224.6	&	Flat			\\
AllWISE J221504.18+613917.1	&	Class II\\
2MASS J22151924+6136523 	&			Class II \\
SSTSL2 J221521.04+614827.3 	&		Class II 	\\
SSTSL2 J221559.25+613536.3 	&		Class II  \\
SSTSL2 J221605.94+613553.9 	&		Class II  \\
SSTSL2 J221627.88+614047.7 	&		Class II  \\
SSTSL2 J221643.39+612417.5 &		Class II  \\
SSTSL2 J221658.12+614620.4 	&		Class II 	\\
SSTSL2 J221703.22+613714.8 	&		Class II 	\\
SSTSL2 J221709.59+614111.5 	&		Flat	\\
SSTSL2 J221725.49+614302.0 	&			Class II \\
SSTSL2 J221728.74+614150.8 	&	Class II 	\\
SSTSL2 J221742.77+613133.2 	&	Class II  \\
AllWISE J221744.97+615537.0	&		Class II 	\\
AllWISE J221746.09+613925.1	&		Class II	\\
AllWISE J221801.53+614030.1	&		Class II 	\\
AllWISE J221906.32+613553.1	&		Class II	\\
AllWISE J221913.72+613745.6	&	Class II 	\\
AllWISE J221914.68+613847.3	&		Class II 	\\
AllWISE J222300.37+615721.2	&		Flat	\\

\hline
\end{tabular}
\end{table*}

\begin{table*}
\caption{Young star candidates, possibly associated with L1188}
\label{possible}
\begin{tabular}{llll}
\hline
Name&Distance (pc) & SED  \\
\hline	

SSTSL2 J221723.20+614953.2	&     $\cdots$						&Class I    \\
SSTSL2 J221726.28+614218.9	&   $\cdots$	&Class I	\\
SSTSL2 J221732.42+614152.6	&    $\cdots$		&Class I				    \\

\hline
\end{tabular}
\end{table*}
\begin{table*}
\caption{Young star candidates, not associated with L1188}
\label{kimaradt}
\begin{tabular}{llll}
\hline
Name&Distance (pc) & SED & Cross Id. \\
\hline
J22140180+6129208	&	    1164   $\sb{-626}\sp{+      3387	    }$ && L1188 H$\alpha$\,1\\
J22141165+6156281	&	    3753    $\sb{-1290}\sp{+     2518	    }$ && L1188 H$\alpha$\,4\\
J22142611+6127246	&	    2997    $\sb{-1598}\sp{+     3010	    }$&Flat&\\
J22142908+6126025	&	    3134	    $\sb{-1765}\sp{+     3046	    }$ &Class I& L1188	H$\alpha$\,9\\
AllWISE J221429.21+612741.9	&   	    3844	    $\sb{-1539}\sp{+      2891 }$&Flat&\\
AllWISE J221429.73+612539.2	&	    3567    $\sb{-1484}\sp{+     2876	    }$&Flat&\\
J22143129+6127587	&	    1298	    $\sb{-150}\sp{+      193	    }$&Flat &  L1188 H$\alpha$\,12\\
J22143312+6129374	&	    1598	    $\sb{-729}\sp{+      2692    }$ &&  L1188 H$\alpha$\,14\\
J22143682+6150250	&	    1292	    $\sb{-184}\sp{+	    256	    }$ &&  L1188 H$\alpha$\,16\\
J22143714+6126246	&	    2334	    $\sb{-1701}\sp{+     3007	    }$&Class I &  L1188 H$\alpha$\,17\\
J22143859+6124377	&	    625	    $\sb{-66.}\sp{+	    84	    }$&Flat&  L1188 H$\alpha$\,21\\
J22143956+6127511	&	    3300    $\sb{-1665}\sp{+ 3060     }$&Class II& L1188 H$\alpha$\,24\\
J22145152+6140127	&	     $\cdots$	&Class II& L1188 H$\alpha$\,30  \\
J22150364+6129297	&	    502	     $\sb{-111}\sp{+198	     }$&Flat& L1188 H$\alpha$\,34 \\
J22153114+6155278	&	    3312  $\sb{-897}\sp{+  1678         }$	&& L1188 H$\alpha$\,37\\
SSTSL2 J221543.96+615228.0	&    1253	   $\sb{-37}\sp{+   39		   }$&Class II& \\
J22155762+6135463	&	    3931	   $\sb{-1878}\sp{+  3110        }$&Flat& L1188 H$\alpha$\,39\\
J22154943+6136464	&	    1188	   $\sb{-397}\sp{+  1125         }$	&Class II& L1188 H$\alpha$\,38\\
J22160406+6136151	&	    1243	   $\sb{-158}\sp{+  211		   }$ &Class II& L1188 H$\alpha$\,40 \\
J22161536+6133101	&	    4000	   $\sb{-1305}\sp{+ 2453         }$ && L1188 H$\alpha$\,41\\
SSTSL2 J221615.78+612948.6	&    $\cdots$	&Class II &					    \\
SSTSL2 J221619.76+612948.3	&    $\cdots$						  &Class II&  \\
SSTSL2 J221643.09+613819.7	&   3466	    $\sb{-1746}\sp{+3040	    }$&Class II&\\
AllWISE J221646.89+613336.4	&   4707    $\sb{-1726   }\sp{+2952	    }$&Class II&\\
SSTSL2 J221653.13+614315.5	&    $\cdots$	&Class II&					    \\
AllWISE J221703.85+620455.9	&   $\cdots$&Class II&						    \\
SSTSL2 J221707.87+615039.4	&   3346	   $\sb{-1718}\sp{+ 3037         }$&&\\
SSTSL2 J221708.55+614127.1	&   2639	   $\sb{-1725}\sp{+ 2992        }$&Class II&\\
SSTSL2 J221715.75+615208.0	&   3348	   $\sb{-799}\sp{+ 1387         }$&Class II\\
SSTSL2	J221716.31+614110.2	&   1740	   $\sb{-793}\sp{+  2683         }$&Class II&\\
J22171752+6138199	&	    3277  $\sb{-683}\sp{+  1103         }$&& L1188 H$\alpha$\,45\\
AllWISE J221718.87+614232.7	&    469	   $\sb{-58}\sp{+   77		   }$&Class II&\\
SSTSL J221719.56+614303.7	&    1014	   $\sb{-517}\sp{+  3303		   }$&Class II&\\
J22172039+6116469	&	    2942	   $\sb{-811}\sp{+  1559        }$&& L1188 H$\alpha$\,46\\
SSTSL2 J221723.32+614231.3	&    $\cdots$						&Class II&    \\
J22172625+6139303	&	    3225     $\sb{-1743}\sp{+ 3048     }$&Class II& L1188 H$\alpha$\,49\\
SSTSL2 J221726.53+614239.9	&   $\cdots$	&Class II&					    \\
SSTSL2 J221726.56+614814.6	&    3054	     $\sb{-1510}\sp{+ 2986		     }$&Class II&\\
SSTSL2	J221726.92+614811.0	&   2305	    $\sb{- 1406}\sp{+ 3022    }$&Class II&\\
AllWISE J221727.07+614217.7	&     $\cdots$	&Class II&					    \\
J22172826+6145172	&	    658.	 $\sb{-55}\sp{+   66	 }$&Class II& L1188 H$\alpha$\,50 \\
SSTSL2 J221732.19+613854.7	&    1823	 $\sb{-208}\sp{+  268	 }$&Class II\\
SSTSL2 J221734.54+613928.3	&    3453 $\sb{-997}\sp{+  1895		 }$&Class II&\\
SSTSL2 J221736.38+612715.1	&   1444	 $\sb{-529}\sp{+  1641		 }$& Class II&\\
SSTSL2 J221737.18+614312.4	&   $\cdots$		&Class II&				    \\
SSTSL2 J221738.26+614146.2	&   		    $\cdots$						&Class II&    \\
SSTSL2 J221743.66+612657.7	&   	    1464    $\sb{-150}\sp{+      187	    }$&Class II&\\
AllWISE J221745.90+615433.3	&   		    $\cdots$				&Class II&	     \\

\hline
\end{tabular}
\end{table*}
\begin{table*}
\contcaption{}

\begin{tabular}{llll}
\hline
Name&Distance (pc)& SED & Comment \\
\hline
AllWISE J221748.98+615514.9	&   2963	    $\sb{-1664}\sp{+ 3038     }$&Class I&\\
SSTSL2 J221808.26+613132.5	&  1506	    $\sb{-616}\sp{+  2171     }$&Class II& \\
J22185005+6122062	&	   4942	    $\sb{-1868}\sp{+ 3037     }$ &&L1188 H$\alpha$\,59 \\
J22190065+6136384	&	    $\cdots$& &L1188 H$\alpha$\,61   \\
J22190317+6137563	&	   581	 $\sb{-100 }\sp{+ 153		 }$ &Class II&L1188 H$\alpha$\,63 \\
J22190443+6136486	&	   627	 $\sb{-54 }\sp{+ 66	 }$&Class II& L1188 H$\alpha$\,65 \\
J22190876+6137075	&	   718	 $\sb{-315 }\sp{+ 2370		 }$&& L1188 H$\alpha$\,67\\
J22190879+6137329	&	   1121	 $\sb{-162 }\sp{+ 227		 }$&Class II& L1188 H$\alpha$\,68\\
J22191296+6137305	&	   1338	 $\sb{-414 }\sp{+ 1018		 }$&Class II& L1188 H$\alpha$\,70\\
J22192789+6158582	&	   1891	$\sb{- 644 }\sp{+ 1653		 }$&Class II& L1188 H$\alpha$\,71\\
AllWISE J221940.96+612119.8	&  1061	 $\sb{-27  }\sp{+ 28	 }$&Flat&\\
AllWISE J221945.24+612437.1	&  6507	$\sb{- 1628}\sp{+ 2529		 }$&Class II&\\
J22201557+6159403	&	   654  	 $\sb{-69  }\sp{+ 87	 }$&& L1188 H$\alpha$\,73\\
J22210934+6120172	&	   2494	 $\sb{-839 }\sp{+ 1901		 }$&& L1188 H$\alpha$\,74\\
AllWISE J222142.33+613058.6	&  3497	 $\sb{-940 }\sp{+ 1714		 }$&Class II&\\
AllWISE J222223.60+613252.0 	&  1364	 $\sb{-87   }\sp{+ 99	 }$&Class II&\\
J22222816+6154302	&	   236	 $\sb{-6   }\sp{+ 6	 }$& &L1188 H$\alpha$\,76\\

\hline
\end{tabular}
\end{table*}

\label{lastpage}
\end{document}